\documentclass[journal,10pt,draftclsnofoot,onecolumn]{IEEEtran}

\usepackage[utf8]{inputenc} 
\usepackage[T1]{fontenc}    
\usepackage{url}            
\usepackage{booktabs}       
\usepackage{amsmath}
\usepackage{amssymb}
\usepackage{amsfonts}       
\usepackage{amsthm}
\usepackage{bbm}
\usepackage{nicefrac}       
\usepackage{microtype}      
\usepackage{graphicx}
\usepackage{cite}
\usepackage{bm}
\usepackage{mathtools}
\usepackage{multirow}
\usepackage{enumitem}
\usepackage{algorithm}
\usepackage{algpseudocode}
\usepackage{xcolor}
\usepackage{hyperref}       
\hypersetup{colorlinks=true,allcolors=blue!50!black}

\theoremstyle{definition}
\newtheorem{lem}{Lemma}
\newtheorem{theorem}{Theorem}
\newtheorem{proposition}{Proposition}
\newtheorem{corollary}{Corollary}

\newtheorem*{definition*}{Definition}

\DeclareMathOperator*{\argmin}{arg\,min}

\begin{document}

\title{MML Probabilistic Principal Component Analysis}

\author{Enes~Makalic
        and~Daniel~F.~Schmidt%
\thanks{E. Makalic is with the Faculty of Information Technology, Monash University, e-mail: enes.makalic@monash.edu}%
\thanks{D. F. Schmidt is with the Faculty of Information Technology, Monash University, e-mail: daniel.schmidt@monash.edu}%
}

\maketitle

\begin{abstract}
Principal component analysis (PCA) is perhaps the most widely used method for data dimensionality reduction. A key question in PCA is deciding how many components to retain. This manuscript describes a new approach to automatically selecting the number of principal components based on the  Bayesian minimum message length method of inductive inference. We derive new estimates of the isotropic residual variance and loadings, and establish their bias and asymptotic distributions. For moderate latent dimensions, the residual variance estimate improves on maximum likelihood, and the implied eigenvalue shrinkage approaches the Stein-optimal shrinkage for strong signals.
\end{abstract}


\IEEEpeerreviewmaketitle

\section{Introduction}
\label{sec:intro}
%
Principal component analysis (PCA) is a fundamental tool for dimension reduction and covariance modelling in multivariate statistics~\cite{Jolliffe02}. In the probabilistic formulation of PCA introduced by Tipping and Bishop~\cite{TippingBishop99}, the  observations ${\bf x}_1,\ldots,{\bf x}_N \in \mathbb{R}^K$ are modelled as
\begin{equation}
\label{eqn:mfa}
    {\bf x}_n = \bm{\mu} + {\bf A}{\bf v}_n + \bm{\epsilon}_n,
    \qquad
    {\bf v}_n \sim \mathcal{N}({\bf 0}_J,{\bf I}_J),
    \qquad
    \bm{\epsilon}_n \sim \mathcal{N}({\bf 0}_K,\sigma^2{\bf I}_K),
\end{equation}
where $\bm{\mu} \in \mathbb{R}^K$ is the population mean, ${\bf A}\in\mathbb{R}^{K\times J}$ is the loading matrix, ${\bf v}_n\in\mathbb{R}^J$ are latent scores, and $\sigma^2>0$ is the isotropic residual variance. The latent scores and errors are mutually independent, and the pairs $({\bf v}_n,\bm{\epsilon}_n)$ are independent across $n=1,\ldots,N$. Marginalising over the latent scores yields the Gaussian probabilistic PCA model
\begin{equation}
\label{eqn:y:marginal}
    {\bf x}_n \sim \mathcal{N}(\bm{\mu},\bm{\Sigma}),
    \qquad
    \bm{\Sigma} = {\bf A}{\bf A}^\prime + \sigma^2{\bf I}_K.
\end{equation}
The idea is to model a potentially high-dimensional covariance
matrix $\bm{\Sigma}$ through a low-rank matrix ${\bf A}$ plus an isotropic noise term. The term ${\bf A}{\bf A}^\prime$ captures the covariation shared across variables along $J$ latent directions, while $\sigma^2{\bf I}_K$ describes variation that is specific to each variable and of equal magnitude in every direction. This structure is parsimonious since an unrestricted covariance matrix has $K(K+1)/2$ free parameters, whereas the covariance in \eqref{eqn:y:marginal} grows only linearly in $K$ for fixed $J$.  Given data $\textbf{X} = ({\bf x}_1,\ldots,{\bf x}_N)^\prime \in \mathbb{R}^{N\times K}$, the task is to estimate the
population mean $\bm{\mu}$, the loading matrix ${\bf A}$, the residual
variance $\sigma^2$ and the latent dimension $J$. 

It is well known that the loading matrix $\textbf{A}$ is identifiable only up to right multiplication by an orthogonal matrix, since 
\[ 
\mathbf A\mathbf A^\prime = (\mathbf A\mathbf H)(\mathbf A\mathbf H)^\prime, \qquad 
\mathbf H\in O(J),
\qquad
O(J) = 
\{ 
\textbf{H} \in \mathbb{R}^{J \times J}: 
\textbf{H}^\prime \textbf{H} = \textbf{H} \textbf{H}^\prime = \textbf{I}_J 
\}
\] 
where $O(J)$ is the orthogonal group consisting of all $(J \times J)$ matrices whose columns and rows are orthonormal vectors.
Consequently, we parameterise the identifiable covariance contribution $\mathbf{A}\mathbf{A}^\prime$ by its ordered spectral decomposition: 
\begin{align} 
\mathbf{A}\mathbf{A}^\prime 
= \mathbf{R}\boldsymbol{\Lambda}^2\mathbf{R}^\prime, 
\qquad 
\mathbf{R} \in\operatorname{St}(K,J) ,
\label{eqn:loading-spectral} 
\end{align}
where
\begin{align}
    \bm{\Lambda} = \operatorname{diag}(\bm{\lambda}) = \operatorname{diag}(\lambda_1, \ldots, \lambda_J) \in \mathbb{R}^{J \times J}, 
    \qquad\lambda_j > 0,
\end{align}
and $\operatorname{St}(K,J)$ denotes the Stiefel manifold 
\begin{align}
    \operatorname{St}(K,J) = \{ {\bf R} \in \mathbb{R}^{K \times J} : {\bf R}^\prime {\bf R} = {\bf I}_J \} ,
\end{align}
which is the set of all $(K \times J)$ matrices whose columns are orthonormal. Equivalently, we select $\mathbf A=\mathbf R\boldsymbol\Lambda$ as a canonical local representative.
The matrix $\mathbf{R}$ specifies the ordered signal directions, while the diagonal entries of $\boldsymbol{\Lambda}$ are their positive loading magnitudes.
With this parameterisation, the population covariance matrix can  be written as  
\begin{align}
\label{eqn:Sigma}
    \bm{\Sigma} = {\bf R} \bm{\Lambda}^2 {\bf R}^\prime + \sigma^2 {\bf I}_K = {\bf R} \,\textbf{L} \,{\bf R}^\prime + \sigma^2 {\bf I}_K,
    \qquad
    \textbf{L} = \bm{\Lambda}^2 = \operatorname{diag}(\lambda_1^2,\ldots,\lambda_J^2) .
\end{align}
%
%
The population eigenvalues of $\bm{\Sigma}$ are 
\[
(\lambda_1^2 + \sigma^2, \ldots, \lambda_J^2 + \sigma^2, \underbrace{\sigma^2,  \ldots, \sigma^2}_{(K-J) \, {\rm times}}) , 
\]
suggesting that probabilistic PCA may be viewed as a low-rank perturbation of a spherical covariance model, or equivalently as a spiked covariance model with $J$ signal eigenvalues and $(K-J)$ noise eigenvalues. We assume that $1\leq J<K$ (the zero-dimensional model is treated separately in Section~\ref{sec:mml:concentrated}), $\sigma>0$, and $\lambda_1>\cdots>\lambda_J>0$ so that all signal eigenvectors are locally identifiable.

As discussed above, a central inferential problem in PCA is determination of the latent dimension $J$. This issue is practically important because underestimation of $J$ discards signal, whereas overestimation weakens interpretability and may substantially inflate estimation error in finite samples. In the classical regime with fixed $K$ and $N\to\infty$, the sample eigenvalues consistently estimate the population eigenvalues; however, for small to moderate sample sizes the empirical noise eigenvalues can exhibit substantial dispersion, making separation of signal and noise difficult~\cite{KritchmanNadler08}. A large literature has therefore developed around estimating the number of retained components, including likelihood-based criteria, scree-plot based methods, and procedures motivated by random matrix theory; see, among others,~\cite{Minka00,HungEtAl22,CaiEtAl20,HongEtAl23}.

In this paper we study probabilistic PCA under the minimum message length (MML) principle of inductive inference~\cite{WallaceFreeman87,Wallace05}. MML treats statistical inference as a coding problem and balances model complexity against goodness of fit through a two-part message. Although MML has been used previously in factor-analytic settings~\cite{WallaceFreeman92,Wallace98b}, a corresponding treatment of probabilistic PCA that works directly with the Gaussian marginal model~\eqref{eqn:y:marginal} and respects the geometry of the loading matrix has not, to our knowledge, been developed.

The main contributions of this paper are as follows.
\begin{itemize}
    \item We derive the MML87 codelength for probabilistic PCA from the marginal Gaussian model~\eqref{eqn:y:marginal}, using an identifiable ordered spectral parameterisation, its expected Fisher information, and a rotation-invariant matrix-variate Cauchy prior pushed forward to that parameterisation. The resulting estimators share the sample mean and signal directions with maximum likelihood, but shrink each fitted signal eigenvalue and increase the residual variance estimate; they are computed by a one-dimensional fixed-point iteration with an exact existence test, and the same codelength selects the latent dimension.

    \item We give an explicit finite-sample condition for the existence of a regular $J$-component MML87 solution. In the proportional high-dimensional regime its limit lies strictly above the upper edge of the Marchenko--Pastur bulk, so a component is retained only when its sample eigenvalue is separated from the noise spectrum.

    \item We characterise the residual variance and loading-strength estimators in both asymptotic regimes. For fixed dimension, MML87 replaces the downward first-order bias of maximum likelihood with a positive bias, which is smaller in magnitude unless $J$ is a substantial fraction of $K$; we also give the second-order bias of both estimators (Corollary~\ref{cor:bias:second}), and both estimators have the same first-order normal limit. In the proportional regime, the MML87 residual variance estimator has the same asymptotic variance but a typically much smaller centring, and its loading shrinkage has smaller limiting Stein loss than maximum likelihood at every signal strength above the existence threshold, with excess over the optimal shrinker of~\cite{DonohoGavishJohnstone18} of order $\rho_J^{-2}$, where $\rho_J$ is the signal-to-noise ratio of the weakest component (Proposition~\ref{prop:stein}, Corollary~\ref{cor:stein:improvement} and Figure~\ref{fig:shrinkers}).
\end{itemize}

The proposed methodology is particularly useful when the latent dimension must be selected automatically and accurate estimation of the residual variance is important. Numerical experiments show that, for a fixed latent dimension, MML87 improves residual variance estimation relative to maximum likelihood and gives smaller Stein loss, while maximum likelihood gives slightly smaller entropy loss when the signal is strong, in agreement with Proposition~\ref{prop:stein}. When the latent dimension is unknown, MML87 retains a component essentially only when its sample eigenvalue is far enough above the noise level for a regular solution to exist. It almost never selects spurious components, but it is conservative near the detection limit. The code implementing the MML methodology in this article is available in the GitHub repository 
\href{https://github.com/EnesMakalic/MML-PCA}{https://github.com/EnesMakalic/MML-PCA}.

The remainder of the paper is organised as follows. Section~\ref{sec:mml:pca} develops the MML formulation of probabilistic PCA, including the identifiable spectral parameterisation, Fisher information, prior specification, and concentrated codelength. Section~\ref{sec:mml:sigma2} derives the MML estimator of the residual variance, describes its computation, and establishes the threshold for the existence of a regular solution. Section~\ref{sec:finite-sample-bias} analyses the bias and asymptotic distribution of the estimator for fixed dimension, and Section~\ref{sec:stein} studies its asymptotic loss and distribution in the proportional high-dimensional regime. Section~\ref{sec:experiments} reports simulation studies of parameter estimation, a comparison with alternative estimators, and latent-dimension selection. Section~\ref{sec:discussion} concludes with a discussion of the main findings and directions for future work. Appendix~\ref{sec:supp:fisher} derives the expected Fisher information, Appendix~\ref{sec:supp:proofs} gives the proofs and numerical checks of the theoretical results, and Appendix~\ref{sec:supp:compare} gives the full results of the comparison with alternative estimators. 
\section{Minimum message length analysis of the PCA model}
\label{sec:mml:pca}
%
We analyse probabilistic PCA under the minimum message length (MML) principle of inductive inference; see \cite{WallaceFreeman87,Wallace05}. MML treats inference as a coding problem and selects the model that minimises the total codelength required to encode both the model and the data. In the present setting, this yields a penalised likelihood criterion that simultaneously estimates the loading matrix, the residual variance, and the latent dimension.

The gold standard MML codelength is strict MML~\cite{WallaceBoulton75}, which is unfortunately intractable for general model classes~\cite{FarrWallace02}. In this manuscript, we utilise a computationally tractable approximation to strict MML, referred to as MML87~\cite{WallaceFreeman87}, that closely approximates strict MML for sufficiently regular models. Its derivation assumes that the prior and likelihood vary slowly over the region of parameter uncertainty. This holds in the interior of the parameter space of the PCA model, but fails near the singular boundaries where a loading magnitude vanishes or two coincide; Section~\ref{sec:mml:concentrated} explains how such boundaries are handled. Under standard regularity conditions, the MML87~\cite{WallaceFreeman87} codelength for data ${\bf X}$ and parameter vector $\bm{\theta}\in\mathbb{R}^P$ is
\begin{equation}
\label{eqn:mml87:codelength}
I_{87}({\bf X},\bm{\theta})
=
-\log \pi(\bm{\theta})
+\frac{1}{2}\log |\mathcal{F}(\bm{\theta})|
+\frac{P}{2}\log \kappa_P
+\frac{P}{2}
-\log p({\bf X}\mid \bm{\theta}),
\end{equation}
where $\pi(\bm{\theta})$ is the prior density, $|\mathcal{F}(\bm{\theta})|$ is the determinant of the expected Fisher information matrix, and $\kappa_P$ is the lattice quantisation constant. For $P\leq16$ we use the tabulated normalised second moments of the best known lattice quantisers~\cite{ConwaySloane98,AgrellAllen23}, and for $P>16$ the approximation 
\begin{equation} 
\label{eqn:kappa:approx} 
\frac{P}{2}\{\log\kappa_P+1\} \approx -\frac{P}{2}\log(2\pi) +\frac{1}{2}\log(P\pi) -\gamma_{\mathrm{E}}, 
\end{equation} 
where \(\gamma_{\mathrm{E}}\approx0.5772\) is the Euler--Mascheroni constant. The same convention is used for every candidate rank, since this term depends on $J$ through $P$. 
%
%
Our task is to derive the codelength \eqref{eqn:mml87:codelength} for the probabilistic PCA model. We work throughout with the identifiable parameter vector 
\begin{equation} \boldsymbol{\theta} = (\boldsymbol{\mu},\boldsymbol{\lambda},\sigma,\mathbf{R}), \qquad \mathbf{R}\in\mathcal{R}_{K,J} 
= \operatorname{St}(K,J)/\mathcal{D}_J, 
\label{eqn:theta:parameterisation} \end{equation} 
where 
\[ 
\boldsymbol{\mu}\in\mathbb{R}^K, \qquad \lambda_1>\cdots>\lambda_J>0, \qquad \sigma>0, 
\] 
and 
\begin{equation} \mathcal{D}_J = \left\{ \operatorname{diag}(d_1,\ldots,d_J): d_j\in\{-1,1\},\quad j=1,\ldots,J \right\} \label{eqn:sign-change-group} \end{equation} 
is the group of independent sign changes of the columns of $\mathbf{R}$. Thus, $\mathcal{R}_{K,J}$ identifies $\mathbf{R}$ and $\mathbf{R}\mathbf{D}$ for every $\mathbf{D}\in\mathcal{D}_J$. The ordering of the loading magnitudes removes the permutation ambiguity, while the quotient by $\mathcal{D}_J$ removes the remaining sign ambiguity. The number of free covariance parameters is 
\begin{align} 
P_{\Sigma} 
&= J+1+\dim\{\operatorname{St}(K,J)\} 
= KJ-\frac{J(J-1)}{2}+1. 
\label{eqn:covariance:dimension} 
\end{align} 
Including the $K$ mean parameters, the total parameter dimension is 
\begin{equation} 
\label{eqn:full:dimension} 
P = K(J+1)-\frac{J(J-1)}{2}+1. 
\end{equation} 
%
%
%
%
\subsection{Maximum likelihood estimates}
\label{sec:ml}
We first record the standard maximum likelihood estimates for
probabilistic PCA; see \cite{TippingBishop99}. Define 
\begin{align}
\mathbf{S}_{\boldsymbol{\mu}}
=
\frac{1}{N}
\sum_{n=1}^{N}
(\mathbf{x}_n-\boldsymbol{\mu})
(\mathbf{x}_n-\boldsymbol{\mu})^{\prime}.
\end{align}
For the population covariance model \eqref{eqn:Sigma}
the log-likelihood of data $\textbf{X}=(\textbf{x}_1,\ldots,\textbf{x}_N)^\prime \in \mathbb{R}^{N \times K}$ is
\begin{align}
\log p(\mathbf{X}\mid\boldsymbol{\mu},
\boldsymbol{\lambda},\sigma,\mathbf{R})
=
-\frac{NK}{2}\log(2\pi)
-\frac{N}{2}\log|\boldsymbol{\Sigma}|
-\frac{N}{2}
\operatorname{tr}
\left(
\boldsymbol{\Sigma}^{-1}
\mathbf{S}_{\boldsymbol{\mu}}
\right).
\label{eqn:ml:loglikelihood}
\end{align}
Let
\begin{align}
\overline{\mathbf{x}}
=
\frac{1}{N}\sum_{n=1}^{N}\mathbf{x}_n,
\qquad
\mathbf{S}
=
\frac{1}{N}
\sum_{n=1}^{N}
(\mathbf{x}_n-\overline{\mathbf{x}})
(\mathbf{x}_n-\overline{\mathbf{x}})^{\prime},
\end{align}
denote the sample mean and sample covariance, and write the eigendecomposition of the sample covariance as
\begin{align}
\mathbf{S}
=
\mathbf{U}\operatorname{diag}(s_1,\ldots,s_K)\mathbf{U}^{\prime},
\qquad
\mathbf{U}=[\mathbf{u}_1,\ldots,\mathbf{u}_K],
\qquad
s_1\geq\cdots\geq s_K\geq0.
\end{align}
For a fixed latent dimension \(J<K\), and writing \(\tau=\sigma^2\), the maximum likelihood
estimates are
\begin{align}
\widehat{\boldsymbol{\mu}}_{\mathrm{ML}}
&=
\overline{\mathbf{x}},
\label{eqn:ml:mu}
\\
\widehat{\sigma}_{\mathrm{ML}}^2
&=
\widehat{\tau}_{\mathrm{ML}}
=
\frac{1}{K-J}
\sum_{j=J+1}^{K}s_j,
\label{eqn:ml:tau}
\\
\widehat{\mathbf{A}}_{\mathrm{ML}}
&=
\mathbf{U}_J
\left(
\mathbf{S}_J
-
\widehat{\tau}_{\mathrm{ML}}\mathbf{I}_J
\right)^{1/2}
\mathbf{O},
\label{eqn:ml:A}
\end{align}
where
\[
\mathbf{U}_J
=
[\mathbf{u}_1,\ldots,\mathbf{u}_J],
\qquad
\mathbf{S}_J
=
\operatorname{diag}(s_1,\ldots,s_J),
\]
and \(\mathbf{O}\in O(J)\) is arbitrary. 
In the ordered identifiable parameterisation $\mathbf{A}=\mathbf{R}\boldsymbol{\Lambda}$, these estimates are equivalently
\begin{align}
\widehat{\mathbf{R}}_{\mathrm{ML}}
=
\mathbf{U}_J,
\qquad
\widehat{\lambda}_{j,\mathrm{ML}}^2
=
s_j-\widehat{\tau}_{\mathrm{ML}},
\qquad j=1,\ldots,J.
\label{eqn:ml:R:lambda}
\end{align}
The regular \(J\)-component solution requires
\(\widehat{\tau}_{\mathrm{ML}}<s_J\). Among the stationary choices of $J$ sample eigenvectors, the likelihood is maximised by pairing the $J$ fitted signal eigenvalues with the $J$ largest sample eigenvalues. Hence the maximising signal subspace is spanned by $\mathbf u_1,\ldots,\mathbf u_J$; all other stationary choices have no larger likelihood and are generally saddle points. 

\subsection{Fisher information}
\label{sec:fisher}

For the multivariate Gaussian model, the mean and covariance parameters are orthogonal under the expected Fisher information. Hence,
\begin{equation}
\mathcal{F}_{\boldsymbol{\mu},\boldsymbol{\Sigma}}
=
\mathbf{0},
\qquad
\left|\mathcal{F}\right|
=
\left|\mathcal{F}_{\mu\mu}\right|
\left|\mathcal{F}_{\Sigma\Sigma}\right|.
\label{eqn:fisher:factorisation}
\end{equation}
The contribution from the mean parameters is
\begin{align}
\left|\mathcal{F}_{\mu\mu}\right|
=
N^K|\boldsymbol{\Sigma}|^{-1}
=
\frac{N^K}{\sigma^{2(K-J)}}
\prod_{j=1}^J
\frac{1}{\lambda_j^2+\sigma^2}.
\label{eqn:fisher:mean}
\end{align}
For a general covariance parameterisation, the expected Fisher information has entries
\begin{equation}
\mathcal{F}_{a,b}
=
\frac{N}{2}
\operatorname{tr}
\left(
\boldsymbol{\Sigma}^{-1}
\frac{\partial\boldsymbol{\Sigma}}{\partial\theta_a}
\boldsymbol{\Sigma}^{-1}
\frac{\partial\boldsymbol{\Sigma}}{\partial\theta_b}
\right).
\label{eqn:fisher:general}
\end{equation}
Under the ordered spectral parameterisation
\[
\boldsymbol{\Sigma}
=
\mathbf{R}\boldsymbol{\Lambda}^2\mathbf{R}^{\prime}
+
\sigma^2\mathbf{I}_K,
\qquad
\boldsymbol{\Lambda}
=
\operatorname{diag}(\lambda_1,\ldots,\lambda_J),
\]
the covariance perturbations separate into three mutually orthogonal classes: changes in the loading magnitudes and residual scale, rotations among the signal directions, and rotations between the signal and noise subspaces. Relative to the corresponding invariant local volume element on
$\mathcal{R}_{K,J}=\operatorname{St}(K,J)/\mathcal{D}_J$, the
covariance-parameter Fisher determinant is
\begin{align}
\left|\mathcal{F}_{\Sigma\Sigma}\right|
&=
\frac{
N^{P_{\Sigma}}2^{J+1}(K-J)
}{
\sigma^{2\{J(K-J)+1\}}
}
\prod_{j=1}^J
\frac{
\lambda_j^{4(K-J)+2}
}{
(\lambda_j^2+\sigma^2)^{K+1}
}
\prod_{1\leq j<k\leq J}
(\lambda_j^2-\lambda_k^2)^2,
\label{eqn:fisher:covariance}
\end{align}
where
\[
P_{\Sigma}
=
KJ-\frac{J(J-1)}{2}+1
\]
is the number of free covariance parameters. The complete derivation of \eqref{eqn:fisher:covariance}, including the three orthogonal Fisher blocks, is given in Appendix~\ref{sec:supp:fisher}.
Combining \eqref{eqn:fisher:mean} and
\eqref{eqn:fisher:covariance} gives
\begin{align}
\left|\mathcal{F}\right|
&=
\frac{
N^P2^{J+1}(K-J)
}{
\sigma^{2\{(K-J)(J+1)+1\}}
}
\prod_{j=1}^J
\frac{
\lambda_j^{4(K-J)+2}
}{
(\lambda_j^2+\sigma^2)^{K+2}
}
\prod_{1\leq j<k\leq J}
(\lambda_j^2-\lambda_k^2)^2,
\label{eqn:fisher:full}
\end{align}
where
\[
P
=
K(J+1)-\frac{J(J-1)}{2}+1.
\]

The determinant vanishes if $\lambda_j=0$ or if
$\lambda_j=\lambda_k$ for some $j\neq k$. In the first case, the
corresponding signal component is indistinguishable from the
isotropic noise space. In the second, the associated signal
eigenvectors are not separately identifiable, although their joint
eigenspace may remain identifiable.

\subsection{Prior specification}
\label{sec:priors}
We specify the prior distribution over all parameters relative to the same product reference measure used in the Fisher determinant,
\[
d\boldsymbol{\mu}\,
d\lambda_1\cdots d\lambda_J\,
d\sigma\,
dV_{K,J}(\mathbf{R}),
\]
where \(dV_{K,J}(\mathbf{R})\) is the invariant volume element on $\mathcal{R}_{K,J}$.
The prior distribution factorises as
\begin{equation}
\pi(\boldsymbol{\mu},\boldsymbol{\lambda},\sigma,\mathbf{R})
=
\pi_{\boldsymbol{\mu}}(\boldsymbol{\mu})\,
\pi_{\sigma}(\sigma)\,
\pi_{\boldsymbol{\lambda}\mid\sigma}
(\boldsymbol{\lambda}\mid\sigma)\,
\pi_{\mathbf{R}}(\mathbf{R}).
\label{eqn:prior:factorisation}
\end{equation}

\subsubsection{Mean and residual scale}
For the mean and residual scale, we use
\begin{equation}
\pi_{\boldsymbol{\mu}}(\boldsymbol{\mu})
=
\Omega_{\mu}^{-1},
\qquad
\pi_{\sigma}(\sigma)
=
(\Omega_{\sigma}\sigma)^{-1},
\label{eqn:prior:mu:sigma}
\end{equation}
on specified bounded supports \(\boldsymbol{\mu}\in\mathcal{M}_{\mu}\) and \(\sigma\in\mathcal{S}_{\sigma}\), with the support of \(\sigma\) bounded away from zero. The constants \(\Omega_{\mu}\) and \(\Omega_{\sigma}\) do not depend on \(J\), provided the same support and reference-measure conventions are used for every candidate model. Throughout, we assume that these supports are wide enough not to bind at the estimates, so that all formulas below concern interior solutions, and asymptotic statements assume that the true $\boldsymbol{\mu}$ and $\sigma$ lie in the interiors of the supports.

\subsubsection{Loading magnitudes}

A prior imposed directly on the orthogonal loading representation should not be interpreted as assuming that the underlying loading vectors are orthogonal a priori. Instead, we first assign a rotation-invariant prior to an unrestricted loading matrix and then push this prior forward to the identifiable ordered spectral parameterisation.
Let
$\widetilde{\mathbf{A}}\in\mathbb{R}^{K\times J}$
denote an unrestricted (true) loading matrix and define the dimensionless (true) scaled loading matrix
$\widetilde{\mathbf{B}} = \widetilde{\mathbf{A}} /\sigma$.
We assign \(\widetilde{\mathbf{B}}\) the matrix-variate Cauchy density
\begin{equation}
\pi_{\widetilde{\mathbf{B}}}
(\widetilde{\mathbf{B}})
=
c_{K,J}
\left|
\mathbf{I}_J+
\widetilde{\mathbf{B}}^{\prime}
\widetilde{\mathbf{B}}
\right|^{-(K+J)/2},
\qquad
c_{K,J}
=
\frac{
\Gamma_J\{(K+J)/2\}
}{
\pi^{KJ/2}\Gamma_J(J/2)
}.
\label{eqn:prior:matrix:cauchy}
\end{equation}
where $\Gamma_J(\cdot)$ is the multivariate gamma function~\cite{Muirhead82}. This prior is proper, heavy-tailed, and invariant under
\[
\widetilde{\mathbf{B}}
\longmapsto
\mathbf{Q}\widetilde{\mathbf{B}}\mathbf{O},
\qquad
\mathbf{Q}\in O(K),
\quad
\mathbf{O}\in O(J).
\]
It therefore assigns no preferred directions in either the observation or latent spaces. Scaling by \(\sigma\) also makes the prior invariant to a common change in the measurement units.
Let
\[
\widetilde{\mathbf{B}}^{\prime}\widetilde{\mathbf{B}}
=
\mathbf{W}
\operatorname{diag}(\vartheta_1^2,\ldots,\vartheta_J^2)
\mathbf{W}^{\prime},
\qquad
\vartheta_1>\cdots>\vartheta_J>0 ,
\]
where \(\vartheta_j\) are the ordered singular values of
\(\widetilde{\mathbf{B}}\). Write the singular value decomposition as $\widetilde{\mathbf B}=\mathbf R\operatorname{diag}(\vartheta_1,\ldots,\vartheta_J)\mathbf W^{\prime}$, with $\mathbf R\in\operatorname{St}(K,J)$. Then $\widetilde{\mathbf B}\widetilde{\mathbf B}^{\prime}=\mathbf R\operatorname{diag}(\vartheta_1^2,\ldots,\vartheta_J^2)\mathbf R^{\prime}$, so the covariance contribution depends on $\widetilde{\mathbf B}$ only through $(\mathbf R,\vartheta_1,\ldots,\vartheta_J)$, and $\widetilde{\mathbf B}\mathbf W=\mathbf R\operatorname{diag}(\vartheta_1,\ldots,\vartheta_J)$ is a canonical representative of its right-orthogonal equivalence class.
The induced density of the ordered scaled loading magnitudes is
\begin{align}
\pi_{\boldsymbol{\vartheta}}
(\vartheta_1,\ldots,\vartheta_J)
&=
C_{K,J}^{\mathrm{ord}}
\prod_{j=1}^J
\frac{
\vartheta_j^{K-J}
}{
(1+\vartheta_j^2)^{(K+J)/2}
}
\prod_{1\leq j<k\leq J}
(\vartheta_j^2-\vartheta_k^2),
\label{eqn:prior:rho}
\\
C_{K,J}^{\mathrm{ord}}
&=
\frac{
2^J\pi^{J^2/2}
}{
\Gamma_J(J/2)\mathrm{B}_J(K/2,J/2)
},
\label{eqn:prior:rho:constant}
\end{align}
for \(\vartheta_1>\cdots>\vartheta_J>0\), where
\[
\mathrm{B}_J(a,b)
=
\frac{\Gamma_J(a)\Gamma_J(b)}{\Gamma_J(a+b)}
\]
is the multivariate beta function. The density \eqref{eqn:prior:rho} follows from the Jacobian of the singular value decomposition with ordered singular values \cite{Muirhead82}, integrated over the right singular vectors in $O(J)$, and it integrates to one over the ordered region. No additional labelling factor arises, because ordering the singular values already removes the permutation ambiguity. The Vandermonde factor reflects the loss of identifiability when two signal magnitudes coincide.
Since
$\lambda_j=\sigma\vartheta_j$, $\vartheta_j=\lambda_j / \sigma$,
the induced conditional prior on the unscaled loading magnitudes is
\begin{align}
\pi_{\boldsymbol{\lambda}\mid\sigma}
(\lambda_1,\ldots,\lambda_J\mid\sigma)
&=
C_{K,J}^{\mathrm{ord}}\,
\sigma^{J^2}
\prod_{j=1}^J
\frac{
\lambda_j^{K-J}
}{
(\sigma^2+\lambda_j^2)^{(K+J)/2}
}
\prod_{1\leq j<k\leq J}
(\lambda_j^2-\lambda_k^2),
\label{eqn:prior:lambda}
\end{align}
for $\lambda_1>\cdots>\lambda_J>0$.

\subsubsection{Signal orientation}

The ordered loading magnitudes remove the permutation ambiguity, but each column of \(\mathbf{R}\) remains identifiable only up to sign. The orientation parameter therefore belongs to $\mathcal{R}_{K,J}$.
Under the standard invariant-volume convention,
\[
\operatorname{Vol}\{\operatorname{St}(K,J)\}
=
\frac{2^J\pi^{KJ/2}}{\Gamma_J(K/2)}.
\]
Since \(\mathcal{D}_J\) contains \(2^J\) elements, the quotient has
volume
\begin{equation}
\operatorname{Vol}(\mathcal{R}_{K,J})
=
\frac{\pi^{KJ/2}}{\Gamma_J(K/2)}.
\label{eqn:orientation:volume}
\end{equation}
We assign a uniform prior to the signal orientation and, relative to \(dV_{K,J}(\mathbf{R})\), its density is 
\begin{equation}
\pi_{\mathbf{R}}(\mathbf{R})
=
\frac{\Gamma_J(K/2)}{\pi^{KJ/2}},
\qquad
\mathbf{R}\in\mathcal{R}_{K,J}.
\label{eqn:prior:R}
\end{equation}

\subsubsection{Complete prior}
Combining \eqref{eqn:prior:mu:sigma},
\eqref{eqn:prior:lambda}, and \eqref{eqn:prior:R} gives the complete prior distribution over all parameters
\begin{align}
\pi(\boldsymbol{\mu},\boldsymbol{\lambda},\sigma,\mathbf{R})
&=
\frac{
C_{K,J}^{\mathrm{ord}}
}{
\Omega_{\mu}\Omega_{\sigma}
}
\frac{
\Gamma_J(K/2)
}{
\pi^{KJ/2}
}
\sigma^{J^2-1}
\prod_{j=1}^J
\frac{
\lambda_j^{K-J}
}{
(\sigma^2+\lambda_j^2)^{(K+J)/2}
}
\prod_{1\leq j<k\leq J}
(\lambda_j^2-\lambda_k^2),
\label{eqn:prior:complete}
\end{align}
for
$\boldsymbol{\mu}\in\mathcal{M}_{\mu}$,
$\sigma\in\mathcal{S}_{\sigma}$,
$\lambda_1>\cdots>\lambda_J>0$, and
$\mathbf{R}\in\mathcal{R}_{K,J}$.

\subsection{Concentrated MML87 codelength}
\label{sec:mml:concentrated}
As the prior on $\boldsymbol{\mu}$ is uniform and the remaining terms in the MML87 codelength do not depend on $\boldsymbol{\mu}$, the MML87 estimate of the population mean is
\begin{equation}
\widehat{\boldsymbol{\mu}}_{\mathrm{MML}}
=
\overline{\mathbf{x}}.
\end{equation}
Moreover, the prior and Fisher information terms are invariant under changes in the signal orientation. Optimisation over $\mathbf R$ is therefore determined by the likelihood. By the trace rearrangement inequality, the fitted signal eigenvalues are paired with the leading sample eigenvalues, giving
\begin{equation}
\widehat{\mathbf R}_{\mathrm{MML}}
=
\mathbf U_J
=
[\mathbf u_1,\ldots,\mathbf u_J],
\label{eqn:mml:mu:R}
\end{equation}
up to independent sign changes of its columns.
Let
\[
x_j=\lambda_j^2,
\qquad
j=1,\ldots,J,
\qquad
\mathbf{x}=(x_1,\ldots,x_J)^{\prime},
\]
and call $x_j$ the loading strengths.
After substituting \eqref{eqn:mml:mu:R} into the complete MML87 codelength and cancelling the Vandermonde terms in the prior and Fisher determinant, we obtain
\begin{align}
I_{87}(\mathbf{X},\mathbf{x},\tau;J)
&=
C_{K,J,N}^{\mathrm{MML}}
+
\frac{A}{2}\log\tau
+
\frac{N}{2\tau}\sum_{k=J+1}^K s_k
+
\frac{1}{2}\sum_{j=1}^J
\left[
(N+J-2)\log(\tau+x_j)
+
\frac{Ns_j}{\tau+x_j}
+
C\log x_j
\right],
\label{eqn:mml:preprofile}
\end{align}
where
\begin{equation}
A=N(K-J)-K(J+1)+J,
\qquad
C=K-J+1,
\end{equation}
and 
\begin{align} 
C_{K,J,N}^{\mathrm{MML}} 
&= \log(\Omega_{\mu}\Omega_{\sigma}) +\frac{P}{2}\log N +\frac{NK}{2}\log(2\pi) +\frac{P}{2}\{\log\kappa_P+1\} \nonumber\\ &\quad 
+\frac{1-J}
{2}\log 2 +\frac{1}{2}\log(K-J) +\frac{J(K-J)}{2}\log\pi \nonumber\\ &\quad 
+2\log\Gamma_J(J/2) -\log\Gamma_J\{(K+J)/2\}, \label{eqn:mml:constant} 
\end{align} 
with  \[ P = K(J+1)-\frac{J(J-1)}{2}+1. \]
The zero-dimensional model is the special case $J=0$ with $P=K+1$ and
\begin{align*}
I_{87}(\mathbf{X}, \tau;0)&=C_{K,0,N}^{\mathrm{MML}}+\frac{K(N-1)}{2}\log\tau+\frac{N}{2\tau}\operatorname{tr}(\mathbf S),\\
C_{K,0,N}^{\mathrm{MML}}&=\log(\Omega_\mu\Omega_\sigma)+\frac{K+1}{2}\log N+\frac{NK}{2}\log(2\pi)+\frac{K+1}{2}\{\log\kappa_{K+1}+1\}+\frac12\log(2K),
\end{align*}
which is minimised at $\widehat\tau_0=N\operatorname{tr}(\mathbf S)/\{K(N-1)\}$. When the latent dimension is selected, we place a uniform prior on the candidate set $\{0,1,\ldots,J_{\max}\}$. Its codelength $\log(J_{\max}+1)$ is common to all ranks and is omitted, as are the support constants $\log(\Omega_\mu\Omega_\sigma)$.
For fixed $\tau$, the loading magnitudes separate in \eqref{eqn:mml:preprofile}. The stationary equation for $x_j$ is
\begin{equation}
(N+K-1)x_j^2
+
\{(N+2K-J)\tau-Ns_j\}x_j
+
(K-J+1)\tau^2
=
0.
\label{eqn:mml:loading:quadratic}
\end{equation}
Define
\begin{align}
d_j(\tau)
&=
Ns_j-(N+2K-J)\tau,
\\
\Delta_j(\tau)
&=
d_j(\tau)^2
-
4(N+K-1)(K-J+1)\tau^2.
\end{align}
When $d_j(\tau)>0$ and $\Delta_j(\tau)>0$, the regular conditional MML87 loading estimate is the larger root
\begin{equation}
\widehat{x}_j(\tau)
=
\frac{
d_j(\tau)+\sqrt{\Delta_j(\tau)}
}{
2(N+K-1)
}.
\label{eqn:mml:loading:profile}
\end{equation}
The smaller root is a local maximum of the codelength. In contrast to maximum likelihood, for which $x_j=s_j-\tau$, the MML87 solution satisfies
\begin{equation}
0<\widehat{x}_j(\tau)<s_j-\tau.
\label{eqn:mml:loading:shrinkage}
\end{equation}
Thus, MML87 shrinks each fitted signal eigenvalue relative to its maximum likelihood value. The difference is of order $N^{-1}$ under fixed-dimensional asymptotics.
Substitution of \eqref{eqn:mml:loading:profile} into \eqref{eqn:mml:preprofile} gives the regular concentrated codelength
\begin{align}
I_{87}^{\mathrm{conc}}(\mathbf{X}, \tau;J)
=
I_{87}
\left(\mathbf{X},
\widehat{\mathbf{x}}(\tau),\tau;J
\right),
\qquad
\tau\in\mathcal T_J,
%
\qquad
\mathcal T_J
=
\left\{
\tau>0:
d_J(\tau)>0,\;
\Delta_J(\tau)>0
\right\}.
\end{align}
Since $s_1\geq\cdots\geq s_J$, the set $\mathcal T_J$ is an interval $(0,\tau_{J,\max})$, with $\tau_{J,\max}$ given explicitly in \eqref{eqn:mml:tau:interval} below. The regular MML87 estimate $\widehat{\tau}_{\mathrm{MML}}$ is the interior local minimum of
$I_{87}^{\mathrm{conc}}(\mathbf{X},\tau;J)$ over $\mathcal T_J$, which is unique when it exists (Lemma~\ref{lem:mml:unique}). The remaining estimates are
\begin{align}
\widehat{\sigma}_{\mathrm{MML}}^2
&=
\widehat{\tau}_{\mathrm{MML}},
\\
\widehat{\lambda}_{j,\mathrm{MML}}^2
&=
\widehat{x}_j(\widehat{\tau}_{\mathrm{MML}}),
\qquad j=1,\ldots,J,
\\
\widehat{\mathbf A}_{\mathrm{MML}}
&=
\mathbf U_J
\operatorname{diag}
\left\{
\sqrt{\widehat{x}_1(\widehat{\tau}_{\mathrm{MML}})},
\ldots,
\sqrt{\widehat{x}_J(\widehat{\tau}_{\mathrm{MML}})}
\right\}.
\label{eqn:mml:remaining:estimates}
\end{align}
Consequently, MML87 and maximum likelihood estimate the same sample mean and signal directions, but generally give different residual variances and loading magnitudes. 
The unrestricted expression \eqref{eqn:mml:preprofile} has no global minimum, because it tends to $-\infty$ as any $x_j\downarrow0$ with the other parameters fixed. The MML87 approximation is not valid near this singular boundary, where the Fisher information vanishes, and the procedure therefore compares the regular local minima of the candidate ranks.
The MML87 estimates are scale equivariant, and the selected rank is invariant to the units of measurement. Replacing ${\bf x}_n$ by $c\,{\bf x}_n$ for $c>0$ multiplies every $s_k$ by $c^2$, and \eqref{eqn:mml:preprofile} is then minimised at $c^2\widehat\tau_{\mathrm{MML}}$ and $c^2\widehat x_j$, $j=1,\ldots,J$. The coefficient of $\log\tau$ after profiling, counting the $\log(\tau+x_j)$ and $\log x_j$ terms, is
\[
\frac{A+J(N+J-2)+J(K-J+1)}{2}=\frac{K(N-1)}{2},
\]
which does not depend on $J$. Every codelength therefore shifts by the same amount, $K(N-1)\log c$, provided the supports of $\boldsymbol\mu$ and $\sigma$ are rescaled with the data so that the estimates remain interior. This changes $\log\Omega_\mu$ by the same amount for every rank and leaves $\Omega_\sigma$ unchanged. The ordering of the candidate ranks is unchanged.

\section{MML estimation of the residual variance}
\label{sec:mml:sigma2}
For a fixed latent dimension $J$, define
\begin{align}
A
&=
N(K-J)-K(J+1)+J,
&
B
&=
N\sum_{k=J+1}^{K}s_k
=
N(K-J)\widehat{\tau}_{\mathrm{ML}},
\nonumber\\
C
&=
K-J+1,
\label{eqn:mml:ABC}
\end{align}
and assume $A>0$; otherwise no regular solution exists (Lemma~\ref{lem:mml:unique}). For fixed $\tau$, the regular conditional MML87 estimate of $x_j=\lambda_j^2$ is given in \eqref{eqn:mml:loading:profile}. 
From the loading stationarity equations, we have
\begin{equation}
\frac{dI_{87}^{\mathrm{conc}}(\mathbf{X}, \tau;J)}{d\tau}
=
\frac{A}{2\tau}
-
\frac{B}{2\tau^2}
-
\frac{C}{2}
\sum_{j=1}^{J}
\frac{1}{\widehat{x}_j(\tau)},
\label{eqn:mml:tau:derivative}
\end{equation}
so that every regular interior estimate satisfies
\begin{equation}
A\tau-B
=
C\tau^2
\sum_{j=1}^{J}
\frac{1}{\widehat{x}_j(\tau)}.
\label{eqn:mml:tau:stationary}
\end{equation}
The sign of \eqref{eqn:mml:tau:derivative} is that of the scaled score
\[
H(\tau)=2\tau^2\frac{dI_{87}^{\mathrm{conc}}(\mathbf{X}, \tau;J)}{d\tau}
=A\tau-B-C\tau^2\sum_{j=1}^J\widehat x_j(\tau)^{-1},
\qquad \tau\in\mathcal T_J,
\]
which determines the existence and uniqueness of the estimate.

\begin{lem}
\label{lem:mml:unique}
Let $J\geq1$, $B>0$ and $s_J>0$. Then $H$ is strictly concave on $\mathcal T_J$, with $H(0^+)=-B$, $H^{\prime}(0^+)=A$, and $H^{\prime}(\tau)\rightarrow-\infty$ as $\tau\uparrow\tau_{J,\max}$. Consequently, if $A\leq0$, no regular interior local minimum exists. If $A>0$, $H$ has a unique maximiser $\tau^\ast\in\mathcal T_J$, and a regular interior local minimum of $I_{87}^{\mathrm{conc}}(\mathbf{X}, \tau;J)$ exists if and only if $H(\tau^\ast)>0$. It is then unique and is the zero of $H$ in $(0,\tau^\ast)$.
\end{lem}

\begin{proof}
Write $a=N+K-1$, $p=N+2K-J=a+C$ and $q_j=Ns_j$, so that
$d_j(\tau)=q_j-p\tau$ and $\Delta_j(\tau)=d_j(\tau)^2-4aC\tau^2$. The two roots of
\eqref{eqn:mml:loading:quadratic} have product $C\tau^2/a$, so
$C\tau^2/\widehat x_j(\tau)=\{d_j(\tau)-\Delta_j(\tau)^{1/2}\}/2$. Hence
\[
H(\tau)=\Bigl(A+\frac{Jp}{2}\Bigr)\tau-B-\frac12\sum_{j=1}^Jq_j
+\frac12\sum_{j=1}^J\Delta_j(\tau)^{1/2}.
\]
Since $\Delta_j$ is quadratic in $\tau$, direct differentiation gives
$H^{\prime\prime}(\tau)=-2aC\sum_jq_j^2\Delta_j(\tau)^{-3/2}<0$. As
$\tau\downarrow0$, $\Delta_j\rightarrow q_j^2$ and
$\{\Delta_j^{1/2}\}^{\prime}\rightarrow-p$, giving $H(0^+)=-B$ and
$H^{\prime}(0^+)=A$. As $\tau\uparrow\tau_{J,\max}$, $\Delta_J\rightarrow0$
with $\Delta_J^{\prime}<0$, so $H^{\prime}\rightarrow-\infty$. Regular interior
local minima are the points where $H$ changes sign from negative to
positive. If $A\leq0$, concavity gives $H^{\prime}<0$, so $H<-B<0$
throughout. If $A>0$, $H$ increases on $(0,\tau^\ast)$ and decreases on
$(\tau^\ast,\tau_{J,\max})$. It therefore has an upward zero crossing if
and only if $H(\tau^\ast)>0$, in which case the crossing is unique and
lies in $(0,\tau^\ast)$. A second zero in $(\tau^\ast,\tau_{J,\max})$, if
present, is a local maximum of the codelength.
\end{proof}
Equation~\eqref{eqn:mml:tau:stationary} suggests the fixed-point map
\begin{equation}
g_J(\tau)
=
\frac{1}{A}
\left[
B+
C\tau^2
\sum_{j=1}^{J}
\frac{1}{\widehat{x}_j(\tau)}
\right],
\label{eqn:mml:tau:fixed:map}
\end{equation}
for computing the MML estimate of the residual variance.
Starting from an admissible value $\tau^{(0)}$, the iteration is
\begin{align}
x_j^{(t+1)}
&=
\widehat{x}_j\bigl(\tau^{(t)}\bigr),
\qquad j=1,\ldots,J,
\label{eqn:mml:fixed:x:update}
\\
\tau^{(t+1)}
&=
(1-\omega)\tau^{(t)}
+
\frac{\omega}{
A}
\left(
B+C(\tau^{(t)})^2
\displaystyle\sum_{j=1}^{J}
\{x_j^{(t+1)}\}^{-1}
\right)
,
\label{eqn:mml:fixed:tau:update}
\end{align}
where $0<\omega\leq1$ is a damping parameter.
The regular loading profile exists when $d_j(\tau)>0$ and $\Delta_j(\tau)>0$ for every $j\leq J$. Since $s_1\geq\cdots\geq s_J$, the admissible interval is
\begin{equation}
\mathcal T_J=(0,\tau_{J,\max}),
\qquad
\tau_{J,\max}
=
\frac{
Ns_J
}{
N+2K-J+
2\sqrt{(N+K-1)(K-J+1)}
}.
\label{eqn:mml:tau:interval}
\end{equation}
A convenient initial value is
\begin{equation}
\tau^{(0)}
=
\min\left\{
\widehat{\tau}_{\mathrm{ML}},
\frac{1}{2}\tau_{J,\max}
\right\}.
\label{eqn:mml:tau:initial}
\end{equation}
A procedure for implementing this fixed-point update is given in~Algorithm~\ref{alg:mml:estimation}.

\begin{algorithm}[t]
\caption{MML87 estimation for a fixed latent dimension $J\geq1$}
\label{alg:mml:estimation}
\begin{algorithmic}[1]
\Require Sample eigenvalues $s_1\geq\cdots\geq s_K$ with $s_J>0$ and $B>0$, relative tolerance
$\epsilon$, initial damping $\omega_0\in(0,1]$, iteration cap $T_{\max}$.
\State Compute $A$, $B$, $C$ and $\tau_{J,\max}$. If $A\leq0$,
\Return no regular $J$-component solution.
\State Set $\tau^{(0)}=\min\{\widehat{\tau}_{\mathrm{ML}},\tau_{J,\max}/2\}$
and $\omega=\omega_0$.
\For{$t=0,1,\ldots,T_{\max}-1$}
    \State Set $x_j^{(t+1)}=\widehat{x}_j(\tau^{(t)})$, $j=1,\ldots,J$,
    using \eqref{eqn:mml:loading:profile}, and $g=g_J(\tau^{(t)})$ using
    \eqref{eqn:mml:tau:fixed:map}.
    \If{$|g-\tau^{(t)}|\leq\epsilon\,\tau^{(t)}$}
        \State Stop the iteration.
    \EndIf
    \State Set $\tau^{(t+1)}=(1-\omega)\tau^{(t)}+\omega g$. While
    $\tau^{(t+1)}\notin\mathcal T_J$, halve $\omega$ and recompute.
\EndFor
\State Certify the result: compute the maximiser $\tau^\ast$ of $H$ on
$\mathcal T_J$ by bracketing the zero of $H^{\prime}$.
\If{$H(\tau^\ast)\leq0$}
    \State \Return no regular $J$-component solution.
\EndIf
\State \Return the zero of $H$ in $(0,\tau^\ast)$, located within a
sign-changing bracket that contains the final iterate when it lies in
$(0,\tau^\ast)$, as $\widehat{\tau}_{\mathrm{MML}}$, and
$\widehat{\lambda}_{j,\mathrm{MML}}^2
=\widehat{x}_j(\widehat{\tau}_{\mathrm{MML}})$.
\end{algorithmic}
\end{algorithm}
The stopping rule uses the undamped residual relative to $\tau$, so the computed estimate is scale equivariant. A small residual, even with $H^{\prime}>0$, does not by itself show that $H$ crosses zero: near tangency, $\max H$ can be slightly negative while the iteration stalls close to the maximiser. The fixed-point iterate is therefore used only as a starting value, and every fit is certified by Lemma~\ref{lem:mml:unique}. A regular solution is accepted only when $\max H>0$, and the estimate is the zero of $H$ in $(0,\tau^\ast)$, located within a sign-changing bracket. Only this test is used to conclude that no regular solution exists. In floating-point arithmetic, a value of $\max H$ within a small multiple of the rounding error of its terms is reported as numerically tangent rather than as either outcome. In checks on simulated data sets, the iteration converged whenever a regular solution existed, with a median of about 40 iterations, and certification changed the estimate only at the level of the tolerance.

The discriminant condition, $d_j(\tau)>0$ and $\Delta_j(\tau)>0$ for every $j\leq J$, also gives the necessary spectral condition
\begin{equation}
\frac{s_J}{\tau}
>
\left[
\sqrt{1+\frac{K-1}{N}}
+
\sqrt{\frac{K-J+1}{N}}
\right]^2
\label{eqn:mml:component:threshold}
\end{equation}
for a regular $J$-component loading profile. Suppose that $N\rightarrow\infty$ with $J$ fixed and $K/N\rightarrow\gamma>0$, the proportional regime studied in Section~\ref{sec:stein}. The right-hand side of \eqref{eqn:mml:component:threshold} then tends to
\begin{equation}
\left(\sqrt{1+\gamma}+\sqrt{\gamma}\right)^2.
\label{eqn:mml:threshold:limit}
\end{equation}
This has the same form as the phase transition of Baik, Ben Arous and P\'ech\'e (BBP)~\cite{BaikEtAl05}, under which a sample spike separates from the noise bulk only if $s_J/\tau$ exceeds the upper Marchenko--Pastur edge $(1+\sqrt{\gamma})^2$, that is, only if the signal-to-noise ratio $\rho_J=\lambda_J^2/\tau$ exceeds $\sqrt\gamma$. The MML87 threshold \eqref{eqn:mml:threshold:limit} replaces $1$ by $\sqrt{1+\gamma}$ and therefore lies strictly above the bulk edge for every $\gamma>0$. Under pure noise, the largest sample eigenvalue tends to the bulk edge, so asymptotically noise alone does not satisfy \eqref{eqn:mml:component:threshold}. In terms of the population spike, \eqref{eqn:mml:threshold:limit} requires $\rho_J>\rho^\ast(\gamma)$, defined in Section~\ref{sec:stein}. For example, for $\gamma=1/2$ the MML87 and BBP thresholds are $3.73$ and $2.91$ for $s_J/\tau$, and $1.98$ and $0.71$ for $\rho_J$.

The latent dimension is selected by comparing the complete MML87 codelengths over all candidate ranks admitting regular solutions. Condition \eqref{eqn:mml:component:threshold} is necessary but not sufficient for a regular solution, which exists if and only if $\max H>0$ (Lemma~\ref{lem:mml:unique}). In our experiments, the selected rank coincided with the largest rank admitting a regular solution in at least $96.9\%$ of data sets (see Section~\ref{sec:exp:model-selection}), so the existence of a regular solution almost always determines whether a component is retained.
At the resulting estimate,
\begin{equation}
0<
\widehat{\lambda}_{j,\mathrm{MML}}^2
<
s_j-\widehat{\tau}_{\mathrm{MML}},
\qquad j=1,\ldots,J,
\end{equation}
so MML87 shrinks the fitted signal eigenvalues relative to maximum likelihood. A vanishing loading magnitude is a singular boundary of the $J$-component model and is handled by comparison with the corresponding lower-rank model.

\section{Fixed dimension: bias and asymptotic distribution}
\label{sec:finite-sample-bias}
We now compare the finite-sample biases of the maximum likelihood and MML87 estimators of the residual variance. Let
\[
\rho_j=\frac{\lambda_j^2}{\tau},
\qquad
\eta_j=\tau+\lambda_j^2=\tau(1+\rho_j),
\qquad
j=1,\ldots,J,
\]
where $\eta_j$ is the $j$-th population signal eigenvalue and $\rho_j$ is its signal-to-noise ratio.

\begin{theorem}
\label{thm:mml:bias}

Assume that $K$ and $J\geq1$ are fixed, the population signal eigenvalues are distinct, and $N\rightarrow\infty$. Let $\widehat{\tau}_{\mathrm{MML}}$ denote the regular MML87 estimate when it exists, and $\widehat{\tau}_{\mathrm{ML}}$ otherwise. Then
\begin{align}
\mathbb{E}
\left(
\widehat{\tau}_{\mathrm{ML}}-\tau
\right)
&=
-\frac{\tau}{N}
\left(
J+1+\sum_{j=1}^J\rho_j^{-1}
\right)
+
O(N^{-2}),
\label{eqn:ml:bias:first-order}
\\
\mathbb{E}
\left(
\widehat{\tau}_{\mathrm{MML}}-\tau
\right)
&=
\frac{\tau}{N(K-J)}
\left(
J^2+\sum_{j=1}^J\rho_j^{-1}
\right)
+
O(N^{-2}).
\label{eqn:mml:bias:first-order}
\end{align}
Thus maximum likelihood has a negative first-order bias, whereas the MML87 estimator has a positive one. The MML87 first-order bias is the smaller in magnitude if and only if $(K-J)(J+1+\sum_j\rho_j^{-1})>J^2+\sum_j\rho_j^{-1}$. In particular, this inequality holds whenever $K-J\geq J\geq1$.
\end{theorem}

\begin{proof}[Proof (sketch)]
For $j\leq J$, $\mathbb{E}(s_j)$ has the classical first-order expansion
of~\cite{Lawley56}, adjusted for the estimated mean. Combining it with
$\widehat{\tau}_{\mathrm{ML}}=\{\operatorname{tr}(\mathbf S)-\sum_{j\leq J}s_j\}/(K-J)$
gives \eqref{eqn:ml:bias:first-order}. For MML87, a one-step expansion of
the penalised score about the maximum likelihood estimate (equivalently,
the stationarity equation \eqref{eqn:mml:tau:stationary}) gives
$\widehat{\tau}_{\mathrm{MML}}-\widehat{\tau}_{\mathrm{ML}}
=\tau\{D+C\sum_j\rho_j^{-1}\}/\{N(K-J)\}+O_p(N^{-3/2})$, with
$D=K(J+1)-J$. The identities $D-(K-J)(J+1)=J^2$ and $C-(K-J)=1$ then give
\eqref{eqn:mml:bias:first-order}. Replacing the sample-dependent
coefficient by its population value introduces fluctuations of order
$N^{-3/2}$, so the $O(N^{-2})$ remainder in expectation needs a separate
argument. This argument, a localisation on a neighbourhood of
$\boldsymbol\Sigma$, is given with the full proof in Appendix~\ref{sec:supp:proofs}.
\end{proof}

The case $J=0$ is excluded because both estimators are then explicit. The maximum likelihood estimate $\operatorname{tr}(\mathbf S)/K$ has exact bias $-\tau/N$, and the zero-dimensional MML87 estimate $\widehat\tau_0=N\operatorname{tr}(\mathbf S)/\{K(N-1)\}$ of Section~\ref{sec:mml:concentrated} is exactly unbiased, in agreement with \eqref{eqn:ml:bias:first-order} and \eqref{eqn:mml:bias:first-order} evaluated with empty sums.

Theorem~\ref{thm:mml:bias} shows that the MML87 correction reverses the characteristic downward bias of maximum likelihood. The ratio of the leading absolute biases is
\begin{equation}
\left|
\frac{
\mathbb{E}(\widehat{\tau}_{\mathrm{ML}}-\tau)
}{
\mathbb{E}(\widehat{\tau}_{\mathrm{MML}}-\tau)
}
\right|
=
(K-J)
\frac{
J+1+\sum_{j=1}^J\rho_j^{-1}
}{
J^2+\sum_{j=1}^J\rho_j^{-1}
}
+
O(N^{-1}).
\label{eqn:bias:ratio:general}
\end{equation}
As the signals strengthen, the leading ratio tends to
\begin{equation}
\left|
\frac{
\mathbb{E}(\widehat{\tau}_{\mathrm{ML}}-\tau)
}{
\mathbb{E}(\widehat{\tau}_{\mathrm{MML}}-\tau)
}
\right|
=
\frac{(K-J)(J+1)}{J^2}.
\label{eqn:bias:ratio:strong}
\end{equation}
Here $N\rightarrow\infty$ first and then $\rho_J\rightarrow\infty$, with $\rho_1>\cdots>\rho_J$; the expansion \eqref{eqn:bias:ratio:general} is not uniform in the signal strengths. For weak signals, the approximation must be interpreted cautiously because a regular $J$-component MML87 solution may cease to exist.

\paragraph{Single principal component}
When $J=1$, write
$
\rho=\lambda_1^2 / \tau.
$
The leading biases reduce to
\begin{align}
\mathbb{E}
\left(
\widehat{\tau}_{\mathrm{ML}}-\tau
\right)
&=
-\frac{\tau}{N}
\left(
2+\rho^{-1}
\right)
+
O(N^{-2}),
\label{eqn:ml:bias:J1}
\\
\mathbb{E}
\left(
\widehat{\tau}_{\mathrm{MML}}-\tau
\right)
&=
\frac{\tau}{N(K-1)}
\left(
1+\rho^{-1}
\right)
+
O(N^{-2}).
\label{eqn:mml:bias:J1}
\end{align}
Thus the ratio of the leading absolute biases is
\begin{align}
\left|
\frac{
\mathbb{E}(\widehat{\tau}_{\mathrm{ML}}-\tau)
}{
\mathbb{E}(\widehat{\tau}_{\mathrm{MML}}-\tau)
}
\right|
&=
(K-1)
\frac{2+\rho^{-1}}{1+\rho^{-1}}
+
O(N^{-1})
=
(K-1)
\frac{2\rho+1}{\rho+1}
+
O(N^{-1}).
\label{eqn:bias:ratio:J1}
\end{align}
For a strong signal, $\rho\rightarrow\infty$, the leading ratio approaches $2(K-1)$; as $\rho\rightarrow0^+$, it approaches $K-1$. In the latter regime, however, the regular one-component loading profile may not exist, and the zero-component model must be considered directly.

\paragraph{Second-order bias}
The first-order terms of Theorem~\ref{thm:mml:bias} can be extended to second order. Put $m=K-J$, $C=m+1$, $D=K(J+1)-J$ and $R_r=\sum_{j=1}^J\rho_j^{-r}$ for $r=1,2,3$.

\begin{corollary}
\label{cor:bias:second}
Under the assumptions of Theorem~\ref{thm:mml:bias},
\begin{align}
\mathbb{E}(\widehat\tau_{\mathrm{ML}}-\tau)
&=
-\frac{\tau}{N}(J+1+R_1)
-\frac{\tau}{N^2}(R_1+2R_2+R_3)
+O(N^{-3}),
\label{eqn:bias:ml:second-order}\\
\mathbb{E}(\widehat\tau_{\mathrm{MML}}-\tau)
&=
\frac{\tau}{Nm}(J^2+R_1)
+\frac{\tau}{N^2}\mathcal B_2
+O(N^{-3}),
\label{eqn:bias:mml:second-order}
\end{align}
where
\[
\mathcal B_2=\frac1{m^2}\Big[
(D+CR_1)\{J^2+(m+2)R_1+CR_2\}
+Cm\{(K-1)R_1+KR_2+R_3\}
+(3m+2)(R_1+2R_2+R_3)
\Big]>0 .
\]
\end{corollary}

The proof, which carries the expansions in the proof of Theorem~\ref{thm:mml:bias} two orders further, is in Appendix~\ref{sec:supp:proofs}. The second-order term is small for maximum likelihood but large and positive for MML87. For example, with $K=10$ and $\rho=(10,5)$ the second-order coefficients are $-0.41$ and $7.35$, against first-order coefficients $-3.30$ and $0.54$. In the simulations of Appendix~\ref{sec:supp:proofs}, the first-order term of \eqref{eqn:mml:bias:first-order} misses $8\%$--$23\%$ of the MML87 bias at $N=100$ and up to about $40\%$ at $N=50$, whereas the first- and second-order terms together are within $5\%$ of the simulated bias at $N=100$. For large $K$ with $J$ and the $\rho_j$ fixed, $\mathcal B_2=K(R_1+R_2)+O(1)$, whereas the first-order coefficient $(J^2+R_1)/m$ is $O(K^{-1})$. The second-order term is therefore comparable to the first once $K^2$ is of order $N$, and the fixed-dimension expansion should not be relied on there; in the proportional regime of Section~\ref{sec:stein} the MML87 centring is of order $N^{-1}$ (Theorem~\ref{thm:prop:clt}). The expansion describes the regular estimator; when a regular solution often fails to exist, the completion by maximum likelihood dominates.

\paragraph{Asymptotic distribution}
The bias expansion of Theorem~\ref{thm:mml:bias} is complemented by the
following distributional result. Write $\widehat x_j=\widehat\lambda_{j,\mathrm{MML}}^2$
for the MML87 loading strengths, with $x_j$ and $\mathbf x$ as in
Section~\ref{sec:mml:concentrated}.

\begin{theorem}
\label{thm:fixed:clt}
Under the assumptions of Theorem~\ref{thm:mml:bias},
\[
\sqrt N
\begin{pmatrix}
\widehat\tau_{\mathrm{MML}}-\tau\\
\widehat{\mathbf x}-\mathbf x
\end{pmatrix}
\Rightarrow \mathcal{N}_{J+1}(\mathbf 0,\mathbf V),
\qquad
\mathbf V=
\begin{pmatrix}
v & -v\mathbf 1^{\prime}\\
-v\mathbf 1 & 2\operatorname{diag}(\eta_1^2,\ldots,\eta_J^2)+v\mathbf 1\mathbf 1^{\prime}
\end{pmatrix},
\qquad
v=\frac{2\tau^2}{K-J},
\]
with $\eta_j$ the population signal eigenvalues defined above. The maximum likelihood estimators have
the same limit, and $\mathbf V$ is the inverse of the per-observation
Fisher information for $(\tau,\mathbf x)$.
\end{theorem}

\begin{proof}[Proof (sketch)]
For maximum likelihood, the central limit theorem for
$\sqrt N(\mathbf S-\boldsymbol\Sigma)$ and first-order perturbation of the
simple signal eigenvalues give asymptotically independent limits
$\sqrt N(s_j-\eta_j)\Rightarrow \mathcal{N}(0,2\eta_j^2)$. The residual variance
$\widehat\tau_{\mathrm{ML}}=\{\operatorname{tr}(\mathbf S)-\sum_{j\leq J}s_j\}/(K-J)$
depends to first order only on the noise block of $\mathbf S$, so
$\sqrt N(\widehat\tau_{\mathrm{ML}}-\tau)\Rightarrow \mathcal{N}(0,2\tau^2/(K-J))$,
independently of the signal eigenvalues. Since
$\widehat x_{\mathrm{ML},j}=s_j-\widehat\tau_{\mathrm{ML}}$, this gives
$\mathbf V$. For MML87, the codelength is the negative log-likelihood plus a
penalty with bounded derivatives near the true parameter, so the MML87
estimates differ from their maximum likelihood counterparts by
$O_p(N^{-1})$ and have the same limit. The Fisher information for
$(\tau,\eta_1,\ldots,\eta_J)$ is diagonal, and the map
$x_j=\eta_j-\tau$ shows that $\mathbf V$ is the inverse Fisher information
for $(\tau,\mathbf x)$. The full proof is in Appendix~\ref{sec:supp:proofs}.
\end{proof}

For fixed dimension, MML87 and maximum likelihood are therefore
equivalent to first order. The two estimators differ by $O_p(N^{-1})$,
and this difference is the bias correction quantified by
Theorem~\ref{thm:mml:bias}. Theorem~\ref{thm:fixed:clt} gives plug-in
standard errors
$\operatorname{se}(\widehat\tau_{\mathrm{MML}})=\widehat\tau_{\mathrm{MML}}\{2/(N(K-J))\}^{1/2}$
and
$\operatorname{se}(\widehat x_j)=[\{2\widehat\eta_j^2+2\widehat\tau_{\mathrm{MML}}^2/(K-J)\}/N]^{1/2}$,
with $\widehat\eta_j=\widehat\tau_{\mathrm{MML}}+\widehat x_j$, and the
asymptotic $(1-\alpha)$ interval
$\widehat\tau_{\mathrm{MML}}\exp[\pm \Phi^{-1}(1-\alpha/2)\{2/(N(K-J))\}^{1/2}]$ for $\tau$, where $\Phi$ is the standard normal distribution function.

\section{Proportional regime: asymptotic loss and distribution}
\label{sec:stein}

Theorem~\ref{thm:mml:bias} describes the regime in which $K$ is fixed and $N\rightarrow\infty$, where maximum likelihood and MML87 are asymptotically equivalent. We now consider the high-dimensional regime in which $K$ grows proportionally with $N$. There the two estimators differ at first order, and the difference is best understood through the fitted signal eigenvalues.
Write the population covariance as $\boldsymbol\Sigma=\tau\bigl(\mathbf I_K+\sum_{j=1}^J\rho_j\mathbf r_j\mathbf r_j^{\prime}\bigr)$, where $\mathbf r_j$ is the $j$-th column of $\mathbf R$, $\rho_j=\lambda_j^2/\tau$ as in Section~\ref{sec:finite-sample-bias} and the population spikes are $\ell_j=1+\rho_j$. Consider $N\rightarrow\infty$ with $J$ and $\rho_1>\cdots>\rho_J>0$ fixed and $K/N\rightarrow\gamma\in(0,1]$. This covers asymptotically $K\leq N$. Theorem~\ref{thm:prop:clt} below additionally requires $\gamma<1$, because its loading-strength limit uses the outlier central limit theorem of~\cite{Paul07}, which is established for $\gamma<1$. When $\gamma>1$ the sample covariance matrix is singular, with $K-N+1$ zero eigenvalues; the MML87 estimator can still be computed, but the limits in this section are not established there, and we do not consider that case. The MML87 and maximum likelihood estimators are scale equivariant, so we may take $\tau=1$. Define
\begin{align}
b(\rho)&=\rho-\gamma+\frac{\gamma}{\rho},
\qquad
\eta_{\mathrm{MML}}(\ell)=1+\frac{b(\rho)+\sqrt{b(\rho)^2-4\gamma(1+\gamma)}}{2(1+\gamma)},
\label{eqn:stein:etaMML}\\
\eta_{\mathrm{St}}(\ell)&=\frac{\ell}{c^2(\ell)+\ell\{1-c^2(\ell)\}}
=\frac{(1+\rho)(\rho+\gamma)}{\rho(1+\gamma)+2\gamma},
\qquad
\eta_{\mathrm{Ent}}(\ell)=1+(\ell-1)c^2(\ell),
\label{eqn:stein:etaDGJ}
\end{align}
where $c^2(\ell)=\{1-\gamma/(\ell-1)^2\}/\{1+\gamma/(\ell-1)\}$ is the limiting squared cosine between the population and sample signal eigenvectors. For a generic spike $\ell=1+\rho$ with $\rho>\sqrt\gamma$, the corresponding sample eigenvalue converges to $\bar s(\ell)=\ell(1+\gamma/\rho)$~\cite{BaikSilverstein06,Paul07}. The functions $\eta_{\mathrm{St}}$ and $\eta_{\mathrm{Ent}}$ are the asymptotically optimal eigenvalue shrinkers of \cite{DonohoGavishJohnstone18} for Stein loss
\[
L^{\mathrm{St}}(\boldsymbol\Sigma,\widehat{\boldsymbol\Sigma})
=\operatorname{tr}(\boldsymbol\Sigma^{-1}\widehat{\boldsymbol\Sigma}-\mathbf I_K)
-\log\det(\boldsymbol\Sigma^{-1}\widehat{\boldsymbol\Sigma})
=2\,\operatorname{KL}(\widehat{\boldsymbol\Sigma},\boldsymbol\Sigma)
\]
and for entropy loss
$L^{\mathrm{Ent}}(\boldsymbol\Sigma,\widehat{\boldsymbol\Sigma})
=L^{\mathrm{St}}(\widehat{\boldsymbol\Sigma},\boldsymbol\Sigma)
=2\,\operatorname{KL}(\boldsymbol\Sigma,\widehat{\boldsymbol\Sigma})$,
respectively, where $\operatorname{KL}(\boldsymbol\Sigma_1,\boldsymbol\Sigma_2)$ is the Kullback--Leibler divergence defined in \eqref{eqn:simulation:KL}. Let $\rho^\ast(\gamma)$ denote the larger root of
$\rho+\gamma/\rho=\gamma+2\sqrt{\gamma(1+\gamma)}$, and let
$h(r)=r-1-\log r$.

\begin{proposition} \label{prop:stein} 
Suppose that $\rho_J>\rho^\ast(\gamma)$. Then, with probability tending to one, the regular $J$-component profile admits an interior local minimum. Every sequence of regular interior stationary solutions satisfies 
\begin{align} 
\widehat\tau_{\mathrm{MML}} &\longrightarrow1, \qquad 
\widehat\tau_{\mathrm{MML}}+\widehat x_j \longrightarrow \eta_{\mathrm{MML}}(\ell_j), \qquad j=1,\ldots,J. 
\end{align} 
Moreover: 
\begin{enumerate} 
\item[(i)] Let $\gamma_0=1-2/\sqrt5$, and, for $0<\gamma<1$, let
$\rho_0(\gamma)=\{\gamma+\sqrt{\gamma(2-\gamma)}\}/(1-\gamma)$ be the
unique positive solution of $(1-\gamma)\rho^2-2\gamma\rho-2\gamma=0$.
On the regular domain $\rho>\rho^\ast(\gamma)$, the following hold.
If $\gamma_0<\gamma<1$, then $\eta_{\mathrm{MML}}(\ell)=\eta_{\mathrm{St}}(\ell)$
exactly at $\rho=\rho_0(\gamma)$, with $\eta_{\mathrm{MML}}<\eta_{\mathrm{St}}$
for $\rho<\rho_0(\gamma)$ and $\eta_{\mathrm{MML}}>\eta_{\mathrm{St}}$ for
$\rho>\rho_0(\gamma)$. If $0<\gamma\leq\gamma_0$, then
$\eta_{\mathrm{MML}}(\ell)>\eta_{\mathrm{St}}(\ell)$. If $\gamma=1$, then
$\eta_{\mathrm{MML}}(\ell)<\eta_{\mathrm{St}}(\ell)$. As $\rho\rightarrow\infty$, 
\[ 
\frac{\eta_{\mathrm{MML}}(\ell)} {\eta_{\mathrm{St}}(\ell)} = 1+ \frac{\gamma(1-\gamma)} {(1+\gamma)\rho} + O(\rho^{-2}). 
\] 
\item[(ii)] The limiting excess Stein loss of MML87 over the optimal sample-eigenvector shrinker is 
\[ \sum_{j=1}^J h\left\{ \frac{\eta_{\mathrm{MML}}(\ell_j)} {\eta_{\mathrm{St}}(\ell_j)} \right\}. 
\]
In the subsequent strong-spike limit $\rho_J\rightarrow\infty$, this is $O(\rho_J^{-2})$.
\item[(iii)] The limiting excess Stein loss of maximum likelihood tends to 
$ J\{\gamma-\log(1+\gamma)\}  $
as $\rho_J\rightarrow\infty$. The same limit holds for a sample-eigenvector estimator whose fitted signal eigenvalues converge to the population values $\ell_j$ and whose fitted residual variance $\widetilde\tau$ satisfies $\sqrt K(\widetilde\tau-1)\rightarrow0$ in probability. \item[(iv)] Under entropy loss, the limiting excess loss relative to the optimal sample-eigenvector shrinker is, in the subsequent strong-spike limit $\rho_J\rightarrow\infty$, $O(\rho_J^{-2})$ for maximum likelihood, whereas for MML87 it tends to $J\{\gamma-\log(1+\gamma)\}$. \end{enumerate} 
\end{proposition}

\begin{figure}[t]
\centering
\includegraphics[width=\textwidth]{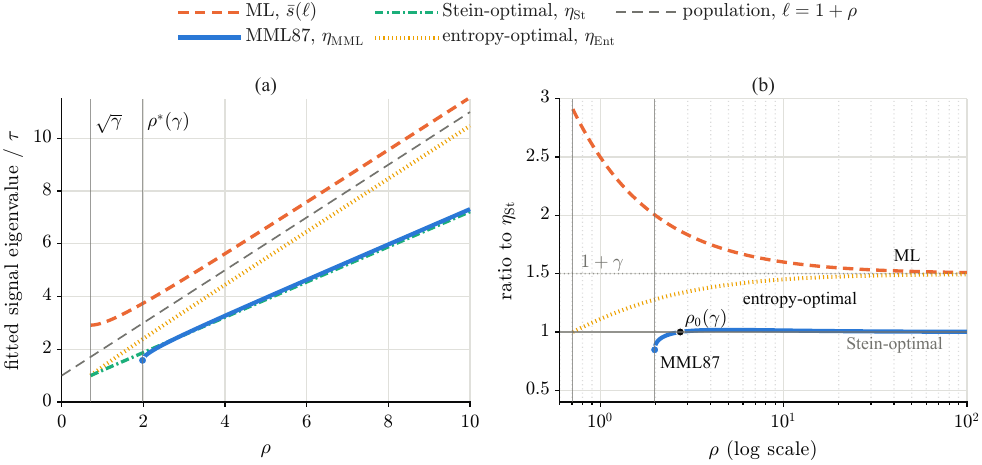}
\caption{Limiting fitted signal eigenvalues in the proportional regime with
$\gamma=1/2$, in units of $\tau$. (a) Maximum likelihood $\bar s(\ell)$, MML87
$\eta_{\mathrm{MML}}(\ell)$, and the Stein- and entropy-optimal shrinkers
$\eta_{\mathrm{St}}(\ell)$ and $\eta_{\mathrm{Ent}}(\ell)$
of~\cite{DonohoGavishJohnstone18}, against the signal-to-noise ratio $\rho$;
the grey dashed line is the population eigenvalue $\ell=1+\rho$. Vertical lines
mark the BBP threshold $\sqrt\gamma$ and the MML87 existence threshold
$\rho^\ast(\gamma)$; the maximum likelihood and optimal curves are shown above
$\sqrt\gamma$, and MML87 above $\rho^\ast(\gamma)$. (b) The same quantities
divided by $\eta_{\mathrm{St}}(\ell)$. MML87 lies below the Stein-optimal value
for $\rho<\rho_0(\gamma)$ and above it beyond, approaching it at rate
$\rho^{-1}$ (Proposition~\ref{prop:stein}), whereas maximum likelihood and the
entropy-optimal shrinker both tend to $1+\gamma$.}
\label{fig:shrinkers}
\end{figure}

\begin{proof}[Proof (sketch)]
Since $\rho^\ast(\gamma)>\sqrt\gamma$, every spike is supercritical, so
$s_j\rightarrow\bar s(\ell_j)=1+\rho_j+\gamma+\gamma/\rho_j$
\cite{BaikSilverstein06,Paul07,DonohoGavishJohnstone18}. The stationarity equation, together with
the bound $\widehat x_j(\tau)\geq\tau\{C/(N+K-1)\}^{1/2}$, gives
$\widehat\tau_{\mathrm{MML}}\rightarrow1$. A regular solution exists
because the upper end of the admissible interval $\mathcal T_J$ converges
to $\bar s(\ell_J)/(\sqrt{1+\gamma}+\sqrt\gamma)^2$, which exceeds one
precisely when $\rho_J>\rho^\ast(\gamma)$, so a fixed neighbourhood of one
is admissible with probability tending to one. Moreover,
$H(\tau)/(NK)\rightarrow \tau-1$ uniformly in probability on this neighbourhood.
Hence $H$ changes sign from negative to positive there, and
Lemma~\ref{lem:mml:unique} gives the unique regular minimum. Dividing
\eqref{eqn:mml:loading:quadratic} by $N$ gives the limiting quadratic
$(1+\gamma)x^2-b(\rho)x+\gamma=0$, whose larger root yields
$\eta_{\mathrm{MML}}$. Part (i) follows by evaluating this quadratic at
$\eta_{\mathrm{St}}-1$, where it factorises. Parts (ii)--(iv) follow from the blockwise
decomposition of the Stein and entropy losses~\cite{DonohoGavishJohnstone18},
after replacing the fitted residual variance by one, which changes these
unnormalised losses by $o_p(1)$ because $\sqrt K(\widehat\tau-1)\rightarrow0$.
The full proof is in Appendix~\ref{sec:supp:proofs}.
\end{proof}

Figure~\ref{fig:shrinkers} illustrates these limits for $\gamma=1/2$. Proposition~\ref{prop:stein} compares each estimator with the optimal shrinker. Comparing MML87 directly with maximum likelihood gives the following.

\begin{corollary}
\label{cor:stein:improvement}
Under the assumptions of Proposition~\ref{prop:stein}, let
$\widehat{\boldsymbol\Sigma}_{\mathrm{MML}}$ and
$\widehat{\boldsymbol\Sigma}_{\mathrm{ML}}$ denote the MML87 and maximum likelihood covariance estimators at the true rank $J\geq1$, where $\widehat{\boldsymbol\Sigma}_{\mathrm{MML}}$ exists with probability tending to one. Then
\[
L^{\mathrm{St}}
(\boldsymbol\Sigma,\widehat{\boldsymbol\Sigma}_{\mathrm{ML}})
-
L^{\mathrm{St}}
(\boldsymbol\Sigma,\widehat{\boldsymbol\Sigma}_{\mathrm{MML}})
\longrightarrow
\Delta_{\mathrm{St}}
=\sum_{j=1}^J
\left[
h\left\{\frac{\bar s(\ell_j)}{\eta_{\mathrm{St}}(\ell_j)}\right\}
-
h\left\{\frac{\eta_{\mathrm{MML}}(\ell_j)}{\eta_{\mathrm{St}}(\ell_j)}\right\}
\right]>0
\]
in probability. In particular,
$\mathbb P\{L^{\mathrm{St}}(\boldsymbol\Sigma,\widehat{\boldsymbol\Sigma}_{\mathrm{MML}})
<L^{\mathrm{St}}(\boldsymbol\Sigma,\widehat{\boldsymbol\Sigma}_{\mathrm{ML}})\}\rightarrow1$.
\end{corollary}

\begin{proof}[Proof (sketch)]
The limit follows from the per-spike excess losses in the proof of Proposition~\ref{prop:stein}, since the maximum likelihood signal eigenvalues are $s_j\rightarrow\bar s(\ell_j)$. For positivity, the larger root $z$ of the limiting quadratic satisfies $z>\rho/(\rho+2)$, and the ratios $\bar s(\ell)/\eta_{\mathrm{St}}(\ell)=1+\gamma+2\gamma/\rho$ and $\bar s(\ell)/\eta_{\mathrm{MML}}(\ell)=1+\gamma+\gamma/z$ then make each summand of $\Delta_{\mathrm{St}}$ positive. The full proof is in Appendix~\ref{sec:supp:proofs}.
\end{proof}

Proposition~\ref{prop:stein} shows that, in the proportional regime, the MML87 shrinker approaches the Stein-optimal shrinker in the subsequent strong-spike limit, with excess Stein loss of order $\rho_J^{-2}$, although for a fixed spike the two differ. Corollary~\ref{cor:stein:improvement} shows that MML87 nevertheless has smaller limiting Stein loss than maximum likelihood for every fixed spike configuration, not only for strong spikes. Maximum likelihood has the analogous strong-spike property for entropy loss, but the reverse of Corollary~\ref{cor:stein:improvement} does not hold: for moderate spikes, MML87 can also have the smaller limiting entropy loss. For example, with $\gamma=0.25$ and $\rho=2$, the per-spike limiting excess entropy losses are $0.026$ for maximum likelihood and $0.008$ for MML87 (Appendix~\ref{sec:supp:proofs}). The advantage of maximum likelihood under entropy loss is therefore a strong-spike property. The two estimators target different losses. The fixed-dimension regime of Theorem~\ref{thm:mml:bias} corresponds to $\gamma\rightarrow0$, in which
all of the excess losses above vanish.

Proposition~\ref{prop:stein} describes the limits of the estimates; their
fluctuations are as follows. Write $z_j=\eta_{\mathrm{MML}}(\ell_j)-1$
for the limiting MML87 loading strength, which satisfies $z_j<\rho_j$, and
$\xi_j=\{b(\rho_j)^2-4\gamma(1+\gamma)\}^{1/2}$.

\begin{theorem}
\label{thm:prop:clt}
Suppose that the assumptions of Proposition~\ref{prop:stein} hold with $0<\gamma<1$, and let $\tau>0$ be arbitrary. Then
\[
N(\widehat\tau_{\mathrm{MML}}-\tau)\Rightarrow
\mathcal{N} \bigl(\tau\beta_\gamma,\,2\tau^2/\gamma\bigr),
\qquad
\beta_\gamma=\sum_{j=1}^J(z_j^{-1}-\rho_j^{-1})>0,
\]
whereas
$N(\widehat\tau_{\mathrm{ML}}-\tau)\Rightarrow
\mathcal{N}\{-\tau(1+J+\sum_j\rho_j^{-1}),\,2\tau^2/\gamma\}$.
If, in addition, $\sqrt N(K/N-\gamma)\rightarrow0$, then
\[
\sqrt N(\widehat x_j-\tau z_j)\Rightarrow
\mathcal{N}\bigl(0,\,\tau^2z_j^2v_j/\xi_j^2\bigr),
\qquad
v_j=2(1+\rho_j)^2(1-\gamma/\rho_j^2).
\]
\end{theorem}

\begin{proof}[Proof (sketch)]
In the population eigenbasis, the noise diagonal entries of $\mathbf S$
contribute $\tau\chi^2_{(N-1)(K-J)}/N$ to $\operatorname{tr}(\mathbf S)$, so
the chi-squared central limit theorem gives
$\operatorname{tr}(\mathbf S)-K\tau\Rightarrow \mathcal{N}(\tau\sum_j\rho_j-\gamma\tau,\,2\gamma\tau^2)$, where the term
$-\gamma\tau$ comes from estimating the mean. Subtracting the $J$ leading
sample eigenvalues, whose limits are $\tau\bar s(\ell_j)$, gives the
maximum likelihood limit. The stationarity equation
\eqref{eqn:mml:tau:stationary} gives
$N(\widehat\tau_{\mathrm{MML}}-\widehat\tau_{\mathrm{ML}})
\rightarrow\tau(J+1+\sum_jz_j^{-1})$ in probability. This deterministic
shift turns the maximum likelihood limit into the MML87 limit, and
$\beta_\gamma>0$ because $z_j<\rho_j$. For the loading strengths,
$\widehat x_j/\tau$ is a smooth function of $s_j/\tau$ up to an error
$O_p(N^{-1}+|K/N-\gamma|)$, which is $o_p(N^{-1/2})$ under the additional
assumption. The central limit theorem for a simple outlier
eigenvalue~\cite{Paul07} and the delta method then give the stated
variance. The full proof is in Appendix~\ref{sec:supp:proofs}.
\end{proof}

In this regime the residual variance converges at the faster rate
$N^{-1}$, and the two estimators have the same asymptotic variance. They
differ in their centring. The maximum likelihood centring is negative and
grows with $J$, whereas the MML87 centring is positive and typically much
smaller. For example, with $\gamma=1/2$ and $(\rho_1,\rho_2)=(8,4)$, the
centrings are $-3.38\tau$ for maximum likelihood and $0.27\tau$ for
MML87. Because the centring and the standard deviation are both of order
$N^{-1}$, the smaller MML87 centring matters at first order. The limiting
mean squared error of $N(\widehat\tau-\tau)/\tau$ is
$2/\gamma+\beta_\gamma^2$ for MML87, against
$2/\gamma+(1+J+\sum_j\rho_j^{-1})^2$ for maximum likelihood; the former is
smaller because $\beta_\gamma<J+\sum_j\rho_j^{-1}$, which follows from the
bound $z_j>\rho_j/(\rho_j+2)$ in the proof of
Corollary~\ref{cor:stein:improvement}. In the example above, these are
$4.07$ and $15.39$. The loading strengths fluctuate at rate $N^{-1/2}$ about the
shrinkage limits $\tau z_j$ of Proposition~\ref{prop:stein}, not about
the population values $\tau\rho_j$, so intervals centred at $\widehat x_j$
target $\tau z_j$. Both results concern the regular estimate at the true
rank, and they need not hold near the regular threshold
$\rho^\ast(\gamma)$, where $\xi_j\rightarrow0$. Numerical
checks are in Appendix~\ref{sec:supp:proofs}.

\section{Simulation experiments}
\label{sec:experiments}

This section examines the finite-sample performance of the proposed MML87 procedure in three parts: estimation for a fixed latent dimension, comparison with alternative estimators, and selection of the latent dimension. The number of Monte Carlo replications is given in each table caption.

\subsection{Parameter estimation}
\label{sec:exp:pest}

We first compare the MML87 and maximum likelihood estimators for a fixed latent dimension $J$. Both methods estimate the population mean by $\overline{\mathbf{x}}$ and the signal directions by the leading $J$ eigenvectors of the sample covariance matrix. They differ in the residual variance and in the loading magnitudes. Maximum likelihood uses $\widehat{\lambda}_j^2=s_j-\widehat{\tau}_{\mathrm{ML}}$, whereas MML87 uses the shrunken estimates \eqref{eqn:mml:loading:profile} evaluated at $\widehat{\tau}_{\mathrm{MML}}$.
We fixed the ambient dimension at $K=10$ and the residual variance at $\tau=1$, and considered sample sizes $N\in\{25,50,100\}$, latent dimensions $J\in\{1,2,4\}$, and per-component signal-to-noise ratios
\begin{equation}
\rho_j=\frac{\lambda_j^2}{\tau}=\rho\in\{2,5,20\},
\qquad j=1,\ldots,J.
\label{eqn:simulation:SNR}
\end{equation}
Defining the signal strength per component makes the population variance ratio of each component the same for every $J$, although detectability in finite samples still depends on $J$ through the threshold \eqref{eqn:mml:component:threshold}. For each configuration, the signal orientation $\mathbf{R}\in\operatorname{St}(K,J)$ was generated uniformly with respect to the invariant measure on the Stiefel manifold, and data were generated independently from
\[
\mathbf{x}_n
\sim \mathcal{N}_K\left(
\boldsymbol{\mu},
\mathbf{R}\boldsymbol{\Lambda}^2\mathbf{R}^{\prime}
+\tau\mathbf{I}_K
\right),
\qquad
n=1,\ldots,N,
\qquad
\boldsymbol{\Lambda}
=
\operatorname{diag}(\lambda_1,\ldots,\lambda_J),
\]
with the mean treated as unknown. A regular $J$-component MML87 solution does not exist for every data set (Section~\ref{sec:mml:sigma2}). When it does not, MML87 was fitted at the largest rank $J'<J$ that admits a regular solution, with the closed form $\widehat\tau=NK^{-1}(N-1)^{-1}\sum_{k}s_k$ when $J'=0$. To compare estimators rather than ranks, maximum likelihood was fitted at the same rank. Table~\ref{tab:results:pest} reports the percentage of data sets with a regular $J$-component solution. For $J>1$, equal loadings give coinciding population signal eigenvalues. This lies outside the distinct-eigenvalue assumptions of Theorem~\ref{thm:mml:bias} and Proposition~\ref{prop:stein}, so these settings are empirical comparisons rather than direct tests of those results.

For each estimator, we recorded the signed log-scale error and squared log-scale error,
\begin{align}
S_1
=
\log\left(
\frac{\widehat{\sigma}}{\sigma}
\right),
\qquad
S_2
=
\left[
\log\left(
\frac{\widehat{\sigma}}{\sigma}
\right)
\right]^2,
\label{eqn:simulation:S1:S2}
\end{align}
and the Kullback--Leibler divergences in both directions between the true covariance matrix $\boldsymbol{\Sigma}_0$ and the fitted covariance matrix
$\widehat{\boldsymbol{\Sigma}}
=\widehat{\mathbf{R}}\widehat{\boldsymbol{\Lambda}}^2\widehat{\mathbf{R}}^{\prime}
+\widehat{\tau}\mathbf{I}_K$:
\begin{equation}
\operatorname{KL}
\left(
\boldsymbol{\Sigma}_0,
\widehat{\boldsymbol{\Sigma}}
\right)
=
\frac{1}{2}
\left[
\operatorname{tr}
\left(
\widehat{\boldsymbol{\Sigma}}^{-1}
\boldsymbol{\Sigma}_0
\right)
+
\log
\left\{
\frac{
|\widehat{\boldsymbol{\Sigma}}|
}{
|\boldsymbol{\Sigma}_0|
}
\right\}
-K
\right],
\label{eqn:simulation:KL}
\end{equation}
and $\operatorname{KL}(\widehat{\boldsymbol{\Sigma}},\boldsymbol{\Sigma}_0)$,
obtained by exchanging the two matrices. The forward divergence \eqref{eqn:simulation:KL} is half the entropy loss and measures predictive performance. The reverse divergence is half the Stein loss (Section~\ref{sec:stein}). Both divergences compare covariance matrices only. All methods use the same mean estimate $\overline{\bf x}$, and a full predictive divergence would add a term involving its estimation error. The averages of $S_1$ and $S_2$ measure signed and squared log-scale error.

Table~\ref{tab:results:pest} reports the averages. Consider first the settings in which a regular $J$-component solution exists for most data sets. There, MML87 removes most of the downward bias of the maximum likelihood estimator of the residual scale and gives smaller squared log-error. The remaining MML87 bias is positive and grows with $J$. Since $S_1$ is on the log scale, Theorems~\ref{thm:mml:bias} and~\ref{thm:fixed:clt} and the delta method give $\mathbb E(S_1)=(J^2+\sum_j\rho_j^{-1}-1)/\{2N(K-J)\}+o(N^{-1})$ for MML87 and $-\{J+1+\sum_j\rho_j^{-1}+1/(K-J)\}/(2N)+o(N^{-1})$ for maximum likelihood. For $N=100$ and $\rho=20$, evaluated formally at the equal ratios, these give $0.000$, $0.002$ and $0.013$ for MML87 with $J=1,2,4$, against the observed $0.000$, $0.002$ and $0.014$. MML87 also gives the smaller reverse KL divergence in these settings. Under the forward KL divergence, maximum likelihood is slightly better when the signal is strong, by at most about three percent, because MML87 shrinks the fitted signal eigenvalues. This is the trade-off described by Proposition~\ref{prop:stein}.

When the signal is weak and a regular $J$-component solution often does not exist, both estimators are fitted at a lower rank. The residual variance then absorbs the variance of the omitted components, so $S_1$ and $S_2$ measured against the true $\sigma$ say little about the estimators themselves. In these settings MML87 gives the smaller forward KL divergence and usually also the smaller reverse KL
divergence. All differences diminish as $N$ increases. 

\begin{table*}[tbph]
\scriptsize
\begin{center}
\caption{Average performance over $10^5$ simulated data sets for maximum likelihood (MLE) and MML87 at a matched rank, with $K=10$ and equal per-component signal-to-noise ratio $\rho$. Reg.\ is the percentage of data sets with a regular $J$-component MML87 solution. Bold marks the better value (smaller $|S_1|$, $S_2$ and KL divergence). The largest Monte Carlo standard errors are $0.0007$ ($S_1$), $0.0003$ ($S_2$), $0.002$ (forward KL) and $0.002$ (reverse KL).
}
\label{tab:results:pest}
\begin{tabular}{ccccccccccccccc}
\toprule
$N$ & $\rho$ & $J$ & Reg. & \multicolumn{2}{c}{$S_1$} & & \multicolumn{2}{c}{$S_2$} & & \multicolumn{2}{c}{$\operatorname{KL}(\boldsymbol{\Sigma}_0,\widehat{\boldsymbol{\Sigma}})$} & & \multicolumn{2}{c}{$\operatorname{KL}(\widehat{\boldsymbol{\Sigma}},\boldsymbol{\Sigma}_0)$} \\
    &        &     & (\%) & MLE & MML87 & ~ & MLE & MML87 & ~ & MLE & MML87 & ~ & MLE & MML87 \\
\cmidrule{1-15}
\multirow{9}{*}{25} & \multirow{3}{*}{2} & 1 & 45 & {\bf -0.002} & 0.037 & & 0.0060 & {\bf 0.0054} & & 0.339 & {\bf 0.284} & & 0.255 & {\bf 0.192} \\
 &  & 2 & 10 & {\bf 0.052} & 0.100 & & {\bf 0.0123} & 0.0155 & & 0.617 & {\bf 0.521} & & 0.458 & {\bf 0.386} \\
 &  & 4 & 0 & {\bf 0.216} & 0.254 & & {\bf 0.0558} & 0.0699 & & 0.848 & {\bf 0.759} & & 0.807 & {\bf 0.793} \\
 & \multirow{3}{*}{5} & 1 & 97 & -0.045 & {\bf 0.007} & & 0.0050 & {\bf 0.0029} & & 0.304 & {\bf 0.291} & & 0.251 & {\bf 0.190} \\
 &  & 2 & 84 & -0.056 & {\bf 0.034} & & 0.0096 & {\bf 0.0062} & & 0.683 & {\bf 0.622} & & 0.482 & {\bf 0.382} \\
 &  & 4 & 17 & {\bf 0.137} & 0.267 & & {\bf 0.0646} & 0.0941 & & 1.876 & {\bf 1.522} & & {\bf 1.319} & 1.341 \\
 & \multirow{3}{*}{20} & 1 & 100 & -0.045 & {\bf 0.001} & & 0.0045 & {\bf 0.0025} & & {\bf 0.289} & 0.297 & & 0.239 & {\bf 0.197} \\
 &  & 2 & 100 & -0.069 & {\bf 0.011} & & 0.0077 & {\bf 0.0030} & & {\bf 0.578} & 0.589 & & 0.440 & {\bf 0.368} \\
 &  & 4 & 100 & -0.121 & {\bf 0.079} & & 0.0189 & {\bf 0.0108} & & {\bf 1.243} & 1.247 & & 0.800 & {\bf 0.737} \\
\vspace{-2mm} \\
\cmidrule{2-15}
\vspace{-2mm} \\
\multirow{9}{*}{50} & \multirow{3}{*}{2} & 1 & 91 & -0.021 & {\bf 0.007} & & 0.0022 & {\bf 0.0016} & & 0.146 & {\bf 0.131} & & 0.130 & {\bf 0.100} \\
 &  & 2 & 69 & {\bf -0.019} & 0.028 & & 0.0039 & {\bf 0.0034} & & 0.309 & {\bf 0.270} & & 0.251 & {\bf 0.200} \\
 &  & 4 & 12 & {\bf 0.084} & 0.149 & & {\bf 0.0208} & 0.0294 & & 0.700 & {\bf 0.586} & & 0.558 & {\bf 0.504} \\
 & \multirow{3}{*}{5} & 1 & 100 & -0.024 & {\bf 0.001} & & 0.0017 & {\bf 0.0012} & & 0.124 & 0.124 & & 0.116 & {\bf 0.100} \\
 &  & 2 & 100 & -0.036 & {\bf 0.007} & & 0.0027 & {\bf 0.0014} & & {\bf 0.237} & 0.238 & & 0.212 & {\bf 0.184} \\
 &  & 4 & 99 & -0.063 & {\bf 0.040} & & 0.0060 & {\bf 0.0038} & & {\bf 0.467} & 0.469 & & 0.378 & {\bf 0.349} \\
 & \multirow{3}{*}{20} & 1 & 100 & -0.022 & {\bf 0.000} & & 0.0016 & {\bf 0.0012} & & {\bf 0.126} & 0.129 & & 0.115 & {\bf 0.104} \\
 &  & 2 & 100 & -0.033 & {\bf 0.004} & & 0.0024 & {\bf 0.0014} & & {\bf 0.240} & 0.246 & & 0.210 & {\bf 0.191} \\
 &  & 4 & 100 & -0.057 & {\bf 0.031} & & 0.0051 & {\bf 0.0028} & & {\bf 0.461} & 0.474 & & 0.373 & {\bf 0.354} \\
\vspace{-2mm} \\
\cmidrule{2-15}
\vspace{-2mm} \\
\multirow{9}{*}{100} & \multirow{3}{*}{2} & 1 & 100 & -0.013 & {\bf 0.001} & & 0.0007 & {\bf 0.0006} & & 0.059 & {\bf 0.057} & & 0.058 & {\bf 0.050} \\
 &  & 2 & 100 & -0.021 & {\bf 0.005} & & 0.0011 & {\bf 0.0007} & & 0.110 & {\bf 0.106} & & 0.106 & {\bf 0.092} \\
 &  & 4 & 95 & -0.034 & {\bf 0.024} & & 0.0024 & {\bf 0.0018} & & 0.213 & {\bf 0.207} & & 0.190 & {\bf 0.174} \\
 & \multirow{3}{*}{5} & 1 & 100 & -0.012 & {\bf 0.000} & & 0.0007 & {\bf 0.0006} & & {\bf 0.058} & 0.059 & & 0.056 & {\bf 0.052} \\
 &  & 2 & 100 & -0.018 & {\bf 0.003} & & 0.0010 & {\bf 0.0007} & & 0.109 & 0.109 & & 0.103 & {\bf 0.096} \\
 &  & 4 & 100 & -0.031 & {\bf 0.016} & & 0.0018 & {\bf 0.0012} & & {\bf 0.200} & 0.202 & & 0.181 & {\bf 0.173} \\
 & \multirow{3}{*}{20} & 1 & 100 & -0.011 & {\bf 0.000} & & 0.0007 & {\bf 0.0006} & & {\bf 0.059} & 0.060 & & 0.056 & {\bf 0.053} \\
 &  & 2 & 100 & -0.017 & {\bf 0.002} & & 0.0009 & {\bf 0.0006} & & {\bf 0.110} & 0.111 & & 0.102 & {\bf 0.098} \\
 &  & 4 & 100 & -0.028 & {\bf 0.014} & & 0.0016 & {\bf 0.0011} & & {\bf 0.201} & 0.205 & & 0.181 & {\bf 0.176} \\
\vspace{-3mm} \\
\bottomrule
\vspace{+1mm}
\end{tabular}

\end{center}
\end{table*}

\subsection{Comparison with alternative estimators}
\label{sec:exp:compare}

We also compared MML87 with alternatives that, like it, use the sample eigenvectors and differ only in the residual variance and the fitted signal eigenvalues. These are maximum likelihood with the first-order bias correction implied by Theorem~\ref{thm:mml:bias} (BC), $\widehat\tau_{\mathrm{BC}}=\widehat\tau_{\mathrm{ML}}\{1+[J+1+\sum_j\widehat\tau_{\mathrm{ML}}/(s_j-\widehat\tau_{\mathrm{ML}})]/N\}$; the bias-corrected residual variance of \cite{PassemierLiYao17} together with their spike estimates (PLY); and the optimal eigenvalue shrinkers of \cite{DonohoGavishJohnstone18} for entropy loss (DGJ-E) and Stein loss (DGJ-S), computed with the PLY residual variance. All estimators were fitted at the matched rank of Section~\ref{sec:exp:pest}. We used three designs with $10^5$ replications each: the design of Table~\ref{tab:results:pest}; the same design with geometric loadings, in which the weakest component has signal-to-noise ratio $\rho$ and $\rho_j/\rho_{j+1}=1/0.75$; and a higher-dimensional design with equal loadings, $K\in\{20,50,100\}$, $J\in\{1,\lceil K/10\rceil,\lceil K/4\rceil\}$, and either $N=10K$ with $\rho\in\{5,20\}$ or $N=2K$ with $\rho\in\{10,20\}$. Table~\ref{tab:results:compare} counts the settings in which each estimator attains the smallest average loss. The full results are in Appendix~\ref{sec:supp:compare}.

The results agree with the theory of Sections~\ref{sec:finite-sample-bias} and~\ref{sec:stein}. In the $77$ settings in which a regular $J$-component solution exists for at least $90\%$ of the data sets, MML87 estimates the residual scale more accurately than maximum likelihood in every case, in both $|S_1|$ and $S_2$. PLY and BC, which remove the first-order bias of maximum likelihood, are more accurate still. The MML87 squared error is at most about twice that of PLY for $K=10$, and up to about eight times when $J=K/4$, which reflects the positive MML87 bias of order $J^2/\{N(K-J)\}$ in Theorem~\ref{thm:mml:bias}. Under the forward KL divergence, PLY or BC is best in $79$ of the $90$ settings, because MML87 shrinks the fitted signal eigenvalues. Under the reverse KL divergence, the Stein loss, MML87 is best in $66$ of the $90$ settings and the Stein-optimal shrinker DGJ-S in the remainder. In the regular settings the two differ by a median of $0.4\%$ or less, and by at most $7.1\%$, $4.9\%$ and $1.3\%$ in the three designs. Although MML87 is not designed for any particular loss, its loading shrinkage is thus close to Stein-optimal in finite samples, as Proposition~\ref{prop:stein} predicts in the proportional regime.

\begin{table}[tb]
\scriptsize
\begin{center}
\caption{Number of settings in which each estimator attains the smallest average loss at a matched rank, over $10^5$ simulated data sets per setting. The DGJ shrinkers use the PLY residual variance, so their $S_2$ equals that of PLY. Full results are in Tables~\ref{tab:supp:equal:S2}--\ref{tab:supp:highdim:rKL} of Appendix~\ref{sec:supp:compare}.}
\label{tab:results:compare}
\begin{tabular}{lcccccccc}
\toprule
Design & Settings & Criterion & MLE & MML87 & BC & PLY & DGJ-E & DGJ-S \\
\midrule
\multirow{3}{*}{Equal} & \multirow{3}{*}{27} & $S_2$ & 4 & 1 & 12 & 10 & -- & -- \\
 &  & Forward KL & 0 & 5 & 8 & 14 & 0 & 0 \\
 &  & Reverse KL & 0 & 18 & 0 & 0 & 0 & 9 \\
\midrule
\multirow{3}{*}{Geometric} & \multirow{3}{*}{27} & $S_2$ & 3 & 1 & 13 & 10 & -- & -- \\
 &  & Forward KL & 0 & 3 & 6 & 18 & 0 & 0 \\
 &  & Reverse KL & 0 & 19 & 0 & 0 & 0 & 8 \\
\midrule
\multirow{3}{*}{Higher-dimensional} & \multirow{3}{*}{36} & $S_2$ & 0 & 0 & 16 & 20 & -- & -- \\
 &  & Forward KL & 0 & 0 & 9 & 24 & 3 & 0 \\
 &  & Reverse KL & 0 & 29 & 0 & 0 & 0 & 7 \\
\bottomrule
\end{tabular}

\end{center}
\end{table}

\subsection{Latent-dimension selection}
\label{sec:exp:model-selection}

\begin{table*}[p]
\scriptsize
\begin{center}
\caption{Latent-dimension selection for pure noise ($J=0$) and equal per-component signal-to-noise ratio $\rho=2$, over $10^5$ simulated data sets with $K=10$ and candidate dimensions $0,\ldots,6$. Bold marks the smallest KL divergences and the highest exact-recovery rate. Excluding SPA, the largest Monte Carlo standard error of the KL divergences is $0.002$.}
\label{tab:results:msel}
\begin{tabular}{ccccccccc}
\toprule
$N$ & $J$ & Method & $\operatorname{KL}(\boldsymbol{\Sigma}_0,\widehat{\boldsymbol{\Sigma}})$ & $\operatorname{KL}(\widehat{\boldsymbol{\Sigma}},\boldsymbol{\Sigma}_0)$ & ~ & \multicolumn{3}{c}{Model selection (\%)} \\
    &     &        &  &  & ~ & $<J$ & $=J$ & $>J$ \\
\cmidrule{1-9}
\multirow{20}{*}{25} & \multirow{5}{*}{0} & MML & {\bf 0.022} & {\bf 0.022} & & -- & 99.32 & 0.68 \\
 &  & BIC & 0.028 & 0.026 & & -- & {\bf 99.77} & 0.23 \\
 &  & Bayes & 0.056 & 0.060 & & -- & 94.07 & 5.93 \\
 &  & GIC & 0.055 & 0.051 & & -- & 96.48 & 3.52 \\
 &  & SPA & 0.032 & 0.032 & & -- & 98.72 & 1.28 \\
 & \multirow{5}{*}{1} & MML & {\bf 0.285} & {\bf 0.193} & & 55.08 & 44.81 & 0.11 \\
 &  & BIC & 0.354 & 0.255 & & 65.01 & 34.77 & 0.22 \\
 &  & Bayes & 0.339 & 0.285 & & 32.60 & {\bf 62.64} & 4.75 \\
 &  & GIC & 0.407 & 0.320 & & 39.48 & 51.44 & 9.07 \\
 &  & SPA & 0.349 & 0.255 & & 61.55 & 38.28 & 0.17 \\
 & \multirow{5}{*}{2} & MML & {\bf 0.522} & {\bf 0.387} & & 90.35 & 9.65 & 0.01 \\
 &  & BIC & 0.622 & 0.467 & & 88.97 & 10.87 & 0.16 \\
 &  & Bayes & 0.618 & 0.481 & & 59.53 & {\bf 37.02} & 3.44 \\
 &  & GIC & 0.752 & 0.562 & & 51.62 & 33.29 & 15.09 \\
 &  & SPA & 0.620 & 0.461 & & 95.29 & 4.71 & 0.00 \\
 & \multirow{5}{*}{4} & MML & {\bf 0.759} & {\bf 0.792} & & 99.98 & 0.02 & 0.00 \\
 &  & BIC & 0.847 & 0.819 & & 99.31 & 0.64 & 0.04 \\
 &  & Bayes & 1.046 & 0.803 & & 88.67 & 9.86 & 1.46 \\
 &  & GIC & 1.258 & 0.906 & & 62.76 & {\bf 18.29} & 18.95 \\
 &  & SPA & 0.821 & 0.812 & & 100.00 & 0.00 & 0.00 \\
\vspace{-2mm} \\
\cmidrule{2-9}
\vspace{-2mm} \\
\multirow{20}{*}{50} & \multirow{5}{*}{0} & MML & {\bf 0.011} & {\bf 0.011} & & -- & 98.82 & 1.18 \\
 &  & BIC & 0.012 & {\bf 0.011} & & -- & {\bf 99.99} & 0.01 \\
 &  & Bayes & 0.021 & 0.024 & & -- & 95.73 & 4.27 \\
 &  & GIC & 0.020 & 0.022 & & -- & 96.79 & 3.21 \\
 &  & SPA & 0.014 & 0.015 & & -- & 98.83 & 1.17 \\
 & \multirow{5}{*}{1} & MML & {\bf 0.134} & {\bf 0.102} & & 10.69 & 88.64 & 0.67 \\
 &  & BIC & 0.210 & 0.160 & & 35.19 & 64.79 & 0.02 \\
 &  & Bayes & 0.148 & 0.139 & & 5.83 & {\bf 89.87} & 4.30 \\
 &  & GIC & 0.163 & 0.152 & & 7.14 & 84.86 & 8.00 \\
 &  & SPA & 0.173 & 0.144 & & 20.12 & 79.61 & 0.27 \\
 & \multirow{5}{*}{2} & MML & {\bf 0.271} & {\bf 0.201} & & 31.01 & 68.79 & 0.20 \\
 &  & BIC & 0.422 & 0.322 & & 63.13 & 36.84 & 0.03 \\
 &  & Bayes & 0.282 & 0.246 & & 15.20 & {\bf 80.44} & 4.37 \\
 &  & GIC & 0.305 & 0.270 & & 11.04 & 74.25 & 14.71 \\
 &  & SPA & 0.406 & 0.311 & & 60.16 & 39.81 & 0.03 \\
 & \multirow{5}{*}{4} & MML & 0.588 & 0.507 & & 88.09 & 11.91 & 0.00 \\
 &  & BIC & 0.751 & 0.742 & & 94.27 & 5.71 & 0.03 \\
 &  & Bayes & 0.552 & {\bf 0.434} & & 37.85 & 57.83 & 4.32 \\
 &  & GIC & {\bf 0.551} & 0.447 & & 15.59 & {\bf 58.54} & 25.87 \\
 &  & SPA & 0.758 & 0.725 & & 99.50 & 0.49 & 0.00 \\
\vspace{-2mm} \\
\cmidrule{2-9}
\vspace{-2mm} \\
\multirow{20}{*}{100} & \multirow{5}{*}{0} & MML & {\bf 0.005} & {\bf 0.005} & & -- & 99.74 & 0.26 \\
 &  & BIC & {\bf 0.005} & {\bf 0.005} & & -- & {\bf 100.00} & 0.00 \\
 &  & Bayes & 0.009 & 0.010 & & -- & 97.14 & 2.86 \\
 &  & GIC & 0.009 & 0.010 & & -- & 96.79 & 3.21 \\
 &  & SPA & 0.007 & 0.007 & & -- & 98.94 & 1.06 \\
 & \multirow{5}{*}{1} & MML & {\bf 0.058} & {\bf 0.051} & & 0.37 & 97.55 & 2.08 \\
 &  & BIC & 0.070 & 0.064 & & 3.91 & 96.09 & 0.00 \\
 &  & Bayes & 0.062 & 0.062 & & 0.07 & 96.79 & 3.15 \\
 &  & GIC & 0.066 & 0.067 & & 0.07 & 93.90 & 6.04 \\
 &  & SPA & 0.062 & 0.060 & & 0.96 & {\bf 98.94} & 0.10 \\
 & \multirow{5}{*}{2} & MML & {\bf 0.107} & {\bf 0.093} & & 0.34 & {\bf 98.55} & 1.11 \\
 &  & BIC & 0.139 & 0.121 & & 9.69 & 90.31 & 0.01 \\
 &  & Bayes & 0.114 & 0.110 & & 0.23 & 96.47 & 3.29 \\
 &  & GIC & 0.121 & 0.117 & & 0.12 & 90.21 & 9.66 \\
 &  & SPA & 0.139 & 0.123 & & 9.07 & 90.93 & 0.01 \\
 & \multirow{5}{*}{4} & MML & {\bf 0.206} & {\bf 0.174} & & 5.04 & 94.83 & 0.13 \\
 &  & BIC & 0.325 & 0.292 & & 28.92 & 71.05 & 0.03 \\
 &  & Bayes & 0.207 & 0.189 & & 1.08 & {\bf 94.87} & 4.05 \\
 &  & GIC & 0.218 & 0.199 & & 0.27 & 81.88 & 17.85 \\
 &  & SPA & 0.480 & 0.423 & & 68.53 & 31.46 & 0.00 \\
\vspace{-3mm} \\
\bottomrule
\vspace{+1mm}
\end{tabular}

\end{center}
\end{table*}

\begin{table*}[p]
\scriptsize
\begin{center}
\caption{Latent-dimension selection as in Table~\ref{tab:results:msel}, with equal per-component signal-to-noise ratio $\rho=5$.}
\label{tab:results:msel:snr8}
\begin{tabular}{ccccccccc}
\toprule
$N$ & $J$ & Method & $\operatorname{KL}(\boldsymbol{\Sigma}_0,\widehat{\boldsymbol{\Sigma}})$ & $\operatorname{KL}(\widehat{\boldsymbol{\Sigma}},\boldsymbol{\Sigma}_0)$ & ~ & \multicolumn{3}{c}{Model selection (\%)} \\
    &     &        &  &  & ~ & $<J$ & $=J$ & $>J$ \\
\cmidrule{1-9}
\multirow{15}{*}{25} & \multirow{5}{*}{1} & MML & {\bf 0.294} & {\bf 0.191} & & 3.16 & {\bf 96.59} & 0.25 \\
 &  & BIC & 0.324 & 0.259 & & 4.94 & 94.59 & 0.47 \\
 &  & Bayes & 0.322 & 0.279 & & 1.02 & 93.43 & 5.55 \\
 &  & GIC & 0.383 & 0.315 & & 1.52 & 89.03 & 9.45 \\
 &  & SPA & 0.345 & 0.266 & & 7.56 & 92.40 & 0.04 \\
 & \multirow{5}{*}{2} & MML & {\bf 0.620} & {\bf 0.380} & & 15.39 & 84.55 & 0.06 \\
 &  & BIC & 0.669 & 0.484 & & 12.60 & 86.40 & 1.00 \\
 &  & Bayes & 0.630 & 0.485 & & 4.10 & {\bf 90.59} & 5.31 \\
 &  & GIC & 0.761 & 0.566 & & 3.03 & 79.19 & 17.78 \\
 &  & SPA & 1.078 & 0.697 & & 57.51 & 42.49 & 0.00 \\
 & \multirow{5}{*}{4} & MML & 1.525 & 1.345 & & 83.39 & 16.61 & 0.00 \\
 &  & BIC & 1.511 & 1.024 & & 35.27 & 62.06 & 2.67 \\
 &  & Bayes & {\bf 1.380} & {\bf 0.862} & & 18.46 & {\bf 76.31} & 5.23 \\
 &  & GIC & 1.502 & 0.943 & & 5.55 & 62.35 & 32.10 \\
 &  & SPA & 2.000 & 2.260 & & 99.96 & 0.04 & 0.00 \\
\vspace{-2mm} \\
\cmidrule{2-9}
\vspace{-2mm} \\
\multirow{15}{*}{50} & \multirow{5}{*}{1} & MML & {\bf 0.125} & {\bf 0.101} & & 0.01 & 99.02 & 0.97 \\
 &  & BIC & {\bf 0.125} & 0.116 & & 0.07 & {\bf 99.90} & 0.04 \\
 &  & Bayes & 0.135 & 0.128 & & 0.00 & 95.76 & 4.24 \\
 &  & GIC & 0.141 & 0.133 & & 0.00 & 94.27 & 5.73 \\
 &  & SPA & 0.126 & 0.117 & & 0.14 & 99.86 & 0.00 \\
 & \multirow{5}{*}{2} & MML & {\bf 0.238} & {\bf 0.184} & & 0.03 & 99.50 & 0.47 \\
 &  & BIC & 0.239 & 0.213 & & 0.24 & {\bf 99.67} & 0.09 \\
 &  & Bayes & 0.249 & 0.223 & & 0.01 & 95.57 & 4.43 \\
 &  & GIC & 0.265 & 0.238 & & 0.01 & 90.61 & 9.38 \\
 &  & SPA & 0.331 & 0.256 & & 9.06 & 90.94 & 0.00 \\
 & \multirow{5}{*}{4} & MML & 0.470 & {\bf 0.350} & & 1.21 & {\bf 98.77} & 0.02 \\
 &  & BIC & {\bf 0.466} & 0.379 & & 0.93 & 98.65 & 0.42 \\
 &  & Bayes & 0.472 & 0.387 & & 0.08 & 94.72 & 5.20 \\
 &  & GIC & 0.501 & 0.409 & & 0.03 & 82.21 & 17.76 \\
 &  & SPA & 1.423 & 1.332 & & 81.02 & 18.98 & 0.00 \\
\vspace{-2mm} \\
\cmidrule{2-9}
\vspace{-2mm} \\
\multirow{15}{*}{100} & \multirow{5}{*}{1} & MML & 0.060 & {\bf 0.053} & & 0.00 & 97.58 & 2.42 \\
 &  & BIC & {\bf 0.058} & 0.056 & & 0.00 & {\bf 100.00} & 0.00 \\
 &  & Bayes & 0.062 & 0.061 & & 0.00 & 96.90 & 3.10 \\
 &  & GIC & 0.065 & 0.064 & & 0.00 & 94.96 & 5.04 \\
 &  & SPA & 0.059 & 0.056 & & 0.00 & {\bf 100.00} & 0.00 \\
 & \multirow{5}{*}{2} & MML & 0.110 & {\bf 0.096} & & 0.00 & 98.54 & 1.46 \\
 &  & BIC & {\bf 0.109} & 0.103 & & 0.00 & {\bf 99.99} & 0.01 \\
 &  & Bayes & 0.112 & 0.107 & & 0.00 & 96.84 & 3.16 \\
 &  & GIC & 0.117 & 0.112 & & 0.00 & 92.99 & 7.01 \\
 &  & SPA & 0.112 & 0.105 & & 0.29 & 99.71 & 0.00 \\
 & \multirow{5}{*}{4} & MML & 0.202 & {\bf 0.173} & & 0.00 & 99.70 & 0.30 \\
 &  & BIC & {\bf 0.200} & 0.181 & & 0.00 & {\bf 99.96} & 0.04 \\
 &  & Bayes & 0.204 & 0.185 & & 0.00 & 96.09 & 3.91 \\
 &  & GIC & 0.212 & 0.193 & & 0.00 & 87.58 & 12.42 \\
 &  & SPA & 0.384 & 0.360 & & 15.03 & 84.97 & 0.00 \\
\vspace{-3mm} \\
\bottomrule
\vspace{+1mm}
\end{tabular}

\end{center}
\end{table*}

We next examine selection of the latent dimension. For each candidate dimension, MML87 minimises the concentrated codelength over the admissible residual variance interval and evaluates the resulting full message length, retaining all terms that depend on the candidate dimension. The selected dimension is
\begin{equation}
\widehat{J}_{\mathrm{MML}}
=
\argmin_{J\in\mathcal{J}_{\mathrm{adm}}}
I_{87}^{\mathrm{conc}}
\left(
\mathbf{X},\widehat{\tau}_{\mathrm{MML},J};J
\right),
\label{eqn:simulation:Jhat}
\end{equation}
where $\mathcal{J}_{\mathrm{adm}}$ is the set of candidate dimensions for which a regular interior MML87 solution exists. In every configuration considered below, $\widehat{J}_{\mathrm{MML}}$ coincided with the largest admissible dimension, $\max\mathcal{J}_{\mathrm{adm}}$, in at least $96.9\%$ of the data sets. The threshold \eqref{eqn:mml:component:threshold} is necessary but not sufficient for a regular solution, and the selection compares full codelengths. Empirically, however, the existence of a regular solution almost always determines whether a component is retained.

We compared MML87 with four established alternatives: the Bayesian
information criterion (BIC), the Laplace approximation to the marginal
likelihood proposed by \cite{Minka00}, denoted Bayes, signflip parallel
analysis (SPA) \cite{HongEtAl23}, and the generalized information
criterion (GIC) \cite{HungEtAl22}. For BIC we used the criterion of
\cite{Minka00}, extended to allow the zero-dimensional model. For SPA
we used pairwise comparison with sequential stopping and compared each
sample singular value with the maximum of $100$ sign-flip draws, the
setting used by \cite{HongEtAl23}. By their Proposition~3.2, the
probability of selecting a component under pure noise is then at most
$1/101$. The
competing methods use maximum likelihood estimates at their selected
dimension, and MML87 uses its own estimates. The data were generated as
in Section~\ref{sec:exp:pest} with equal per-component signal-to-noise
ratio $\rho\in\{2,5,20\}$, $N\in\{25,50,100\}$, $K=10$ and
$J\in\{0,1,2,4\}$, where $J=0$ corresponds to pure noise. Each method
selected among candidate dimensions $0,\ldots,6$. We recorded the KL
divergences in both directions and classified the selected dimension as
underfitting ($\widehat J<J$), exactly recovering ($\widehat J=J$), or
overfitting ($\widehat J>J$) the generating dimension.
Tables~\ref{tab:results:msel} and~\ref{tab:results:msel:snr8} report
the results for pure noise and $\rho=2$, and for $\rho=5$,
respectively.

Under pure noise, MML87, BIC and SPA selected the zero-dimensional model
in at least $98.7\%$ of the data sets, compared with $94$--$97\%$ for
Bayes and GIC. For strong signals ($\rho=20$,
not tabulated), MML87 recovered the generating dimension in at least
$97.5\%$ of the data sets in every configuration, BIC in at least
$96.0\%$, and Bayes in $94$--$97\%$, while GIC and SPA were less
reliable. With several components and $N=100$, for example, GIC
recovered $89\%$ and SPA $97\%$ at $J=4$. With $N=25$ and $J=4$, they
recovered $78\%$ and $2\%$, respectively. SPA was the most conservative
method with several components and a small sample.

The methods differ most near the detection limit. Because the
retention threshold lies above the noise bulk, MML87 is conservative
when the signal is weak. At $\rho=2$ it recovers the generating dimension
considerably more often than BIC once $N\geq50$; for example, $68.8\%$
against $36.8\%$ for $N=50$, $\rho=2$, $J=2$, and $94.8\%$ against
$71.0\%$ for $N=100$, $\rho=2$, $J=4$. With several moderate components
and a small sample, however, MML87 underfits more often than Bayes,
BIC and GIC. For $N=25$, $\rho=5$, $J=4$ it recovered the generating
dimension in $16.6\%$ of the data sets, compared with $76.3\%$ for
Bayes. Bayes was the most successful method in these settings.

Exact recovery of the generating dimension does not necessarily
minimise covariance-estimation loss. Across all thirty configurations,
MML87 gave the smallest reverse KL divergence in $28$ and the smallest
forward KL divergence in $15$ ($16$ including ties at the reported
precision). Its advantage under the reverse
divergence reflects both its retention rule and the loading shrinkage
analysed in Proposition~\ref{prop:stein}. When the signal is strong and
all methods select the generating dimension, maximum likelihood
estimates at the same dimension give a slightly smaller forward
divergence, as in Section~\ref{sec:exp:pest}.

Overall, the experiments support two empirical conclusions. First, for
a fixed latent dimension, MML87 substantially reduces the downward
finite-sample bias of the maximum likelihood residual variance
estimator, and its loading shrinkage lowers Stein loss at a small cost
in entropy loss when the signal is strong. Second, when the latent dimension is unknown, MML87
retains a component essentially only when a regular solution exists. It
almost never selects
spurious components and recovers strong components reliably. Near the
detection limit, it is more conservative than the Laplace approximation
of \cite{Minka00}, particularly with several components and small
samples. Its fitted covariance matrices have the smallest Stein loss in
most configurations.

\section{Discussion}
\label{sec:discussion}

We have developed a minimum message length (MML) formulation of probabilistic principal component analysis under the multivariate Gaussian model of \cite{TippingBishop99}. The resulting criterion provides a unified framework for parameter estimation and latent-dimension selection. Although the MML and maximum likelihood estimators share the same loading directions, the two approaches differ materially in their estimates of the residual variance and of the loading magnitudes. This difference is consequential in finite samples: the MML criterion yields an automatic rule for selecting the number of components and an estimator of the residual variance that is less prone to the downward bias of maximum likelihood.

The theoretical analysis clarifies the nature of this correction. For a fixed residual variance, the MML87 loading estimates solve separate quadratic equations. The codelength therefore reduces to a one-dimensional function of the residual variance, and a regular solution exists only above an explicit spectral threshold. In the fixed-dimension regime, the finite-sample bias analysis shows that MML87 replaces the negative first-order bias of maximum likelihood with a positive one. This positive bias is smaller in magnitude unless the latent dimension is a substantial fraction of the ambient dimension. In the proportional high-dimensional regime, the loading shrinkage implied by the codelength approaches the Stein-optimal shrinkage as the signal strength grows, whereas maximum likelihood approaches the entropy-optimal one. For every fixed signal strength above the spectral threshold, MML87 also has smaller limiting Stein loss than maximum likelihood. The retention threshold lies strictly above the upper edge of the Marchenko--Pastur noise bulk, so the criterion retains a component only when its sample eigenvalue is separated from the noise spectrum.

The simulation results agree with these findings. For a fixed latent dimension, MML87 improves on maximum likelihood in the bias and squared log-error of the residual scale and in Stein loss, while maximum likelihood is slightly better under entropy loss when the signal is strong. In the model-selection experiments, MML87 almost never selects spurious components under pure noise and recovers strong components reliably. Near the detection limit it is more conservative than the Laplace approximation of \cite{Minka00}, particularly with several components and small samples, and this is its main limitation. Its fitted covariance matrices have the smallest Stein loss in most configurations. Exact recovery of the latent dimension is not always synonymous with the best covariance approximation.

A further contribution of the paper is the prior construction for the loading matrix. Rather than imposing orthogonality on the unknown loading vectors a priori, we derive an induced prior from a rotation-invariant, heavy-tailed distribution on the underlying scaled loadings. This leads naturally to a matrix-variate Cauchy prior combined with the uniform distribution on the corresponding Stiefel manifold. In our view, this is a more appropriate default specification for probabilistic PCA than assuming orthogonality of the latent structure at the prior stage.

The present development also suggests several extensions. Because the criterion is expressed as a codelength, it can in principle be used to compare probabilistic PCA with alternative covariance models within a common inferential framework. The same construction may also be embedded in richer models, including mixtures of probabilistic principal component analysers and tree- or partition-based multivariate models. Extensions to heteroscedastic noise structures and non-Gaussian latent-variable models, and a fuller analysis of rank selection in the proportional high-dimensional regime, are natural directions for future work.

Overall, the proposed MML formulation provides a theoretically grounded and practically effective alternative to standard maximum likelihood analysis of probabilistic PCA. It combines automatic rank determination with improved residual variance estimation, while retaining a tractable computational structure and a clear connection to modern spectral theory.
\section*{Data availability}
%
The data and code in this article are available in the GitHub repository:
\href{https://github.com/EnesMakalic/MML-PCA}{https://github.com/EnesMakalic/MML-PCA}.
%

%
\section*{Acknowledgment}
Generative AI tools (Claude Opus 4.5 and Claude Opus 5.5, Anthropic; Microsoft Copilot GPT-5.4) were used in the preparation and revision of this manuscript for generation and exploration of ideas, checking of derivations, development and verification of the accompanying code, and language improvement. The authors reviewed all such material and take full responsibility for the content of the manuscript.

\appendices

\section{Derivation of the expected Fisher information determinant}
\label{sec:supp:fisher}

This appendix derives the expected Fisher information determinant for the probabilistic PCA model
\begin{equation}
\mathbf{x}_n
\sim \mathcal{N}_K(\boldsymbol{\mu},\boldsymbol{\Sigma}),
\qquad
\boldsymbol{\Sigma}
=
\mathbf{R}\boldsymbol{\Lambda}^2\mathbf{R}^{\prime}
+
\sigma^2\mathbf{I}_K,
\qquad
n=1,\ldots,N,
\label{eqn:supp:ppca}
\end{equation}
where
\[
\boldsymbol{\Lambda}
=
\operatorname{diag}(\lambda_1,\ldots,\lambda_J),
\qquad
\lambda_1>\cdots>\lambda_J>0,
\qquad
\sigma>0,
\]
and
\[
\mathbf{R}\in\mathcal{R}_{K,J}
=
\operatorname{St}(K,J)/\mathcal{D}_J.
\]
Here,
\[
\mathcal{D}_J
=
\left\{
\operatorname{diag}(d_1,\ldots,d_J):
d_j\in\{-1,1\}
\right\}
\]
is the group of independent sign changes of the columns of
$\mathbf{R}$. Since $\mathcal{D}_J$ is finite, the quotient has the
same local dimension as the Stiefel manifold.
The covariance parameter vector is
\[
\boldsymbol{\theta}_{\Sigma}
=
(\boldsymbol{\lambda},\sigma,\mathbf{R}),
\]
with dimension
\begin{align}
P_{\Sigma}
&=
J+1+\dim\{\operatorname{St}(K,J)\}
=
KJ-\frac{J(J-1)}{2}+1.
\label{eqn:supp:covariance-dimension}
\end{align}
Including the $K$ mean parameters gives
\begin{equation}
P
=
K+P_{\Sigma}
=
K(J+1)-\frac{J(J-1)}{2}+1.
\label{eqn:supp:full-dimension}
\end{equation}

\subsection{Gaussian Fisher metric}
For a multivariate Gaussian model, the mean and covariance parameters
are orthogonal under the expected Fisher information:
\begin{equation}
\mathcal{F}_{\boldsymbol{\mu},\boldsymbol{\Sigma}}
=
\mathbf{0}.
\label{eqn:supp:mean-covariance-orthogonal}
\end{equation}
Consequently,
\begin{equation}
\left|
\mathcal{F}_{\mathrm{full}}
\right|
=
\left|
\mathcal{F}_{\mu\mu}
\right|
\left|
\mathcal{F}_{\Sigma\Sigma}
\right|.
\label{eqn:supp:fisher-factorization}
\end{equation}
The mean block is
\[
\mathcal{F}_{\mu\mu}
=
N\boldsymbol{\Sigma}^{-1},
\]
and hence
\begin{align}
\left|
\mathcal{F}_{\mu\mu}
\right|
&=
N^K|\boldsymbol{\Sigma}|^{-1}.
\label{eqn:supp:fisher-mean-general}
\end{align}
Since the eigenvalues of $\boldsymbol{\Sigma}$ are
\[
\lambda_1^2+\sigma^2,\ldots,\lambda_J^2+\sigma^2,
\underbrace{\sigma^2,\ldots,\sigma^2}_{K-J\ {\rm times}},
\]
we have
\begin{equation}
\left|
\mathcal{F}_{\mu\mu}
\right|
=
\frac{N^K}{\sigma^{2(K-J)}}
\prod_{j=1}^J
\frac{1}{\lambda_j^2+\sigma^2}.
\label{eqn:supp:fisher-mean}
\end{equation}
For the covariance parameters, the expected Fisher information has
entries
\begin{equation}
\mathcal{F}_{a,b}
=
\frac{N}{2}
\operatorname{tr}
\left(
\boldsymbol{\Sigma}^{-1}
\frac{\partial\boldsymbol{\Sigma}}{\partial\theta_a}
\boldsymbol{\Sigma}^{-1}
\frac{\partial\boldsymbol{\Sigma}}{\partial\theta_b}
\right).
\label{eqn:supp:fisher-general}
\end{equation}
Equivalently, the corresponding Fisher metric is
\begin{equation}
ds_{\mathcal F}^2
=
\frac{N}{2}
\operatorname{tr}
\left(
\boldsymbol{\Sigma}^{-1}d\boldsymbol{\Sigma}
\boldsymbol{\Sigma}^{-1}d\boldsymbol{\Sigma}
\right).
\label{eqn:supp:fisher-metric}
\end{equation}

\subsection{Covariance eigenbasis and tangent decomposition}
Let
\[
\mathbf{L}
=
\boldsymbol{\Lambda}^2
=
\operatorname{diag}(\lambda_1^2,\ldots,\lambda_J^2).
\]
Choose an orthonormal complement
$\mathbf{R}_{\perp}\in\mathbb{R}^{K\times(K-J)}$ and define
\begin{equation}
\mathbf{Q}
=
[\mathbf{R}\ \mathbf{R}_{\perp}]
\in O(K).
\label{eqn:supp:Q}
\end{equation}
In this basis,
\begin{align}
\mathbf{Q}^{\prime}
\boldsymbol{\Sigma}
\mathbf{Q}
&=
\boldsymbol{\Sigma}_{\mathrm{orth}}
=
\begin{pmatrix}
\mathbf{T} & \mathbf{0}\\
\mathbf{0} & \sigma^2\mathbf{I}_{K-J}
\end{pmatrix},
\label{eqn:supp:Sigma-orth}
\\
\mathbf{T}
&=
\operatorname{diag}
\left(
\lambda_1^2+\sigma^2,\ldots,
\lambda_J^2+\sigma^2
\right).
\label{eqn:supp:T}
\end{align}
Therefore,
\begin{equation}
\boldsymbol{\Sigma}_{\mathrm{orth}}^{-1}
=
\begin{pmatrix}
\mathbf{T}^{-1} & \mathbf{0}\\
\mathbf{0} & \sigma^{-2}\mathbf{I}_{K-J}
\end{pmatrix}.
\label{eqn:supp:Sigma-orth-inverse}
\end{equation}
Since $\mathbf{R}^{\prime}\mathbf{R}=\mathbf{I}_J$, every tangent
perturbation of $\mathbf{R}$ can be written as
\begin{equation}
d\mathbf{R}
=
\mathbf{R}\boldsymbol{\Omega}_A
+
\mathbf{R}_{\perp}\boldsymbol{\Omega}_B,
\label{eqn:supp:dR}
\end{equation}
where
\[
\boldsymbol{\Omega}_A^{\prime}
=
-\boldsymbol{\Omega}_A,
\qquad
\boldsymbol{\Omega}_B
\in\mathbb{R}^{(K-J)\times J}.
\]
The skew-symmetric matrix $\boldsymbol{\Omega}_A$ describes rotations
among the ordered signal directions, whereas
$\boldsymbol{\Omega}_B$ describes rotations between the signal and
noise subspaces.
The covariance differential is
\begin{align}
d\boldsymbol{\Sigma}
&=
(d\mathbf{R})\mathbf{L}\mathbf{R}^{\prime}
+
\mathbf{R}\mathbf{L}(d\mathbf{R})^{\prime}
+
\mathbf{R}(d\mathbf{L})\mathbf{R}^{\prime}
+
2\sigma\mathbf{I}_K\,d\sigma,
\label{eqn:supp:dSigma}
\\
d\mathbf{L}
&=
\operatorname{diag}
\left(
2\lambda_1\,d\lambda_1,\ldots,
2\lambda_J\,d\lambda_J
\right).
\end{align}
Substituting \eqref{eqn:supp:dR} into
\eqref{eqn:supp:dSigma} separates the perturbation into
\begin{equation}
d\boldsymbol{\Sigma}
=
d\boldsymbol{\Sigma}_{\mathrm{spec}}
+
d\boldsymbol{\Sigma}_{\mathrm{int}}
+
d\boldsymbol{\Sigma}_{\mathrm{ext}},
\label{eqn:supp:dSigma-decompose}
\end{equation}
where
\begin{align}
d\boldsymbol{\Sigma}_{\mathrm{spec}}
&=
\mathbf{R}(d\mathbf{L})\mathbf{R}^{\prime}
+
2\sigma\mathbf{I}_K\,d\sigma,
\\
d\boldsymbol{\Sigma}_{\mathrm{int}}
&=
\mathbf{R}
\left(
\boldsymbol{\Omega}_A\mathbf{L}
-
\mathbf{L}\boldsymbol{\Omega}_A
\right)
\mathbf{R}^{\prime},
\\
d\boldsymbol{\Sigma}_{\mathrm{ext}}
&=
\mathbf{R}_{\perp}\boldsymbol{\Omega}_B
\mathbf{L}\mathbf{R}^{\prime}
+
\mathbf{R}\mathbf{L}
\boldsymbol{\Omega}_B^{\prime}\mathbf{R}_{\perp}^{\prime}.
\end{align}

In the covariance eigenbasis, these perturbations become
\begin{align}
\mathbf{Q}^{\prime}
d\boldsymbol{\Sigma}_{\mathrm{spec}}
\mathbf{Q}
&=
\begin{pmatrix}
d\mathbf{T} & \mathbf{0}\\
\mathbf{0} & 2\sigma\mathbf{I}_{K-J}\,d\sigma
\end{pmatrix},
\label{eqn:supp:dSigma-spec}
\\
\mathbf{Q}^{\prime}
d\boldsymbol{\Sigma}_{\mathrm{int}}
\mathbf{Q}
&=
\begin{pmatrix}
\boldsymbol{\Omega}_A\mathbf{L}
-\mathbf{L}\boldsymbol{\Omega}_A
&
\mathbf{0}
\\
\mathbf{0} & \mathbf{0}
\end{pmatrix},
\label{eqn:supp:dSigma-int}
\\
\mathbf{Q}^{\prime}
d\boldsymbol{\Sigma}_{\mathrm{ext}}
\mathbf{Q}
&=
\begin{pmatrix}
\mathbf{0} & \mathbf{L}\boldsymbol{\Omega}_B^{\prime}\\
\boldsymbol{\Omega}_B\mathbf{L} & \mathbf{0}
\end{pmatrix},
\label{eqn:supp:dSigma-ext}
\end{align}
with
\begin{equation}
d\mathbf{T}
=
2\operatorname{diag}
\left(
\lambda_1\,d\lambda_1+\sigma\,d\sigma,\ldots,
\lambda_J\,d\lambda_J+\sigma\,d\sigma
\right).
\label{eqn:supp:dT}
\end{equation}

The spectral perturbation is diagonal, the internal-rotation
perturbation has zero diagonal and is confined to the signal block,
and the external-rotation perturbation occupies only the off-diagonal
signal--noise blocks. These three tangent subspaces are therefore
mutually orthogonal under the Fisher metric. Consequently,
\begin{equation}
\left|
\mathcal{F}_{\Sigma\Sigma}
\right|
=
\left|
\mathcal{F}_{\mathrm{spec}}
\right|
\left|
\mathcal{F}_{\mathrm{int}}
\right|
\left|
\mathcal{F}_{\mathrm{ext}}
\right|.
\label{eqn:supp:fisher-cov-factorization}
\end{equation}

\subsection{Spectral block}
The spectral block contains the loading magnitudes
$\boldsymbol{\lambda}$ and the residual scale $\sigma$. Substitution
of \eqref{eqn:supp:dSigma-spec} into the Fisher metric gives
\begin{align}
ds_{\mathrm{spec}}^2
&=
2N
\sum_{j=1}^J
\frac{
(\lambda_j\,d\lambda_j+\sigma\,d\sigma)^2
}{
(\lambda_j^2+\sigma^2)^2
}
+
\frac{2N(K-J)}{\sigma^2}(d\sigma)^2.
\label{eqn:supp:metric-spec}
\end{align}
Thus,
\begin{align}
\mathcal{F}_{\lambda_j,\lambda_k}
&=
2N
\frac{\lambda_j^2}
{(\lambda_j^2+\sigma^2)^2}
\mathbbm{1}_{(j=k)},
\\
\mathcal{F}_{\lambda_j,\sigma}
&=
2N
\frac{\lambda_j\sigma}
{(\lambda_j^2+\sigma^2)^2},
\\
\mathcal{F}_{\sigma,\sigma}
&=
2N
\left[
\sum_{j=1}^J
\frac{\sigma^2}
{(\lambda_j^2+\sigma^2)^2}
+
\frac{K-J}{\sigma^2}
\right].
\end{align}
Applying the Schur-complement determinant formula yields
\begin{equation}
\left|
\mathcal{F}_{\mathrm{spec}}
\right|
=
(2N)^{J+1}
\frac{K-J}{\sigma^2}
\prod_{j=1}^J
\frac{\lambda_j^2}
{(\lambda_j^2+\sigma^2)^2}.
\label{eqn:supp:fisher-spec}
\end{equation}
The factor $K-J$ reflects the fact that $\sigma^2$ cannot be
identified separately from the signal eigenvalues when $J=K$.

\subsection{Internal signal rotations}

Consider an infinitesimal rotation between signal directions $j$ and
$k$, where $1\leq j<k\leq J$. Write
\[
(\boldsymbol{\Omega}_A)_{jk}
=
d\omega_{jk},
\qquad
(\boldsymbol{\Omega}_A)_{kj}
=
-d\omega_{jk}.
\]
The corresponding nonzero entries of $\boldsymbol{\Omega}_A\mathbf{L}
-\mathbf{L}\boldsymbol{\Omega}_A$ are both
\[
(\lambda_k^2-\lambda_j^2)d\omega_{jk}.
\]
Its contribution to the Fisher metric is
\begin{equation}
ds_{jk,\mathrm{int}}^2
=
N
\frac{
(\lambda_j^2-\lambda_k^2)^2
}{
(\lambda_j^2+\sigma^2)
(\lambda_k^2+\sigma^2)
}
(d\omega_{jk})^2.
\label{eqn:supp:metric-int-pair}
\end{equation}
Distinct rotation pairs are orthogonal under the Fisher metric.
Therefore,
\begin{equation}
\left|
\mathcal{F}_{\mathrm{int}}
\right|
=
N^{J(J-1)/2}
\prod_{1\leq j<k\leq J}
\frac{
(\lambda_j^2-\lambda_k^2)^2
}{
(\lambda_j^2+\sigma^2)
(\lambda_k^2+\sigma^2)
}.
\label{eqn:supp:fisher-int}
\end{equation}
This determinant vanishes when two loading magnitudes coincide because
the corresponding individual signal directions are then not
separately identifiable.

\subsection{Signal--noise rotations}

Consider an infinitesimal rotation between signal direction
$\mathbf{r}_j$ and noise-space direction $\mathbf{r}_{\perp,a}$, where
\[
j=1,\ldots,J,
\qquad
a=1,\ldots,K-J.
\]
Let
\[
(\boldsymbol{\Omega}_B)_{aj}
=
d\omega_{aj}.
\]
Substitution of \eqref{eqn:supp:dSigma-ext} into the Fisher metric
gives
\begin{align}
ds_{\mathrm{ext}}^2
&=
\frac{N}{\sigma^2}
\sum_{j=1}^J
\sum_{a=1}^{K-J}
\frac{\lambda_j^4}
{\lambda_j^2+\sigma^2}
(d\omega_{aj})^2.
\label{eqn:supp:metric-ext}
\end{align}
Hence,
\begin{equation}
\mathcal{F}_{\omega_{aj},\omega_{bk}}
=
\mathbbm{1}_{(a=b)}
\mathbbm{1}_{(j=k)}
\frac{N}{\sigma^2}
\frac{\lambda_j^4}
{\lambda_j^2+\sigma^2}.
\end{equation}
For each signal direction $j$, the same information coefficient occurs
for all $K-J$ noise-space directions. It follows that
\begin{align}
\left|
\mathcal{F}_{\mathrm{ext}}
\right|
=
\prod_{j=1}^J
\left[
\frac{N}{\sigma^2}
\frac{\lambda_j^4}
{\lambda_j^2+\sigma^2}
\right]^{K-J}
=
\frac{
N^{J(K-J)}
}{
\sigma^{2J(K-J)}
}
\prod_{j=1}^J
\left[
\frac{\lambda_j^4}
{\lambda_j^2+\sigma^2}
\right]^{K-J}.
\label{eqn:supp:fisher-ext}
\end{align}
This block becomes singular when $\lambda_j=0$, because the
corresponding signal direction is then indistinguishable from the
isotropic noise subspace.

\subsection{Covariance and full Fisher determinants}

Combining
\eqref{eqn:supp:fisher-spec},
\eqref{eqn:supp:fisher-int}, and
\eqref{eqn:supp:fisher-ext} gives
\begin{align}
\left|
\mathcal{F}_{\Sigma\Sigma}
\right|
&=
\frac{
N^{P_{\Sigma}}2^{J+1}(K-J)
}{
\sigma^{2\{J(K-J)+1\}}
}
\prod_{j=1}^J
\frac{
\lambda_j^{4(K-J)+2}
}{
(\lambda_j^2+\sigma^2)^{K+1}
}
\prod_{1\leq j<k\leq J}
(\lambda_j^2-\lambda_k^2)^2,
\label{eqn:supp:fisher-covariance-final}
\end{align}
where
\[
P_{\Sigma}
=
KJ-\frac{J(J-1)}{2}+1.
\]
To verify the exponent of
$\lambda_j^2+\sigma^2$, note that the spectral block contributes
power $2$, the internal-rotation block contributes power $J-1$, and
the external-rotation block contributes power $K-J$. Their sum is
\[
2+(J-1)+(K-J)=K+1.
\]

Finally, multiplying
\eqref{eqn:supp:fisher-covariance-final} by the mean contribution
\eqref{eqn:supp:fisher-mean} gives
\begin{align}
\left|
\mathcal{F}_{\mathrm{full}}
\right|
&=
\frac{
N^P2^{J+1}(K-J)
}{
\sigma^{2\{(K-J)(J+1)+1\}}
}
\prod_{j=1}^J
\frac{
\lambda_j^{4(K-J)+2}
}{
(\lambda_j^2+\sigma^2)^{K+2}
}
\prod_{1\leq j<k\leq J}
(\lambda_j^2-\lambda_k^2)^2,
\label{eqn:supp:fisher-full-final}
\end{align}
where
\[
P
=
K(J+1)-\frac{J(J-1)}{2}+1.
\]

The determinants in
\eqref{eqn:supp:fisher-covariance-final} and
\eqref{eqn:supp:fisher-full-final} are expressed relative to the
invariant local volume element on
$\mathcal{R}_{K,J}$. 

Two singularities are apparent. If $\lambda_j=0$, the corresponding signal direction merges with the isotropic noise subspace. If $\lambda_j=\lambda_k$ for $j\neq k$, the individual signal directions within the repeated eigenspace are no longer separately identifiable. The latter singularity is matched by the Vandermonde factor in the induced prior on the ordered loading magnitudes.

\section{Proofs and numerical checks}
\label{sec:supp:proofs}

This appendix gives the proofs of the results in Sections~\ref{sec:finite-sample-bias} and~\ref{sec:stein}, including Corollary~\ref{cor:bias:second}. The notation is that of the main text; in particular, $\tau=\sigma^2$, $\rho_j=\lambda_j^2/\tau$, $\eta_j=\tau(1+\rho_j)$, $\ell_j=1+\rho_j$ and $\gamma=\lim K/N$. Throughout, the mean is estimated, $m=K-J$,
$D=K(J+1)-J=JK+m$, $A=Nm-D$, $B=N\sum_{k>J}s_k=Nm\widehat\tau_{\mathrm{ML}}$ and
$C=m+1$. When no regular solution exists, the MML87 estimates are completed by the maximum likelihood values, as in Theorem~\ref{thm:mml:bias}. Once the probability of a regular solution tends to one, any two completions agree with probability tending to one. Hence, for every bounded continuous $f$, the expectations of $f$ applied to the two centred and scaled estimates differ by $o(1)$, and all weak limits below are unaffected by the completion.

\subsection{Proof of Theorem~\ref{thm:mml:bias}}

\begin{proof}[Proof of Theorem~\ref{thm:mml:bias}]

Because the covariance matrix is centred about the estimated sample mean,
\[
N \,\mathbf S
\sim
W_K(N-1,\boldsymbol{\Sigma}).
\]
For $j\leq J$, the classical first-order mean expansion of the $j$-th sample signal eigenvalue~\cite{Lawley56}, adjusted for the estimated mean, is
\begin{align}
\mathbb{E}(s_j)
&=
\eta_j+\frac{\widetilde b_j}{N}+O(N^{-2}),
\label{eqn:bias:eigenvalue:expansion}
\\
\frac{\widetilde b_j}{\tau}
&=
(1+\rho_j)
\left[
\sum_{\substack{k=1\\k\neq j}}^J
\frac{1+\rho_k}{\rho_j-\rho_k}
+
\frac{K-J}{\rho_j}
-1
\right].
\label{eqn:bias:eigenvalue:coefficient}
\end{align}
The final term in \eqref{eqn:bias:eigenvalue:coefficient} accounts for estimation of the population mean. Since
\[
\widehat{\tau}_{\mathrm{ML}}
=
\frac{1}{K-J}
\left\{
\operatorname{tr}(\mathbf S)
-
\sum_{j=1}^J s_j
\right\},
\]
the corresponding expansion gives
\begin{equation}
\mathbb{E}
\left(
\widehat{\tau}_{\mathrm{ML}}-\tau
\right)
=
-\frac{\tau}{N}
\left(
J+1+\sum_{j=1}^J\rho_j^{-1}
\right)
+
O(N^{-2}),
\end{equation}
which proves the maximum likelihood bias expression.

For the MML87 bias, write $u=\widehat\tau_{\mathrm{ML}}$ and
parameterise the covariance by the residual variance $t$ and the total
signal variances $y_j=t+x_j$, $j=1,\ldots,J$. Substituting
$x_j=y_j-t$ into \eqref{eqn:mml:preprofile} shows that, up to an
additive constant, the codelength is the negative profile
log-likelihood
\[
\mathcal L(t,\mathbf y)
=
\frac{Nm}{2}\Bigl(\log t+\frac ut\Bigr)
+\frac N2\sum_{j=1}^J\Bigl(\log y_j+\frac{s_j}{y_j}\Bigr)
\]
plus the penalty
\[
\mathcal P(t,\mathbf y)
=
-\frac D2\log t
+\frac{J-2}{2}\sum_{j=1}^J\log y_j
+\frac C2\sum_{j=1}^J\log(y_j-t).
\]
In these coordinates, $\mathcal L$ is separable and is minimised at
$(u,\mathbf s)$, with $\mathbf s=(s_1,\ldots,s_J)^\prime$, and its
Hessian there is diagonal, with entries $Nm/(2u^2)$ and $N/(2s_j^2)$.

Fix $\varepsilon>0$ small enough that, on the closed ball
$\mathcal U=\{\mathbf S:\|\mathbf S-\boldsymbol\Sigma\|\leq\varepsilon\}$,
the signal eigenvalues are simple and separated from the noise
eigenvalues, and $s_j-u$ is bounded away from zero. Then
$s_1,\ldots,s_J$ and $u$ are smooth functions of $\mathbf S$ on
$\mathcal U$. On a fixed neighbourhood of $(u,\mathbf s)$ on which
$y_j-t$ stays bounded away from zero, the Hessian of $\mathcal L$ is
positive definite of order $N$ and its third derivatives are $O(N)$,
uniformly for $\mathbf S\in\mathcal U$, whereas the derivatives of
$\mathcal P$ are bounded. Hence, for all sufficiently large $N$, the
penalised score has a unique zero in this neighbourhood, within
$O(N^{-1})$ of $(u,\mathbf s)$, and it is a strict local minimum of the
codelength. Its loading $x_j=y_j-t$ stays bounded away from zero,
whereas the smaller root of \eqref{eqn:mml:loading:quadratic} is
$O(N^{-1})$, so it lies on the larger-root branch. It is therefore a
regular interior local minimum of the concentrated codelength, which
Lemma~\ref{lem:mml:unique} identifies as the unique regular MML87
estimate. A one-step Taylor expansion of the penalised score about
$(u,\mathbf s)$, using the diagonal Hessian, gives, uniformly on
$\mathcal U$,
\begin{align}
\widehat\tau_{\mathrm{MML}}-\widehat\tau_{\mathrm{ML}}
&=
-\frac{2u^2}{Nm}\,\mathcal P_t(u,\mathbf s)+O(N^{-2})
=
\frac{g(\mathbf S)}{N}+O(N^{-2}),
\label{eqn:bias:mml:correction}
\\
g(\mathbf S)
&=
\frac1m
\left[
Du+Cu^2\sum_{j=1}^J\frac{1}{s_j-u}
\right],
\nonumber
\end{align}
where
$\mathcal P_t(t,\mathbf y)=-D/(2t)-(C/2)\sum_j(y_j-t)^{-1}$ is the
derivative of the penalty with respect to $t$.

To pass to expectations, let
$\mathcal E_N=\{\mathbf S\in\mathcal U\}$.
For fixed $K$, concentration inequalities for Wishart matrices give
$\mathbb P(\mathcal E_N^c)=O(e^{-c_0N})$ for some $c_0>0$.
Every regular stationary solution satisfies
$\widehat\tau_{\mathrm{MML}}\leq B/(A-c_N)$, with
$c_N=J\{C(N+K-1)\}^{1/2}$, as shown in the proof of Proposition~\ref{prop:stein} below.
Since $B/(A-c_N)=Nm\,u/(Nm-D-c_N)$, this bound is at most
$2u$ once $Nm\geq2(D+c_N)$.
Hence, for all sufficiently large $N$, the completed estimators satisfy
$0\leq\widehat\tau_{\mathrm{ML}},\widehat\tau_{\mathrm{MML}}
\leq2u\leq2\operatorname{tr}(\mathbf S)/m$.
Since $\operatorname{tr}(\mathbf S)$ has uniformly bounded second
moments, the Cauchy--Schwarz inequality shows that the contribution of
$\mathcal E_N^c$ to each expectation is exponentially small.
On $\mathcal U$, the function $g$ has bounded second derivatives.
Taylor's formula, together with
\[
\mathbb E(\mathbf S-\boldsymbol\Sigma)
=-\frac{\boldsymbol\Sigma}{N},
\qquad
\mathbb E\|\mathbf S-\boldsymbol\Sigma\|^2=O(N^{-1}),
\]
then yields, up to exponentially small terms,
\[
\mathbb E\!\left[
g(\mathbf S)\mathbf 1_{\mathcal E_N}
\right]
=
g(\boldsymbol\Sigma)+O(N^{-1}).
\]
Since $u(\boldsymbol\Sigma)=\tau$ and
$s_j(\boldsymbol\Sigma)-\tau=\tau\rho_j$, consequently,
\begin{equation}
\mathbb E
\left(
\widehat\tau_{\mathrm{MML}}
-
\widehat\tau_{\mathrm{ML}}
\right)
=
\frac{\tau}{Nm}
\left(
D+C\sum_{j=1}^J\rho_j^{-1}
\right)
+O(N^{-2}).
\label{eqn:bias:mml:mean:correction}
\end{equation}
Combining this with the maximum likelihood bias above,
and using $D-m(J+1)=J^2$ and $C-m=1$, gives
\[
\mathbb E
\left(
\widehat\tau_{\mathrm{MML}}-\tau
\right)
=
\frac{\tau}{Nm}
\left(
J^2+\sum_{j=1}^J\rho_j^{-1}
\right)
+O(N^{-2}),
\]
which proves the MML87 bias.
\end{proof}

\subsection{Proof of Corollary~\ref{cor:bias:second}}

\begin{proof}[Proof of Corollary~\ref{cor:bias:second}]
By scale equivariance, take $\tau=1$, and write
\[
B_0=J+1+R_1,\qquad W=R_1+2R_2+R_3.
\]
We use the localisation from the proof of Theorem~\ref{thm:mml:bias}, with smooth bounded
extensions of the auxiliary functions. Taylor expansion through degree five,
bounded sixth derivatives and centred Wishart moments justify the expectation
remainders below. The contribution outside the localisation neighbourhood is
exponentially small.

\emph{Maximum likelihood.}
Put $\nu=N-1$ and $\mathbf T=N\mathbf S/\nu$, so that
$\nu\mathbf T\sim W_K(\nu,\bm\Sigma)$. In the population eigenbasis write
\[
\bm\Sigma=
\begin{pmatrix}
\mathbf I_J+\mathbf P&\mathbf0\\
\mathbf0&\mathbf I_m
\end{pmatrix},
\qquad
\mathbf P=\operatorname{diag}(\rho_1,\ldots,\rho_J),
\qquad
\mathbf T-\bm\Sigma=
\begin{pmatrix}
\mathbf E_{11}&\mathbf E_{12}\\
\mathbf E_{21}&\mathbf E_{22}
\end{pmatrix}.
\]
Locally, the residual invariant subspace of $\mathbf T$ is the graph of a
$J\times m$ matrix $\mathbf Y=\mathbf Y_1+\mathbf Y_2+\mathbf Y_3+\cdots$, with
$\mathbf Y_r$ homogeneous of degree $r$ in $\mathbf T-\bm\Sigma$:
\[
\mathbf Y_1=-\mathbf P^{-1}\mathbf E_{12},\qquad
\mathbf Y_2=\mathbf P^{-1}(\mathbf Y_1\mathbf E_{22}-\mathbf E_{11}\mathbf Y_1),\qquad
\mathbf Y_3=\mathbf P^{-1}(\mathbf Y_2\mathbf E_{22}-\mathbf E_{11}\mathbf Y_2
+\mathbf Y_1\mathbf E_{21}\mathbf Y_1).
\]
The average $u(\mathbf T)$ of the $m$ residual eigenvalues of $\mathbf T$ is
then exactly $1+m^{-1}\operatorname{tr}(\mathbf E_{22})
+m^{-1}\operatorname{tr}(\mathbf E_{21}\mathbf Y)$, so
\[
u(\mathbf T)=1+q_1+q_2+q_3+q_4+\mathcal R_N,
\qquad
q_1=\frac{\operatorname{tr}\mathbf E_{22}}m,\quad
q_r=\frac{\operatorname{tr}(\mathbf E_{21}\mathbf Y_{r-1})}m\ (r\geq2),
\qquad
\mathbb{E}\mathcal R_N=O(\nu^{-3}).
\]
The centred Wishart moment identities give
\[
\mathbb{E} q_1=0,\qquad
\mathbb{E} q_2=-\frac{J+R_1}{\nu},\qquad
\mathbb{E} q_3=\frac{\mathcal A}{\nu^2},\qquad
\mathbb{E} q_4=-\frac{\mathcal A+W}{\nu^2}+O(\nu^{-3}),
\]
with
\[
\mathcal A
=
(J+R_1)^2+J+2R_1+R_2-C(R_1+R_2).
\]
Here the cubic term uses the third centred Wishart moments, while the leading
quartic term uses the three covariance pairings. Hence
\[
\mathbb{E} u(\mathbf T)
=
1-\frac{J+R_1}{\nu}-\frac{W}{\nu^2}+O(\nu^{-3}).
\]
Since $\widehat\tau_{\mathrm{ML}}=(\nu/N)u(\mathbf T)$,
\[
\mathbb{E}\widehat\tau_{\mathrm{ML}}
=
1-\frac{B_0}{N}-\frac{W}{N^2}+O(N^{-3}).
\]

\emph{MML87 correction.}
Write $u=\widehat\tau_{\mathrm{ML}}$ and $d_j=s_j-u$. Expansion of the
stationarity equations gives
\[
\widehat\tau_{\mathrm{MML}}
=
u+\frac{g(u,\mathbf s)}{N}
+\frac{k(u,\mathbf s)}{N^2}
+O(N^{-3}),
\qquad
g(u,\mathbf s)
=
\frac{Du+Cu^2\sum_jd_j^{-1}}{m},
\]
uniformly on the localisation neighbourhood, with $k$ smooth there. Only the
population value of $k$ is needed. Define
\[
G=g(1,\bm\ell)=\frac{D+CR_1}{m},
\qquad
G'=g_u(1,\bm\ell)
=\frac{D+C(2R_1+R_2)}{m},
\]
where $\bm\ell=(\ell_1,\ldots,\ell_J)$. Differentiating the loading and
residual variance equations once more gives
\[
k(1,\bm\ell)
=
GG'+\frac Cm
\left\{(K-1)(R_1+R_2)+C(R_2+R_3)\right\}
=:k_0,
\]
and consequently $\mathbb{E} k(u,\mathbf s)=k_0+O(N^{-1})$.

To evaluate $\mathbb{E} g(u,\mathbf s)$, recall the first-order mean expansion of the
sample signal eigenvalues used in the proof of Theorem~\ref{thm:mml:bias}~\cite{Lawley56},
adjusted for the estimated mean:
\[
\mathbb{E}(s_j-\ell_j)=\frac{\mu_j}{N}+O(N^{-2}),
\qquad
\mu_j
=
-\ell_j
+\sum_{k\ne j}\frac{\ell_j\ell_k}{\rho_j-\rho_k}
+\frac{m\ell_j}{\rho_j}.
\]
Pairing the signal--signal terms yields
\[
\sum_j\frac{\mu_j}{\rho_j^2}
=
-B_0(R_1+R_2)+m(R_2+R_3)+W.
\]
The leading covariance matrix of $(u,\mathbf s)$ is $N^{-1}\operatorname{diag}(2/m,2\ell_1^2,\ldots,2\ell_J^2)$. A Taylor expansion of $g$ therefore gives
\[
\mathbb{E} g(u,\mathbf s)
=
G+\frac{\Gamma}{N}+O(N^{-2}),
\]
where
\begin{align*}
\Gamma
&=
-B_0G'-\frac Cm\sum_j\frac{\mu_j}{\rho_j^2}
+\frac{2C^2}{m^2}W
=
-B_0G-C(R_2+R_3)+\frac{C(m+2)}{m^2}W.
\end{align*}

Combining these expansions,
\[
\mathbb{E}(\widehat\tau_{\mathrm{MML}}-1)
=
\frac{G-B_0}{N}
+\frac{k_0+\Gamma-W}{N^2}
+O(N^{-3}).
\]
Finally,
\[
G-B_0=\frac{J^2+R_1}{m},
\qquad
k_0+\Gamma-W=\mathcal B_2.
\]
Restoring the scale $\tau$ proves both expansions. Every term of $\mathcal B_2$ is nonnegative and $(D+CR_1)J^2>0$, so $\mathcal B_2>0$ when $J\geq1$.
\end{proof}

\paragraph{Numerical checks}
The algebraic identities in the proof, $G-B_0=(J^2+R_1)/m$, the pairing identity for $\sum_j\mu_j/\rho_j^2$, $k_0+\Gamma-W=\mathcal B_2$ and $\mathcal B_2>0$, were verified in  over $2000$ random designs with $1\leq J\leq6$ and $1\leq m\leq30$. 
%
Finally, the bias was simulated in nine designs with $N=50$ to $3200$ (omitting $N\leq2K$) and $10^7$ replications per setting, using exact Wishart draws and regression control variates built from Wishart entries with known means. Table~\ref{tab:bias:second} compares the second-order coefficients, estimated from these simulations, with Corollary~\ref{cor:bias:second}; all agree to within two Monte Carlo standard errors.

\begin{table}[htbp]
\centering\small
\caption{Second-order bias coefficients $a_2$, where $\mathbb{E}(\widehat\tau-\tau)/\tau=a_1/N+a_2/N^2+O(N^{-3})$: estimates from simulation (intercept of a weighted least-squares fit of $a_2+c/N+d/N^2$ to $N^2\{\mathbb{E}(\widehat\tau-\tau)/\tau-a_1/N\}$ over $N\geq100$, $10^7$ replications per $N$) against Corollary~\ref{cor:bias:second}.}
\label{tab:bias:second}
\begin{tabular}{lrrrr}
\toprule
 & \multicolumn{2}{c}{ML} & \multicolumn{2}{c}{MML87}\\
\cmidrule(lr){2-3}\cmidrule(lr){4-5}
Design & simulated (SE) & Corollary~\ref{cor:bias:second} & simulated (SE) & Corollary~\ref{cor:bias:second}\\
\midrule
$K=10$, $\rho=(5)$ & $-$0.288 (0.009) & $-$0.288 & 3.491 (0.009) & 3.490\\
$K=10$, $\rho=(10, 5)$ & $-$0.418 (0.014) & $-$0.409 & 7.343 (0.015) & 7.350\\
$K=6$, $\rho=(5, 2)$ & $-$1.426 (0.030) & $-$1.413 & 19.731 (0.030) & 19.714\\
$K=20$, $\rho=(20, 10, 5)$ & $-$0.460 (0.013) & $-$0.464 & 13.041 (0.013) & 13.031\\
$K=5$, $\rho=(2)$ & $-$1.097 (0.022) & $-$1.125 & 9.018 (0.022) & 8.977\\
$K=12$, $\rho=(40, 20, 10, 5)$ & $-$0.461 (0.023) & $-$0.490 & 24.371 (0.024) & 24.334\\
$K=50$, $\rho=(10)$ & $-$0.120 (0.005) & $-$0.121 & 5.806 (0.005) & 5.805\\
$K=8$, $\rho=(3, 2.5)$ & $-$1.393 (0.025) & $-$1.377 & 18.280 (0.025) & 18.269\\
$K=3$, $\rho=(4)$ & $-$0.376 (0.019) & $-$0.391 & 4.996 (0.019) & 4.980\\
\bottomrule
\end{tabular}

\end{table}

\subsection{Proof of Proposition~\ref{prop:stein} and Corollary~\ref{cor:stein:improvement}}
\label{sec:supp:stein:proof}

\begin{proof}[Proof of Proposition~\ref{prop:stein}]
Since $\sqrt\gamma$ minimises $\rho+\gamma/\rho$ and $\rho^\ast(\gamma)$
is the larger root of its defining equation,
$\rho^\ast(\gamma)>\sqrt\gamma$, so every spike is supercritical. The
standard spiked-covariance limits
\cite{BaikSilverstein06,Paul07,DonohoGavishJohnstone18} give
\begin{align}
s_j
&\longrightarrow
\bar s(\ell_j)
=(1+\rho_j)(1+\gamma/\rho_j),
\label{eqn:stein:sample:limit}
\\
|\mathbf r_k^\prime\mathbf u_j|^2
&\longrightarrow
\delta_{kj}c^2(\ell_j),
\qquad j,k\leq J.
\label{eqn:stein:cosine:limit}
\end{align}
All sampling limits below are in probability. Estimating the mean
changes the Wishart degrees of freedom to $N-1$ and leaves these limits
unchanged. Moreover, Wishart trace moments give
$\operatorname{tr}(\mathbf S)-K=O_p(1)$, whence
\begin{equation}
u:=\widehat\tau_{\mathrm{ML}}
=
\frac{\operatorname{tr}(\mathbf S)-\sum_{j\leq J}s_j}{m}
=
1+O_p(N^{-1}).
\label{eqn:stein:noise:limit}
\end{equation}

Let $A$, $B$ and $C$ be as defined at the start of this appendix. The
larger root of \eqref{eqn:mml:loading:quadratic} is no smaller than
the geometric mean of its two positive roots, so
$\widehat x_j(t)\geq t\{C/(N+K-1)\}^{1/2}$. Every regular stationary
solution $t$ therefore satisfies
\[
0\leq At-B=Ct^2\sum_{j=1}^J\widehat x_j(t)^{-1}\leq c_Nt,
\qquad
c_N=J\{C(N+K-1)\}^{1/2}=O(N),
\]
and hence $B/A\leq t\leq B/(A-c_N)$. Since $B=Nmu$, $A=Nm-D$ and
$D,c_N=O(N)$, these bounds give $t=u+O_p(N^{-1})=1+O_p(N^{-1})$.
For existence, the upper endpoint of the regular loading profile
satisfies
\[
\tau_{J,\max}
\longrightarrow
\frac{\bar s(\ell_J)}
{(\sqrt{1+\gamma}+\sqrt\gamma)^2}.
\]
Since $\rho_J>\rho^\ast(\gamma)>\sqrt\gamma$,
$\rho_J+\gamma/\rho_J>\gamma+2\sqrt{\gamma(1+\gamma)}$, so this limit
exceeds one. A fixed neighbourhood of one is
therefore admissible with probability tending to one. Uniformly there,
$H(t)/(NK)\rightarrow t-1$, so $H$ changes sign from negative to
positive, and Lemma~\ref{lem:mml:unique} gives the unique regular
minimum.

Dividing \eqref{eqn:mml:loading:quadratic} by $N$ gives the limiting
equation
\[
Q_\rho(x):=(1+\gamma)x^2-b(\rho)x+\gamma=0,
\]
whose roots $z_-<z_+$ are distinct for $\rho>\rho^\ast(\gamma)$.
Continuity of the larger root gives $\widehat x_j\rightarrow z_+(\rho_j)$
and hence
$\widehat\tau_{\mathrm{MML}}+\widehat x_j
\rightarrow1+z_+(\rho_j)=\eta_{\mathrm{MML}}(\ell_j)$.

For part~(i), put
\[
d_\rho=(1+\gamma)\rho+2\gamma,
\qquad
w=\eta_{\mathrm{St}}(\ell)-1=\frac{\rho^2-\gamma}{d_\rho}.
\]
Direct substitution gives
\[
d_\rho^2\,Q_\rho(w)
=
-\frac{\gamma(\rho+1)(\rho+\gamma)}{\rho}
\bigl\{(1-\gamma)\rho^2-2\gamma\rho-2\gamma\bigr\}.
\]
For $0<\gamma<1$, the polynomial in braces has the unique positive zero
$\rho_0$, at which $w=\rho_0/2$. Since $z_-z_+=\gamma/(1+\gamma)$, this
root is the larger one precisely when
$\rho_0/2>\{\gamma/(1+\gamma)\}^{1/2}$, which is equivalent to
$\gamma>\gamma_0$. At $\gamma=\gamma_0$ the roots coincide and
$\rho_0=\rho^\ast$, outside the regular domain. For $\rho>\rho_0$,
$Q_\rho(w)<0$, so $z_-<w<z_+$ and $\eta_{\mathrm{MML}}>\eta_{\mathrm{St}}$.
For $\rho^\ast<\rho<\rho_0$, $Q_\rho(w)>0$, so $w$ lies outside both
roots and, by continuity, stays on the same side. As $\rho\uparrow\rho_0$,
$w$ tends to $z_+$ when $\gamma>\gamma_0$ and to $z_-$ when
$\gamma<\gamma_0$. Hence $w>z_+$, and
$\eta_{\mathrm{MML}}<\eta_{\mathrm{St}}$, in the first case, and $w<z_-$,
and $\eta_{\mathrm{MML}}>\eta_{\mathrm{St}}$, in the second. When
$\gamma=1$, $w=(\rho-1)/2$ and, on the regular domain, $Q_\rho(w)>0$ and
$Q_\rho^\prime(w)=\rho-1-\rho^{-1}>0$, so $w>z_+$ and
$\eta_{\mathrm{MML}}<\eta_{\mathrm{St}}$. Finally, expansion of the
explicit shrinkers gives
\[
\frac{\eta_{\mathrm{MML}}(\ell)}
{\eta_{\mathrm{St}}(\ell)}
=
1+\frac{\gamma(1-\gamma)}{(1+\gamma)\rho}
+O(\rho^{-2}).
\]

We next account for the fitted residual variance, which matters because
the losses are unnormalised. For fixed fitted signal eigenvalues
$y_j>0$, write
\[
\mathbf M(t)
=
\sum_{j=1}^J y_j\mathbf u_j\mathbf u_j^\prime
+t\mathbf P_\perp,
\qquad
\mathbf P_\perp
=
\mathbf I_K-\sum_{j=1}^J\mathbf u_j\mathbf u_j^\prime.
\]
Because
$\operatorname{tr}(\boldsymbol\Sigma^{\pm1}\mathbf P_\perp)=m+O(1)$,
subtraction of traces and log-determinants gives, for either loss and
$t$ near one,
\[
\bigl|
L\{\boldsymbol\Sigma,\mathbf M(t)\}
-
L\{\boldsymbol\Sigma,\mathbf M(1)\}
\bigr|
=
O\{K(t-1)^2+|t-1|\}.
\]
Replacing the fitted residual variance by one therefore changes the loss by
$o_p(1)$ whenever $\sqrt K(t-1)\rightarrow0$ in probability. The rates
established above verify this condition for both maximum likelihood and
MML87.

With unit fitted residual variance, \eqref{eqn:stein:cosine:limit} shows
that the terms of the limiting Stein and entropy losses that depend on
a fitted signal eigenvalue $\eta$ are, respectively,
\[
\frac{\eta}{\eta_{\mathrm{St}}(\ell)}-\log\eta,
\qquad
\frac{\eta_{\mathrm{Ent}}(\ell)}{\eta}+\log\eta.
\]
Subtracting their minima gives the per-spike excess losses
$h\{\eta/\eta_{\mathrm{St}}(\ell)\}$ and
$h\{\eta_{\mathrm{Ent}}(\ell)/\eta\}$. Summing the first over the spikes
proves the expression in part~(ii). Its stated rate follows from the
ratio expansion above and $h(1+v)=v^2/2+O(v^3)$.

For parts~(iii) and~(iv), the exact identities
\begin{align*}
\frac{\bar s(\ell)}{\eta_{\mathrm{St}}(\ell)}
&=1+\gamma+\frac{2\gamma}{\rho},
\qquad
\frac{\ell}{\eta_{\mathrm{St}}(\ell)}
=1+\gamma+\frac{\gamma(1-\gamma)}{\rho+\gamma},
\\
\frac{\eta_{\mathrm{Ent}}(\ell)}{\bar s(\ell)}
&=\left(\frac{\rho}{\rho+\gamma}\right)^2
=1-\frac{2\gamma}{\rho}+O(\rho^{-2})
\end{align*}
give the Stein limit $Jh(1+\gamma)$ for maximum likelihood and for the
population-spike comparator, which also requires the residual variance
condition above, and the $O(\rho_J^{-2})$ entropy excess for maximum
likelihood. Finally, $\eta_{\mathrm{Ent}}(\ell)\sim\rho$ and
$\eta_{\mathrm{MML}}(\ell)\sim\rho/(1+\gamma)$, so the MML87 entropy
excess tends to
\[
Jh(1+\gamma)
=
J\{\gamma-\log(1+\gamma)\}.
\]
All strong-spike limits are taken after the proportional limit, with
$J$ fixed.
\end{proof}

\begin{proof}[Proof of Corollary~\ref{cor:stein:improvement}]
The maximum likelihood estimator is the sample-eigenvector estimator
with fitted signal eigenvalues $s_j\rightarrow\bar s(\ell_j)$ and
residual variance $u$, which satisfies the residual variance condition above
by \eqref{eqn:stein:noise:limit}. The per-spike excess Stein losses
established in the proof of Proposition~\ref{prop:stein} are
therefore $h\{\bar s(\ell_j)/\eta_{\mathrm{St}}(\ell_j)\}$ for maximum
likelihood and $h\{\eta_{\mathrm{MML}}(\ell_j)/\eta_{\mathrm{St}}(\ell_j)\}$
for MML87. The remaining terms of the two losses differ by $o_p(1)$, so
the loss difference converges in probability to $\Delta_{\mathrm{St}}$.

It remains to show that each summand of $\Delta_{\mathrm{St}}$ is
positive. Fix a spike $\ell=1+\rho$ with $\rho>\rho^\ast(\gamma)$, write
$z=z_+(\rho)$, so that $\eta_{\mathrm{MML}}(\ell)=1+z$, and put
$\underline z=\rho/(\rho+2)$. We first show that $z>\underline z$. Direct substitution gives
\[
(\rho+2)^2\,Q_\rho(\underline z)
=
(\rho+1)\bigl\{\gamma(3\rho+2)-\rho^2\bigr\}.
\]
If $\gamma(3\rho+2)<\rho^2$, then $Q_\rho(\underline z)<0$, so $z_-<\underline z<z_+$.
Otherwise
$\gamma\geq\rho^2/(3\rho+2)>\rho^2/\{4(\rho+1)\}=\underline z^2/(1-\underline z^2)$, so
$\underline z^2<\gamma/(1+\gamma)=z_-z_+<z_+^2$. In both cases $z>\underline z$.

Next, $\bar s(\ell)=1+2\gamma+b(\rho)$, and $Q_\rho(z)=0$ gives
$b(\rho)=(1+\gamma)z+\gamma/z$. Hence $\bar s(\ell)=(1+z)(1+\gamma+\gamma/z)$
and, with the first exact identity above,
\[
\zeta=\frac{\bar s(\ell)}{\eta_{\mathrm{St}}(\ell)}
=1+\gamma+\frac{2\gamma}{\rho},
\qquad
\psi=\frac{\bar s(\ell)}{\eta_{\mathrm{MML}}(\ell)}
=1+\gamma+\frac{\gamma}{z}.
\]
The bound $z>\underline z$ gives $\gamma/z<\gamma+2\gamma/\rho$, and so
\[
1<\psi<1+2\gamma+\frac{2\gamma}{\rho}<2\zeta-1.
\]
Since $\eta_{\mathrm{MML}}/\eta_{\mathrm{St}}=\zeta/\psi$, the summand is
\[
h(\zeta)-h(\zeta/\psi)
=\zeta(1-\psi^{-1})-\log\psi
>\frac{\psi-\psi^{-1}}{2}-\log\psi
>0,
\]
where the first inequality uses $\zeta>(\psi+1)/2$ and $1-\psi^{-1}>0$.
The final expression vanishes at $\psi=1$ and has derivative
$(\psi-1)^2/(2\psi^2)>0$ for $\psi>1$, which gives the second inequality.
\end{proof}

No analogue of Corollary~\ref{cor:stein:improvement} holds for
entropy loss. By the proof of Proposition~\ref{prop:stein}, the
per-spike excess entropy losses are
$h\{\eta_{\mathrm{Ent}}(\ell)/\bar s(\ell)\}$ for maximum likelihood and
$h\{\eta_{\mathrm{Ent}}(\ell)/\eta_{\mathrm{MML}}(\ell)\}$ for MML87.
With $\gamma=0.25$ and $\rho=2$, which lies in the regular domain since
$\rho^\ast(0.25)\approx1.15$, these equal $0.0257$ and $0.0082$, so MML87
has the smaller limiting entropy loss. Maximum likelihood has the smaller
entropy loss for sufficiently strong spikes, by part~(iv).

\subsection{Proof of Theorem~\ref{thm:fixed:clt}}

\begin{proof}[Proof of Theorem~\ref{thm:fixed:clt}]
Write $\boldsymbol\phi=(\tau,\eta_1,\ldots,\eta_J)^\prime$, where
$\eta_j=\tau+x_j$, and rotate to the population eigenbasis, so that
$\boldsymbol\Sigma=\operatorname{diag}(\eta_1,\ldots,\eta_J,\tau,\ldots,\tau)$.
Since $\boldsymbol\Sigma=\tau\mathbf P_\perp+\sum_j\eta_j\mathbf r_j\mathbf r_j^\prime$,
with $\mathbf P_\perp=\mathbf I_K-\sum_j\mathbf r_j\mathbf r_j^\prime$, we have
$\partial\boldsymbol\Sigma/\partial\tau=\mathbf P_\perp$ and
$\partial\boldsymbol\Sigma/\partial\eta_j=\mathbf r_j\mathbf r_j^\prime$.
By \eqref{eqn:supp:fisher-general}, the per-observation Fisher
information for $\boldsymbol\phi$ is
\[
\mathcal I_0
=
\frac12\operatorname{diag}
\left(
\frac{m}{\tau^2},
\frac{1}{\eta_1^2},\ldots,\frac{1}{\eta_J^2}
\right).
\]
Changes in $\boldsymbol\phi$ are diagonal in the population
eigenbasis, whereas changes in the signal directions are off-diagonal,
and the mean is orthogonal to the covariance parameters
(Appendix~\ref{sec:supp:fisher}). Hence $\mathcal I_0$ is unaffected by
these nuisance parameters.

\emph{Maximum likelihood limit.} The maximum likelihood estimate is
$\widehat{\boldsymbol\phi}_{\mathrm{ML}}=(\widehat\tau_{\mathrm{ML}},s_1,\ldots,s_J)^\prime$.
The central limit theorem gives
$\mathbf G_N=\sqrt N(\mathbf S-\boldsymbol\Sigma)\Rightarrow\mathbf G$, with
$\operatorname{Cov}(G_{ik},G_{i'k'})=\Sigma_{ii'}\Sigma_{kk'}+\Sigma_{ik'}\Sigma_{ki'}$;
estimating the mean changes $\mathbf S$ by $O_p(N^{-1})$. The diagonal
entries of $\mathbf G$ are independent normal variables, with variances
$2\eta_j^2$ for $j\leq J$ and $2\tau^2$ otherwise. Since the signal
eigenvalues are simple and separated from $\tau$, first-order
perturbation gives $\sqrt N(s_j-\eta_j)=(\mathbf G_N)_{jj}+o_p(1)$, and,
since $\widehat\tau_{\mathrm{ML}}=\{\operatorname{tr}(\mathbf S)-\sum_{j\leq J}s_j\}/m$,
$\sqrt N(\widehat\tau_{\mathrm{ML}}-\tau)=m^{-1}\sum_{k>J}(\mathbf G_N)_{kk}+o_p(1)$.
Therefore
\[
\sqrt N
\left(
\widehat{\boldsymbol\phi}_{\mathrm{ML}}
-\boldsymbol\phi
\right)
\Rightarrow
\mathcal N_{J+1}(\mathbf 0,\mathcal I_0^{-1}).
\]

\emph{Transfer to MML87.} In the notation of the proof of
Theorem~\ref{thm:mml:bias}, with $\boldsymbol\phi$ in place of
$(t,\mathbf y)$, that proof shows that on $\mathcal E_N$, whose
probability tends to one, the regular MML87 estimate exists and is
unique, and the same one-step expansion of the penalised score gives
\[
\widehat{\boldsymbol\phi}_{\mathrm{MML}}
-\widehat{\boldsymbol\phi}_{\mathrm{ML}}
=
-\{\nabla^2\mathcal L(\widehat{\boldsymbol\phi}_{\mathrm{ML}})\}^{-1}
\nabla\mathcal P(\widehat{\boldsymbol\phi}_{\mathrm{ML}})
+O(N^{-2})
=
O(N^{-1}),
\]
uniformly on $\mathcal U$. Slutsky's theorem therefore transfers the
maximum likelihood limit to MML87.

\emph{Covariance.} Applying the linear map $x_j=\eta_j-\tau$ to
$\mathcal I_0^{-1}=\operatorname{diag}(v,2\eta_1^2,\ldots,2\eta_J^2)$,
with $v=2\tau^2/m$, gives
\[
\mathbf V
=
\begin{pmatrix}
v & -v\mathbf1^\prime\\
-v\mathbf1 &
2\operatorname{diag}(\eta_1^2,\ldots,\eta_J^2)
+v\mathbf1\mathbf1^\prime
\end{pmatrix}.
\]
Since the inverse Fisher information transforms in the same way under
this reparameterisation, $\mathbf V$ is the inverse of the
per-observation Fisher information for $(\tau,\mathbf x)$.
\end{proof}

\subsection{Proof of Theorem~\ref{thm:prop:clt}}

\begin{proof}[Proof of Theorem~\ref{thm:prop:clt}]
By scale equivariance, it suffices to take $\tau=1$.
Proposition~\ref{prop:stein} gives existence of the regular estimate
with probability tending to one, together with
$\widehat\tau_{\mathrm{MML}}\xrightarrow{p}1$ and
$\widehat x_j\xrightarrow{p}z_j$.

\emph{Residual variance.} Put $\nu=N-1$ and $R_1=\sum_j\rho_j^{-1}$.
In the population eigenbasis,
$\operatorname{tr}(\mathbf S)=U_N/N+\sum_{j\leq J}S_{jj}$, where the
noise diagonal entries contribute $U_N/N$ with $U_N\sim\chi^2_{\nu m}$,
and the $J$ signal diagonal entries satisfy
$S_{jj}=\ell_j+O_p(N^{-1/2})$. Hence
\[
Z_N=\frac{U_N-\nu m}{m}\Rightarrow\mathcal N(0,2/\gamma),
\]
since $\operatorname{Var}(U_N)/m^2=2\nu/m\rightarrow2/\gamma$. Since
$\widehat\tau_{\mathrm{ML}}=\{\operatorname{tr}(\mathbf S)-\sum_{j\leq J}s_j\}/m$
and $Nm=\nu m+m$,
\[
\begin{aligned}
N(\widehat\tau_{\mathrm{ML}}-1)
&=
Z_N-1+\frac Nm\sum_{j=1}^J(\ell_j-s_j)+o_p(1)\\
&=
Z_N-(1+J+R_1)+o_p(1),
\end{aligned}
\]
using $s_j\xrightarrow{p}\bar s(\ell_j)=\ell_j(1+\gamma/\rho_j)$, so that
$\bar s(\ell_j)-\ell_j=\gamma+\gamma/\rho_j$, and $N/m\rightarrow1/\gamma$.
The term $-1$ arises from estimating the mean. Dividing the
stationarity equation \eqref{eqn:mml:tau:stationary} by $m$ gives
\[
N(\widehat\tau_{\mathrm{MML}}-\widehat\tau_{\mathrm{ML}})
=
\frac{
D\widehat\tau_{\mathrm{MML}}
+C\widehat\tau_{\mathrm{MML}}^2\sum_j\widehat x_j^{-1}
}{m}
=
J+1+\sum_jz_j^{-1}+o_p(1),
\]
because $D/m\rightarrow J+1$, $C/m\rightarrow1$ and
$\widehat x_j\xrightarrow{p}z_j$. Thus
\[
N(\widehat\tau_{\mathrm{MML}}-1)
=
Z_N+\beta_\gamma+o_p(1),
\]
which proves both residual variance limits and
$\widehat\tau_{\mathrm{MML}}-1=O_p(N^{-1})$.

\emph{Positivity of $\beta_\gamma$.} The limiting loading polynomial
$Q_{\rho_j}$ of the proof of Proposition~\ref{prop:stein} factorises as
\[
Q_{\rho_j}(z)
=
(z-\rho_j)(z-\gamma/\rho_j)+\gamma z(z+1).
\]
Since $\rho_j>\sqrt\gamma$, we have $\rho_j>\gamma/\rho_j$, so this
polynomial is positive for $z\geq\rho_j$. Its larger root $z_j$
therefore satisfies $z_j<\rho_j$, giving $\beta_\gamma>0$.

\emph{Loading strengths.} Suppose that
$\sqrt N(\gamma_N-\gamma)\rightarrow0$, where $\gamma_N=K/N$. Dividing
\eqref{eqn:mml:loading:quadratic} by $N$, and using
$b(\rho_j)=\bar s(\ell_j)-1-2\gamma$ and
$\widehat\tau_{\mathrm{MML}}-1=O_p(N^{-1})$, gives
\[
Q_{\rho_j}(\widehat x_j)
=
\widehat x_j\{s_j-\bar s(\ell_j)\}
+O_p\bigl(N^{-1}+|\gamma_N-\gamma|\bigr)
=
\widehat x_j\{s_j-\bar s(\ell_j)\}
+o_p(N^{-1/2}).
\]
Since $Q_{\rho_j}(z_j)=0$, $Q_{\rho_j}^\prime(z_j)=\xi_j>0$ and
$\widehat x_j\xrightarrow{p}z_j$, the mean value theorem and the
outlier central limit theorem~\cite[Theorem~3]{Paul07} give
\[
\sqrt N(\widehat x_j-z_j)
=
\frac{z_j}{\xi_j}
\sqrt N\{s_j-\bar s(\ell_j)\}+o_p(1)
\Rightarrow
\mathcal N\Bigl(0,\frac{z_j^2v_j}{\xi_j^2}\Bigr).
\]
The outlier theorem applies through the Wishart representation with
$\nu=N-1$ observations, since the resulting changes in the aspect
ratio and the divisor are $O(N^{-1})$. Restoring the scale $\tau$ gives
the stated limits.
\end{proof}

\subsection{Numerical checks}

Table~\ref{tab:supp:asym} compares Theorems~\ref{thm:fixed:clt}
and~\ref{thm:prop:clt} with simulation, using Algorithm~\ref{alg:mml:estimation} with an estimated mean and
$2\times10^4$ replications for each sample size. At $N=100$, $99.99\%$ of
the fits in the fixed-dimensional design and $99.76\%$ in the proportional
regime had a regular solution; the remaining fits were completed by the
maximum likelihood values. At larger $N$ every fit had a regular
solution. At $N=1000$ and $N=10000$ every entry lies within about two
Monte Carlo standard errors of its limit. The two covariances in the
fixed-dimensional design have large Monte Carlo errors, because the
loading estimates have large variances. At $N=100$ the variance of the
residual variance estimate and the mean of $N(\widehat\tau-\tau)$ in the
proportional regime are still noticeably above their limits.

\begin{table}[htbp]
\scriptsize
\begin{center}
\caption{Simulation checks of the asymptotic distributions, with $2\times10^4$ replications for each $N$. The Monte Carlo standard errors are about $1\%$ of the value for the variances, $0.043$ and $0.26$ for the covariances of $\widehat\tau$ with $\widehat x_1$ and of $\widehat x_1$ with $\widehat x_2$, $0.002$ for the coverage and $0.015$ for the mean of $N(\widehat\tau-\tau)$.}
\label{tab:supp:asym}
\begin{tabular}{lcccc}
\toprule
Quantity & Limit & $N=100$ & $N=1000$ & $N=10000$ \\
\midrule
\multicolumn{5}{l}{\emph{Fixed dimension:} $K=6$, $J=2$, $\tau=1$, $(\rho_1,\rho_2)=(5,2)$}\\
$\operatorname{Var}\{\sqrt N(\widehat\tau-\tau)\}$ & $0.500$ & $0.555$ & $0.499$ & $0.501$ \\
$\operatorname{Var}\{\sqrt N(\widehat x_1-x_1)\}$ & $72.5$ & $64.0$ & $72.1$ & $72.5$ \\
$\operatorname{Var}\{\sqrt N(\widehat x_2-x_2)\}$ & $18.5$ & $16.8$ & $18.1$ & $18.1$ \\
$\operatorname{Cov}\{\sqrt N(\widehat\tau-\tau),\sqrt N(\widehat x_1-x_1)\}$ & $-0.500$ & $-0.547$ & $-0.542$ & $-0.413$ \\
$\operatorname{Cov}\{\sqrt N(\widehat x_1-x_1),\sqrt N(\widehat x_2-x_2)\}$ & $0.500$ & $1.255$ & $0.274$ & $0.684$ \\
Coverage of the 95\% interval for $\tau$ & $95\%$ & $93.8\%$ & $94.8\%$ & $95.0\%$ \\
\midrule
\multicolumn{5}{l}{\emph{Proportional regime:} $K=N/2$, $J=2$, $\tau=1$, $(\rho_1,\rho_2)=(8,4)$}\\
Mean of $N(\widehat\tau-\tau)$ & $0.267$ & $0.425$ & $0.291$ & $0.275$ \\
$\operatorname{Var}\{N(\widehat\tau-\tau)\}$ & $4.00$ & $4.53$ & $4.07$ & $4.04$ \\
$\operatorname{Var}\{\sqrt N(\widehat x_1-\tau z_1)\}$ & $73.4$ & $75.4$ & $72.4$ & $73.4$ \\
$\operatorname{Var}\{\sqrt N(\widehat x_2-\tau z_2)\}$ & $24.6$ & $24.2$ & $24.3$ & $24.2$ \\
\bottomrule
\end{tabular}

\end{center}
\end{table}

\section{Comparison with alternative estimators}
\label{sec:supp:compare}

This appendix compares MML87 with alternative estimators of the residual
variance and of the covariance matrix in probabilistic PCA. The results
are summarised in Section~\ref{sec:exp:compare} and
Table~\ref{tab:results:compare}; here we give the
details and the full tables. All
estimators use the sample mean and the leading sample eigenvectors, and
are fitted at a matched rank, as in the parameter-estimation experiment of Section~\ref{sec:exp:pest}. In each data set, MML87 is fitted at the generating
dimension $J$ when a regular solution exists, and otherwise at the
largest rank admitting one. Every other estimator is fitted at the same
rank. We compare the following estimators.
\begin{itemize}
\item MLE: maximum likelihood, with $\widehat\lambda_j^2=s_j-\widehat\tau_{\mathrm{ML}}$.
\item MML87: the estimator of Algorithm~\ref{alg:mml:estimation}.
\item BC: maximum likelihood with the first-order bias correction implied
by Theorem~\ref{thm:mml:bias},
$\widehat\tau_{\mathrm{BC}}=\widehat\tau_{\mathrm{ML}}\{1+[J+1+\sum_j\widehat\tau_{\mathrm{ML}}/(s_j-\widehat\tau_{\mathrm{ML}})]/N\}$,
and loadings
$\widehat\lambda_j^2=\max(s_j-\widehat\tau_{\mathrm{BC}},0)$.
\item PLY: the bias-corrected residual variance of \cite{PassemierLiYao17}
(their equation 9), applied to the eigenvalues $Ns_j/(N-1)$ of the
sample covariance with divisor $N-1$ and with $c_n=K/(N-1)$, together with
their spike estimates obtained by inverting the limiting map from
population to sample spikes at the corrected residual variance. Sample
eigenvalues at or below the bulk edge give a zero spike estimate.
\item DGJ-E and DGJ-S: the optimal eigenvalue shrinkers of
\cite{DonohoGavishJohnstone18} for entropy loss and Stein loss,
respectively. We use the PLY residual variance in place of the
median-based estimate of their Section~9.2. With $K=10$, the median
sample eigenvalue is inflated by the spikes. For the unnormalised losses
used here, the noise estimate must satisfy
$\sqrt K(\widetilde\tau/\tau-1)\rightarrow0$ in probability
(proof of Proposition~\ref{prop:stein} in Appendix~\ref{sec:supp:proofs}), which holds for the PLY estimate
because its error is $O_p(N^{-1})$.
\end{itemize}
The James--Stein estimator of \cite{Shkolnik22} is not included. It
targets the leading eigenvector when $K\gg N$ and shrinks it towards the
equal-weight direction, assuming loadings with nonzero mean, whereas in
our designs $K<N$ and the signal directions are uniformly distributed.

We consider three designs, each with $10^5$ replications and $\tau=1$:
(i) equal per-component signal-to-noise ratio $\rho\in\{2,5,20\}$ with
$K=10$, $N\in\{25,50,100\}$ and $J\in\{1,2,4\}$, which is the design of
Table~\ref{tab:results:pest}; (ii) geometric loadings with the same $K$,
$N$ and $J$, in which the weakest component has signal-to-noise ratio
$\rho$ and $\rho_j/\rho_{j+1}=1/0.75$; and (iii) a high-dimensional
design with equal loadings, $K\in\{20,50,100\}$,
$J\in\{1,\lceil K/10\rceil,\lceil K/4\rceil\}$, and either $N=10K$ with
$\rho\in\{5,20\}$ or $N=2K$ with $\rho\in\{10,20\}$. Because $J$ grows with $K$
in design (iii), and equal loadings give coinciding signal eigenvalues,
these designs fall outside the fixed-$J$, distinct-spike setting of
Proposition~\ref{prop:stein}, and their results are empirical. For each design,
Tables~\ref{tab:supp:equal:S2}--\ref{tab:supp:highdim:rKL} report the
squared log-scale error $S_2$ of the residual scale and the forward and
reverse KL divergences. The two DGJ shrinkers use the PLY residual
variance, so their $S_2$ coincides with that of PLY and is not
reported. The column Reg.\ gives the percentage of data sets with a
regular $J$-component MML87 solution.

The results are consistent across the three designs. We summarise them
over the $77$ settings in which a regular solution exists for at least
$90\%$ of the data sets.
\begin{enumerate}
\item \emph{Residual variance.} MML87 has smaller $|S_1|$ and $S_2$ than
maximum likelihood in all $77$ settings. PLY has smaller $S_2$ than
MML87 in all $77$, and BC in $66$. The ratio of the
MML87 to the PLY squared error is at most about two for $K=10$, and
reaches about eight in the high-dimensional design when $J=K/4$. This
reflects the positive MML87 bias of order $J^2/\{N(K-J)\}$ in
Theorem~\ref{thm:mml:bias}.
\item \emph{Forward KL divergence (entropy loss).} PLY or BC gives the
smallest forward divergence in $79$ of the $90$ settings. MML87 is
smallest in $8$ weak-signal settings and DGJ-E in $3$ high-dimensional ones. In the regular settings, the MML87 forward
divergence exceeds that of PLY by $2\%$--$34\%$, because MML87 shrinks
the fitted signal eigenvalues.
\item \emph{Reverse KL divergence (Stein loss).} MML87 gives the
smallest reverse divergence in $18$ of $27$, $19$ of $27$ and $29$ of
$36$ settings in the three designs, respectively, and DGJ-S gives the
smallest in the remainder. In the regular settings, the MML87 reverse
divergence differs from that of DGJ-S by a median of $0.4\%$, $0.4\%$
and $0.14\%$, and by at most $7.1\%$, $4.9\%$ and $1.3\%$. This agrees
with Proposition~\ref{prop:stein}, which shows that the MML87
shrinker approaches the Stein-optimal shrinker as the signal strength
grows.
\item \emph{Weak signals.} When a regular solution often does not exist,
all estimators are fitted at a lower rank. In several of these settings
MML87 gives the smallest forward divergence.
\end{enumerate}
In summary, existing bias corrections estimate the residual variance
better than MML87 at a fixed rank, and they are preferable under
entropy loss. MML87 attains close to the optimal Stein loss without
being designed for it, and it also provides a principled rule for
selecting the latent dimension.

\begin{table}[p]
\scriptsize
\begin{center}
\caption{Squared log-scale error $S_2$ of the residual scale at a matched rank, averaged over $10^5$ simulated data sets with equal per-component signal-to-noise ratio $\rho$ and $K=10$. Bold marks the smallest value in each row. The largest Monte Carlo standard error is $0.00027$.}
\label{tab:supp:equal:S2}
\begin{tabular}{cccccccc}
\toprule
$N$ & $\rho$ & $J$ & Reg.\ (\%) & MLE & MML87 & BC & PLY \\
\cmidrule{1-8}
25 & 2 & 1 & 45 & 0.0060 & {\bf 0.0054} & 0.0056 & 0.0055 \\
 &  & 2 & 10 & {\bf 0.0123} & 0.0155 & 0.0148 & 0.0149 \\
 &  & 4 & 0 & {\bf 0.0558} & 0.0699 & 0.0672 & 0.0680 \\
 & 5 & 1 & 97 & 0.0050 & 0.0029 & 0.0029 & {\bf 0.0028} \\
 &  & 2 & 84 & 0.0096 & 0.0062 & 0.0056 & {\bf 0.0055} \\
 &  & 4 & 17 & {\bf 0.0646} & 0.0941 & 0.0784 & 0.0829 \\
 & 20 & 1 & 100 & 0.0045 & 0.0025 & 0.0025 & {\bf 0.0024} \\
 &  & 2 & 100 & 0.0077 & 0.0030 & 0.0030 & {\bf 0.0029} \\
 &  & 4 & 100 & 0.0189 & 0.0108 & {\bf 0.0050} & 0.0052 \\
\midrule
50 & 2 & 1 & 91 & 0.0022 & 0.0016 & {\bf 0.0015} & {\bf 0.0015} \\
 &  & 2 & 69 & 0.0039 & 0.0034 & {\bf 0.0031} & 0.0032 \\
 &  & 4 & 12 & {\bf 0.0208} & 0.0294 & 0.0254 & 0.0268 \\
 & 5 & 1 & 100 & 0.0017 & {\bf 0.0012} & {\bf 0.0012} & {\bf 0.0012} \\
 &  & 2 & 100 & 0.0027 & {\bf 0.0014} & {\bf 0.0014} & {\bf 0.0014} \\
 &  & 4 & 99 & 0.0060 & 0.0038 & {\bf 0.0022} & 0.0027 \\
 & 20 & 1 & 100 & 0.0016 & {\bf 0.0012} & {\bf 0.0012} & {\bf 0.0012} \\
 &  & 2 & 100 & 0.0024 & 0.0014 & {\bf 0.0013} & {\bf 0.0013} \\
 &  & 4 & 100 & 0.0051 & 0.0028 & {\bf 0.0019} & 0.0023 \\
\midrule
100 & 2 & 1 & 100 & 0.0007 & {\bf 0.0006} & {\bf 0.0006} & {\bf 0.0006} \\
 &  & 2 & 100 & 0.0011 & {\bf 0.0007} & {\bf 0.0007} & {\bf 0.0007} \\
 &  & 4 & 95 & 0.0024 & 0.0018 & {\bf 0.0012} & 0.0016 \\
 & 5 & 1 & 100 & 0.0007 & {\bf 0.0006} & {\bf 0.0006} & {\bf 0.0006} \\
 &  & 2 & 100 & 0.0010 & 0.0007 & {\bf 0.0006} & 0.0007 \\
 &  & 4 & 100 & 0.0018 & 0.0012 & {\bf 0.0009} & 0.0011 \\
 & 20 & 1 & 100 & 0.0007 & {\bf 0.0006} & {\bf 0.0006} & {\bf 0.0006} \\
 &  & 2 & 100 & 0.0009 & {\bf 0.0006} & {\bf 0.0006} & {\bf 0.0006} \\
 &  & 4 & 100 & 0.0016 & 0.0011 & {\bf 0.0009} & 0.0010 \\
\bottomrule
\end{tabular}

\end{center}
\end{table}

\begin{table}[p]
\scriptsize
\begin{center}
\caption{Forward KL divergence (half the entropy loss) at a matched rank, averaged over $10^5$ simulated data sets with equal per-component signal-to-noise ratio $\rho$ and $K=10$. Bold marks the smallest value in each row. The largest Monte Carlo standard error is $0.0018$.}
\label{tab:supp:equal:KL}
\begin{tabular}{cccccccccc}
\toprule
$N$ & $\rho$ & $J$ & Reg.\ (\%) & MLE & MML87 & BC & PLY & DGJ-E & DGJ-S \\
\cmidrule{1-10}
25 & 2 & 1 & 45 & 0.339 & {\bf 0.284} & 0.305 & 0.296 & 0.289 & 0.287 \\
 &  & 2 & 10 & 0.617 & {\bf 0.521} & 0.562 & 0.546 & 0.536 & 0.529 \\
 &  & 4 & 0 & 0.848 & {\bf 0.759} & 0.803 & 0.788 & 0.778 & 0.768 \\
 & 5 & 1 & 97 & 0.304 & 0.291 & 0.260 & {\bf 0.259} & 0.264 & 0.297 \\
 &  & 2 & 84 & 0.683 & 0.622 & 0.569 & {\bf 0.560} & 0.575 & 0.644 \\
 &  & 4 & 17 & 1.876 & {\bf 1.522} & 1.623 & 1.553 & 1.555 & 1.595 \\
 & 20 & 1 & 100 & 0.289 & 0.297 & 0.250 & {\bf 0.249} & 0.251 & 0.300 \\
 &  & 2 & 100 & 0.578 & 0.589 & 0.485 & {\bf 0.475} & 0.483 & 0.604 \\
 &  & 4 & 100 & 1.243 & 1.247 & 0.998 & {\bf 0.939} & 0.980 & 1.322 \\
\midrule
50 & 2 & 1 & 91 & 0.146 & 0.131 & 0.132 & {\bf 0.128} & 0.129 & 0.135 \\
 &  & 2 & 69 & 0.309 & 0.270 & 0.276 & {\bf 0.267} & 0.270 & 0.281 \\
 &  & 4 & 12 & 0.700 & {\bf 0.586} & 0.634 & 0.605 & 0.603 & 0.607 \\
 & 5 & 1 & 100 & 0.124 & 0.124 & {\bf 0.114} & {\bf 0.114} & 0.116 & 0.128 \\
 &  & 2 & 100 & 0.237 & 0.238 & {\bf 0.212} & 0.213 & 0.220 & 0.250 \\
 &  & 4 & 99 & 0.467 & 0.469 & {\bf 0.401} & 0.404 & 0.438 & 0.520 \\
 & 20 & 1 & 100 & 0.126 & 0.129 & {\bf 0.117} & {\bf 0.117} & {\bf 0.117} & 0.130 \\
 &  & 2 & 100 & 0.240 & 0.246 & 0.219 & {\bf 0.217} & 0.218 & 0.250 \\
 &  & 4 & 100 & 0.461 & 0.474 & 0.410 & {\bf 0.398} & 0.404 & 0.488 \\
\midrule
100 & 2 & 1 & 100 & 0.059 & 0.057 & {\bf 0.055} & {\bf 0.055} & 0.056 & 0.059 \\
 &  & 2 & 100 & 0.110 & 0.106 & {\bf 0.101} & 0.102 & 0.106 & 0.114 \\
 &  & 4 & 95 & 0.213 & 0.207 & {\bf 0.190} & 0.196 & 0.219 & 0.241 \\
 & 5 & 1 & 100 & 0.058 & 0.059 & {\bf 0.056} & {\bf 0.056} & {\bf 0.056} & 0.060 \\
 &  & 2 & 100 & 0.109 & 0.109 & {\bf 0.103} & {\bf 0.103} & 0.104 & 0.112 \\
 &  & 4 & 100 & 0.200 & 0.202 & {\bf 0.185} & {\bf 0.185} & 0.191 & 0.214 \\
 & 20 & 1 & 100 & 0.059 & 0.060 & {\bf 0.057} & {\bf 0.057} & {\bf 0.057} & 0.060 \\
 &  & 2 & 100 & 0.110 & 0.111 & {\bf 0.104} & {\bf 0.104} & {\bf 0.104} & 0.112 \\
 &  & 4 & 100 & 0.201 & 0.205 & 0.189 & {\bf 0.186} & 0.187 & 0.208 \\
\bottomrule
\end{tabular}

\end{center}
\end{table}

\begin{table}[p]
\scriptsize
\begin{center}
\caption{Reverse KL divergence (half the Stein loss) at a matched rank, averaged over $10^5$ simulated data sets with equal per-component signal-to-noise ratio $\rho$ and $K=10$. Bold marks the smallest value in each row. The largest Monte Carlo standard error is $0.0022$.}
\label{tab:supp:equal:rKL}
\begin{tabular}{cccccccccc}
\toprule
$N$ & $\rho$ & $J$ & Reg.\ (\%) & MLE & MML87 & BC & PLY & DGJ-E & DGJ-S \\
\cmidrule{1-10}
25 & 2 & 1 & 45 & 0.255 & {\bf 0.192} & 0.245 & 0.227 & 0.210 & 0.193 \\
 &  & 2 & 10 & 0.458 & 0.386 & 0.458 & 0.434 & 0.409 & {\bf 0.384} \\
 &  & 4 & 0 & 0.807 & 0.793 & 0.848 & 0.829 & 0.809 & {\bf 0.787} \\
 & 5 & 1 & 97 & 0.251 & 0.190 & 0.231 & 0.223 & 0.211 & {\bf 0.189} \\
 &  & 2 & 84 & 0.482 & 0.382 & 0.444 & 0.431 & 0.413 & {\bf 0.379} \\
 &  & 4 & 17 & 1.319 & 1.341 & 1.375 & 1.368 & 1.335 & {\bf 1.282} \\
 & 20 & 1 & 100 & 0.239 & {\bf 0.197} & 0.222 & 0.227 & 0.223 & 0.198 \\
 &  & 2 & 100 & 0.440 & {\bf 0.368} & 0.405 & 0.413 & 0.408 & 0.369 \\
 &  & 4 & 100 & 0.800 & 0.737 & 0.722 & 0.740 & 0.740 & {\bf 0.711} \\
\midrule
50 & 2 & 1 & 91 & 0.130 & {\bf 0.100} & 0.123 & 0.113 & 0.105 & {\bf 0.100} \\
 &  & 2 & 69 & 0.251 & 0.200 & 0.238 & 0.220 & 0.208 & {\bf 0.199} \\
 &  & 4 & 12 & 0.558 & 0.504 & 0.557 & 0.530 & 0.511 & {\bf 0.496} \\
 & 5 & 1 & 100 & 0.116 & {\bf 0.100} & 0.111 & 0.109 & 0.106 & {\bf 0.100} \\
 &  & 2 & 100 & 0.212 & {\bf 0.184} & 0.200 & 0.198 & 0.195 & 0.187 \\
 &  & 4 & 99 & 0.378 & {\bf 0.349} & 0.351 & 0.356 & 0.359 & 0.359 \\
 & 20 & 1 & 100 & 0.115 & {\bf 0.104} & 0.111 & 0.112 & 0.111 & {\bf 0.104} \\
 &  & 2 & 100 & 0.210 & {\bf 0.191} & 0.201 & 0.203 & 0.202 & 0.192 \\
 &  & 4 & 100 & 0.373 & {\bf 0.354} & 0.355 & 0.361 & 0.361 & 0.355 \\
\midrule
100 & 2 & 1 & 100 & 0.058 & {\bf 0.050} & 0.056 & 0.054 & 0.052 & {\bf 0.050} \\
 &  & 2 & 100 & 0.106 & {\bf 0.092} & 0.101 & 0.097 & 0.096 & 0.095 \\
 &  & 4 & 95 & 0.190 & {\bf 0.174} & 0.179 & 0.179 & 0.185 & 0.188 \\
 & 5 & 1 & 100 & 0.056 & {\bf 0.052} & 0.055 & 0.055 & 0.054 & {\bf 0.052} \\
 &  & 2 & 100 & 0.103 & {\bf 0.096} & 0.100 & 0.099 & 0.099 & {\bf 0.096} \\
 &  & 4 & 100 & 0.181 & {\bf 0.173} & 0.175 & 0.176 & 0.177 & 0.177 \\
 & 20 & 1 & 100 & 0.056 & {\bf 0.053} & 0.055 & 0.055 & 0.055 & {\bf 0.053} \\
 &  & 2 & 100 & 0.102 & {\bf 0.098} & 0.100 & 0.101 & 0.101 & {\bf 0.098} \\
 &  & 4 & 100 & 0.181 & {\bf 0.176} & {\bf 0.176} & 0.178 & 0.178 & {\bf 0.176} \\
\bottomrule
\end{tabular}

\end{center}
\end{table}

\begin{table}[p]
\scriptsize
\begin{center}
\caption{Squared log-scale error $S_2$ of the residual scale at a matched rank, averaged over $10^5$ simulated data sets with geometric loadings with weakest-component signal-to-noise ratio $\rho$, ratio $0.75$, and $K=10$. Bold marks the smallest value in each row. The largest Monte Carlo standard error is $0.00018$.}
\label{tab:supp:geometric:S2}
\begin{tabular}{cccccccc}
\toprule
$N$ & $\rho$ & $J$ & Reg.\ (\%) & MLE & MML87 & BC & PLY \\
\cmidrule{1-8}
25 & 2 & 1 & 45 & 0.0060 & {\bf 0.0054} & 0.0056 & 0.0055 \\
 &  & 2 & 16 & {\bf 0.0121} & 0.0149 & 0.0140 & 0.0141 \\
 &  & 4 & 0 & {\bf 0.0592} & 0.0849 & 0.0755 & 0.0786 \\
 & 5 & 1 & 97 & 0.0050 & 0.0029 & 0.0029 & {\bf 0.0028} \\
 &  & 2 & 90 & 0.0091 & 0.0051 & 0.0046 & {\bf 0.0045} \\
 &  & 4 & 53 & 0.0288 & 0.0433 & {\bf 0.0287} & 0.0322 \\
 & 20 & 1 & 100 & 0.0045 & 0.0025 & 0.0025 & {\bf 0.0024} \\
 &  & 2 & 100 & 0.0076 & 0.0030 & 0.0030 & {\bf 0.0029} \\
 &  & 4 & 100 & 0.0185 & 0.0099 & {\bf 0.0049} & 0.0050 \\
\midrule
50 & 2 & 1 & 91 & 0.0022 & 0.0016 & {\bf 0.0015} & {\bf 0.0015} \\
 &  & 2 & 80 & 0.0035 & 0.0027 & {\bf 0.0024} & 0.0025 \\
 &  & 4 & 51 & {\bf 0.0086} & 0.0123 & 0.0087 & 0.0100 \\
 & 5 & 1 & 100 & 0.0017 & {\bf 0.0012} & {\bf 0.0012} & {\bf 0.0012} \\
 &  & 2 & 100 & 0.0026 & {\bf 0.0014} & {\bf 0.0014} & {\bf 0.0014} \\
 &  & 4 & 100 & 0.0056 & 0.0032 & {\bf 0.0020} & 0.0024 \\
 & 20 & 1 & 100 & 0.0016 & {\bf 0.0012} & {\bf 0.0012} & {\bf 0.0012} \\
 &  & 2 & 100 & 0.0024 & 0.0014 & {\bf 0.0013} & {\bf 0.0013} \\
 &  & 4 & 100 & 0.0050 & 0.0028 & {\bf 0.0019} & 0.0023 \\
\midrule
100 & 2 & 1 & 100 & 0.0007 & {\bf 0.0006} & {\bf 0.0006} & {\bf 0.0006} \\
 &  & 2 & 100 & 0.0011 & {\bf 0.0007} & {\bf 0.0007} & {\bf 0.0007} \\
 &  & 4 & 100 & 0.0020 & 0.0013 & {\bf 0.0009} & 0.0012 \\
 & 5 & 1 & 100 & 0.0007 & {\bf 0.0006} & {\bf 0.0006} & {\bf 0.0006} \\
 &  & 2 & 100 & 0.0010 & 0.0007 & {\bf 0.0006} & 0.0007 \\
 &  & 4 & 100 & 0.0017 & 0.0011 & {\bf 0.0009} & 0.0011 \\
 & 20 & 1 & 100 & 0.0007 & {\bf 0.0006} & {\bf 0.0006} & {\bf 0.0006} \\
 &  & 2 & 100 & 0.0009 & {\bf 0.0006} & {\bf 0.0006} & {\bf 0.0006} \\
 &  & 4 & 100 & 0.0016 & 0.0011 & {\bf 0.0009} & 0.0010 \\
\bottomrule
\end{tabular}

\end{center}
\end{table}

\begin{table}[p]
\scriptsize
\begin{center}
\caption{Forward KL divergence (half the entropy loss) at a matched rank, averaged over $10^5$ simulated data sets with geometric loadings with weakest-component signal-to-noise ratio $\rho$, ratio $0.75$, and $K=10$. Bold marks the smallest value in each row. The largest Monte Carlo standard error is $0.0025$.}
\label{tab:supp:geometric:KL}
\begin{tabular}{cccccccccc}
\toprule
$N$ & $\rho$ & $J$ & Reg.\ (\%) & MLE & MML87 & BC & PLY & DGJ-E & DGJ-S \\
\cmidrule{1-10}
25 & 2 & 1 & 45 & 0.339 & {\bf 0.284} & 0.305 & 0.296 & 0.289 & 0.287 \\
 &  & 2 & 16 & 0.678 & {\bf 0.570} & 0.610 & 0.592 & 0.582 & 0.580 \\
 &  & 4 & 0 & 1.343 & {\bf 1.148} & 1.221 & 1.187 & 1.176 & 1.175 \\
 & 5 & 1 & 97 & 0.304 & 0.291 & 0.260 & {\bf 0.259} & 0.264 & 0.297 \\
 &  & 2 & 90 & 0.648 & 0.607 & 0.536 & {\bf 0.530} & 0.548 & 0.630 \\
 &  & 4 & 53 & 1.726 & 1.460 & 1.411 & {\bf 1.341} & 1.383 & 1.558 \\
 & 20 & 1 & 100 & 0.289 & 0.297 & 0.250 & {\bf 0.249} & 0.251 & 0.300 \\
 &  & 2 & 100 & 0.579 & 0.591 & 0.487 & {\bf 0.476} & 0.483 & 0.604 \\
 &  & 4 & 100 & 1.241 & 1.249 & 1.004 & {\bf 0.935} & 0.962 & 1.310 \\
\midrule
50 & 2 & 1 & 91 & 0.146 & 0.131 & 0.132 & {\bf 0.128} & 0.129 & 0.135 \\
 &  & 2 & 80 & 0.288 & 0.260 & 0.255 & {\bf 0.249} & 0.256 & 0.273 \\
 &  & 4 & 51 & 0.622 & 0.552 & 0.539 & {\bf 0.525} & 0.552 & 0.598 \\
 & 5 & 1 & 100 & 0.124 & 0.124 & {\bf 0.114} & {\bf 0.114} & 0.116 & 0.128 \\
 &  & 2 & 100 & 0.237 & 0.239 & {\bf 0.213} & {\bf 0.213} & 0.219 & 0.249 \\
 &  & 4 & 100 & 0.459 & 0.467 & 0.400 & {\bf 0.397} & 0.418 & 0.502 \\
 & 20 & 1 & 100 & 0.126 & 0.129 & {\bf 0.117} & {\bf 0.117} & {\bf 0.117} & 0.130 \\
 &  & 2 & 100 & 0.240 & 0.247 & 0.219 & {\bf 0.217} & 0.218 & 0.250 \\
 &  & 4 & 100 & 0.462 & 0.476 & 0.412 & {\bf 0.398} & 0.402 & 0.486 \\
\midrule
100 & 2 & 1 & 100 & 0.059 & 0.057 & {\bf 0.055} & {\bf 0.055} & 0.056 & 0.059 \\
 &  & 2 & 100 & 0.109 & 0.107 & {\bf 0.101} & 0.102 & 0.106 & 0.114 \\
 &  & 4 & 100 & 0.201 & 0.201 & {\bf 0.183} & 0.186 & 0.200 & 0.223 \\
 & 5 & 1 & 100 & 0.058 & 0.059 & {\bf 0.056} & {\bf 0.056} & {\bf 0.056} & 0.060 \\
 &  & 2 & 100 & 0.109 & 0.109 & {\bf 0.103} & {\bf 0.103} & 0.104 & 0.112 \\
 &  & 4 & 100 & 0.200 & 0.203 & 0.186 & {\bf 0.185} & 0.189 & 0.211 \\
 & 20 & 1 & 100 & 0.059 & 0.060 & {\bf 0.057} & {\bf 0.057} & {\bf 0.057} & 0.060 \\
 &  & 2 & 100 & 0.110 & 0.112 & 0.105 & {\bf 0.104} & {\bf 0.104} & 0.112 \\
 &  & 4 & 100 & 0.201 & 0.205 & 0.189 & {\bf 0.186} & 0.187 & 0.208 \\
\bottomrule
\end{tabular}

\end{center}
\end{table}

\begin{table}[p]
\scriptsize
\begin{center}
\caption{Reverse KL divergence (half the Stein loss) at a matched rank, averaged over $10^5$ simulated data sets with geometric loadings with weakest-component signal-to-noise ratio $\rho$, ratio $0.75$, and $K=10$. Bold marks the smallest value in each row. The largest Monte Carlo standard error is $0.0016$.}
\label{tab:supp:geometric:rKL}
\begin{tabular}{cccccccccc}
\toprule
$N$ & $\rho$ & $J$ & Reg.\ (\%) & MLE & MML87 & BC & PLY & DGJ-E & DGJ-S \\
\cmidrule{1-10}
25 & 2 & 1 & 45 & 0.255 & {\bf 0.192} & 0.245 & 0.227 & 0.210 & 0.193 \\
 &  & 2 & 16 & 0.489 & 0.405 & 0.484 & 0.458 & 0.431 & {\bf 0.402} \\
 &  & 4 & 0 & 1.091 & 1.085 & 1.151 & 1.131 & 1.097 & {\bf 1.057} \\
 & 5 & 1 & 97 & 0.251 & 0.190 & 0.231 & 0.223 & 0.211 & {\bf 0.189} \\
 &  & 2 & 90 & 0.469 & 0.373 & 0.429 & 0.420 & 0.404 & {\bf 0.371} \\
 &  & 4 & 53 & 0.986 & 0.981 & 0.968 & 0.981 & 0.968 & {\bf 0.926} \\
 & 20 & 1 & 100 & 0.239 & {\bf 0.197} & 0.222 & 0.227 & 0.223 & 0.198 \\
 &  & 2 & 100 & 0.440 & {\bf 0.369} & 0.405 & 0.414 & 0.410 & 0.370 \\
 &  & 4 & 100 & 0.799 & 0.736 & 0.724 & 0.744 & 0.743 & {\bf 0.711} \\
\midrule
50 & 2 & 1 & 91 & 0.130 & {\bf 0.100} & 0.123 & 0.113 & 0.105 & {\bf 0.100} \\
 &  & 2 & 80 & 0.239 & {\bf 0.193} & 0.225 & 0.211 & 0.202 & 0.194 \\
 &  & 4 & 51 & 0.455 & {\bf 0.414} & 0.436 & 0.428 & 0.424 & 0.417 \\
 & 5 & 1 & 100 & 0.116 & {\bf 0.100} & 0.111 & 0.109 & 0.106 & {\bf 0.100} \\
 &  & 2 & 100 & 0.211 & {\bf 0.185} & 0.200 & 0.199 & 0.196 & 0.187 \\
 &  & 4 & 100 & 0.374 & {\bf 0.349} & 0.351 & 0.356 & 0.358 & 0.355 \\
 & 20 & 1 & 100 & 0.115 & {\bf 0.104} & 0.111 & 0.112 & 0.111 & {\bf 0.104} \\
 &  & 2 & 100 & 0.210 & {\bf 0.192} & 0.201 & 0.204 & 0.203 & {\bf 0.192} \\
 &  & 4 & 100 & 0.373 & 0.356 & {\bf 0.355} & 0.362 & 0.362 & {\bf 0.355} \\
\midrule
100 & 2 & 1 & 100 & 0.058 & {\bf 0.050} & 0.056 & 0.054 & 0.052 & {\bf 0.050} \\
 &  & 2 & 100 & 0.105 & {\bf 0.093} & 0.100 & 0.098 & 0.096 & 0.095 \\
 &  & 4 & 100 & 0.183 & {\bf 0.171} & 0.174 & 0.175 & 0.178 & 0.180 \\
 & 5 & 1 & 100 & 0.056 & {\bf 0.052} & 0.055 & 0.055 & 0.054 & {\bf 0.052} \\
 &  & 2 & 100 & 0.103 & {\bf 0.096} & 0.100 & 0.100 & 0.099 & 0.097 \\
 &  & 4 & 100 & 0.181 & {\bf 0.174} & 0.175 & 0.177 & 0.177 & 0.177 \\
 & 20 & 1 & 100 & 0.056 & {\bf 0.053} & 0.055 & 0.055 & 0.055 & {\bf 0.053} \\
 &  & 2 & 100 & 0.102 & {\bf 0.098} & 0.100 & 0.101 & 0.101 & {\bf 0.098} \\
 &  & 4 & 100 & 0.181 & {\bf 0.176} & {\bf 0.176} & 0.178 & 0.178 & {\bf 0.176} \\
\bottomrule
\end{tabular}

\end{center}
\end{table}

\begin{table}[p]
\scriptsize
\renewcommand{\arraystretch}{0.8}
\begin{center}
\caption{Squared log-scale error $S_2$ of the residual scale at a matched rank, averaged over $10^5$ simulated data sets with equal per-component signal-to-noise ratio $\rho$ and $N\in\{2K,10K\}$. Bold marks the smallest value in each row. The largest Monte Carlo standard error is $0.000019$.}
\label{tab:supp:highdim:S2}
\begin{tabular}{ccccccccc}
\toprule
$K$ & $N$ & $\rho$ & $J$ & Reg.\ (\%) & MLE & MML87 & BC & PLY \\
\cmidrule{1-9}
20 & 40 & 10 & 1 & 100 & 0.0015 & {\bf 0.0007} & {\bf 0.0007} & {\bf 0.0007} \\
 &  &  & 2 & 100 & 0.0026 & {\bf 0.0008} & {\bf 0.0008} & {\bf 0.0008} \\
 &  &  & 5 & 98 & 0.0090 & 0.0026 & {\bf 0.0013} & {\bf 0.0013} \\
 &  & 20 & 1 & 100 & 0.0014 & {\bf 0.0007} & {\bf 0.0007} & {\bf 0.0007} \\
 &  &  & 2 & 100 & 0.0024 & {\bf 0.0008} & {\bf 0.0008} & {\bf 0.0008} \\
 &  &  & 5 & 100 & 0.0084 & 0.0019 & 0.0012 & {\bf 0.0011} \\
 & 200 & 5 & 1 & 100 & 0.0002 & {\bf 0.0001} & {\bf 0.0001} & {\bf 0.0001} \\
 &  &  & 2 & 100 & 0.0002 & {\bf 0.0001} & {\bf 0.0001} & {\bf 0.0001} \\
 &  &  & 5 & 100 & 0.0005 & {\bf 0.0002} & {\bf 0.0002} & {\bf 0.0002} \\
 &  & 20 & 1 & 100 & 0.0002 & {\bf 0.0001} & {\bf 0.0001} & {\bf 0.0001} \\
 &  &  & 2 & 100 & 0.0002 & {\bf 0.0001} & {\bf 0.0001} & {\bf 0.0001} \\
 &  &  & 5 & 100 & 0.0004 & {\bf 0.0002} & {\bf 0.0002} & {\bf 0.0002} \\
\midrule
50 & 100 & 10 & 1 & 100 & 0.0002 & {\bf 0.0001} & {\bf 0.0001} & {\bf 0.0001} \\
 &  &  & 5 & 100 & 0.0013 & {\bf 0.0001} & {\bf 0.0001} & {\bf 0.0001} \\
 &  &  & 13 & 100 & 0.0071 & 0.0016 & {\bf 0.0003} & {\bf 0.0003} \\
 &  & 20 & 1 & 100 & 0.0002 & {\bf 0.0001} & {\bf 0.0001} & {\bf 0.0001} \\
 &  &  & 5 & 100 & 0.0012 & {\bf 0.0001} & {\bf 0.0001} & {\bf 0.0001} \\
 &  &  & 13 & 100 & 0.0065 & 0.0012 & {\bf 0.0003} & {\bf 0.0003} \\
 & 500 & 5 & 1 & 100 & {\bf 0.0000} & {\bf 0.0000} & {\bf 0.0000} & {\bf 0.0000} \\
 &  &  & 5 & 100 & 0.0001 & {\bf 0.0000} & {\bf 0.0000} & {\bf 0.0000} \\
 &  &  & 13 & 100 & 0.0003 & 0.0001 & {\bf 0.0000} & 0.0001 \\
 &  & 20 & 1 & 100 & {\bf 0.0000} & {\bf 0.0000} & {\bf 0.0000} & {\bf 0.0000} \\
 &  &  & 5 & 100 & 0.0001 & {\bf 0.0000} & {\bf 0.0000} & {\bf 0.0000} \\
 &  &  & 13 & 100 & 0.0002 & 0.0001 & {\bf 0.0000} & {\bf 0.0000} \\
\midrule
100 & 200 & 10 & 1 & 100 & 0.0001 & {\bf 0.0000} & {\bf 0.0000} & {\bf 0.0000} \\
 &  &  & 10 & 100 & 0.0010 & 0.0001 & {\bf 0.0000} & {\bf 0.0000} \\
 &  &  & 25 & 100 & 0.0060 & 0.0012 & 0.0002 & {\bf 0.0001} \\
 &  & 20 & 1 & 100 & 0.0001 & {\bf 0.0000} & {\bf 0.0000} & {\bf 0.0000} \\
 &  &  & 10 & 100 & 0.0009 & {\bf 0.0000} & {\bf 0.0000} & {\bf 0.0000} \\
 &  &  & 25 & 100 & 0.0054 & 0.0008 & {\bf 0.0001} & {\bf 0.0001} \\
 & 1000 & 5 & 1 & 100 & {\bf 0.0000} & {\bf 0.0000} & {\bf 0.0000} & {\bf 0.0000} \\
 &  &  & 10 & 100 & {\bf 0.0000} & {\bf 0.0000} & {\bf 0.0000} & {\bf 0.0000} \\
 &  &  & 25 & 100 & 0.0003 & {\bf 0.0000} & {\bf 0.0000} & {\bf 0.0000} \\
 &  & 20 & 1 & 100 & {\bf 0.0000} & {\bf 0.0000} & {\bf 0.0000} & {\bf 0.0000} \\
 &  &  & 10 & 100 & {\bf 0.0000} & {\bf 0.0000} & {\bf 0.0000} & {\bf 0.0000} \\
 &  &  & 25 & 100 & 0.0002 & {\bf 0.0000} & {\bf 0.0000} & {\bf 0.0000} \\
\bottomrule
\end{tabular}

\end{center}
\end{table}

\begin{table}[p]
\scriptsize
\renewcommand{\arraystretch}{0.8}
\begin{center}
\caption{Forward KL divergence (half the entropy loss) at a matched rank, averaged over $10^5$ simulated data sets with equal per-component signal-to-noise ratio $\rho$ and $N\in\{2K,10K\}$. Bold marks the smallest value in each row. The largest Monte Carlo standard error is $0.0017$.}
\label{tab:supp:highdim:KL}
\begin{tabular}{ccccccccccc}
\toprule
$K$ & $N$ & $\rho$ & $J$ & Reg.\ (\%) & MLE & MML87 & BC & PLY & DGJ-E & DGJ-S \\
\cmidrule{1-11}
20 & 40 & 10 & 1 & 100 & 0.301 & 0.322 & {\bf 0.271} & 0.272 & 0.274 & 0.327 \\
 &  &  & 2 & 100 & 0.609 & 0.646 & {\bf 0.532} & {\bf 0.532} & 0.541 & 0.660 \\
 &  &  & 5 & 98 & 1.709 & 1.736 & 1.374 & {\bf 1.358} & 1.428 & 1.836 \\
 &  & 20 & 1 & 100 & 0.306 & 0.334 & {\bf 0.278} & {\bf 0.278} & 0.279 & 0.337 \\
 &  &  & 2 & 100 & 0.619 & 0.671 & 0.547 & {\bf 0.545} & 0.550 & 0.680 \\
 &  &  & 5 & 100 & 1.697 & 1.765 & 1.391 & {\bf 1.355} & 1.388 & 1.837 \\
 & 200 & 5 & 1 & 100 & 0.054 & 0.054 & 0.053 & {\bf 0.052} & 0.053 & 0.055 \\
 &  &  & 2 & 100 & 0.104 & 0.105 & {\bf 0.100} & {\bf 0.100} & 0.101 & 0.107 \\
 &  &  & 5 & 100 & 0.244 & 0.247 & {\bf 0.231} & 0.232 & 0.236 & 0.256 \\
 &  & 20 & 1 & 100 & 0.054 & 0.056 & {\bf 0.053} & {\bf 0.053} & {\bf 0.053} & 0.056 \\
 &  &  & 2 & 100 & 0.104 & 0.107 & {\bf 0.102} & {\bf 0.102} & {\bf 0.102} & 0.108 \\
 &  &  & 5 & 100 & 0.246 & 0.253 & 0.236 & {\bf 0.235} & {\bf 0.235} & 0.255 \\
\midrule
50 & 100 & 10 & 1 & 100 & 0.263 & 0.292 & {\bf 0.251} & {\bf 0.251} & {\bf 0.251} & 0.296 \\
 &  &  & 5 & 100 & 1.370 & 1.488 & 1.234 & {\bf 1.233} & 1.249 & 1.524 \\
 &  &  & 13 & 100 & 3.993 & 4.145 & 3.254 & {\bf 3.224} & 3.390 & 4.419 \\
 &  & 20 & 1 & 100 & 0.268 & 0.304 & {\bf 0.257} & {\bf 0.257} & {\bf 0.257} & 0.306 \\
 &  &  & 5 & 100 & 1.393 & 1.548 & 1.267 & {\bf 1.264} & 1.273 & 1.571 \\
 &  &  & 13 & 100 & 4.046 & 4.285 & 3.365 & {\bf 3.292} & 3.367 & 4.468 \\
 & 500 & 5 & 1 & 100 & 0.051 & 0.052 & 0.051 & {\bf 0.050} & {\bf 0.050} & 0.053 \\
 &  &  & 5 & 100 & 0.246 & 0.250 & {\bf 0.240} & {\bf 0.240} & 0.241 & 0.255 \\
 &  &  & 13 & 100 & 0.609 & 0.617 & {\bf 0.579} & {\bf 0.579} & 0.588 & 0.641 \\
 &  & 20 & 1 & 100 & {\bf 0.051} & 0.053 & {\bf 0.051} & {\bf 0.051} & {\bf 0.051} & 0.054 \\
 &  &  & 5 & 100 & 0.248 & 0.257 & {\bf 0.243} & {\bf 0.243} & 0.244 & 0.258 \\
 &  &  & 13 & 100 & 0.613 & 0.633 & 0.589 & {\bf 0.587} & 0.589 & 0.639 \\
\midrule
100 & 200 & 10 & 1 & 100 & 0.251 & 0.283 & 0.245 & {\bf 0.244} & {\bf 0.244} & 0.286 \\
 &  &  & 10 & 100 & 2.661 & 2.910 & 2.421 & {\bf 2.417} & 2.446 & 2.981 \\
 &  &  & 25 & 100 & 7.409 & 7.736 & 6.113 & {\bf 6.062} & 6.337 & 8.231 \\
 &  & 20 & 1 & 100 & 0.256 & 0.294 & {\bf 0.250} & {\bf 0.250} & {\bf 0.250} & 0.296 \\
 &  &  & 10 & 100 & 2.705 & 3.028 & 2.482 & {\bf 2.478} & 2.493 & 3.072 \\
 &  &  & 25 & 100 & 7.508 & 8.017 & 6.315 & {\bf 6.197} & 6.323 & 8.341 \\
 & 1000 & 5 & 1 & 100 & {\bf 0.050} & 0.051 & {\bf 0.050} & {\bf 0.050} & {\bf 0.050} & 0.052 \\
 &  &  & 10 & 100 & 0.486 & 0.494 & 0.475 & {\bf 0.474} & 0.476 & 0.505 \\
 &  &  & 25 & 100 & 1.163 & 1.178 & {\bf 1.108} & {\bf 1.108} & 1.124 & 1.222 \\
 &  & 20 & 1 & 100 & 0.051 & 0.053 & {\bf 0.050} & {\bf 0.050} & {\bf 0.050} & 0.053 \\
 &  &  & 10 & 100 & 0.490 & 0.508 & {\bf 0.481} & {\bf 0.481} & 0.482 & 0.511 \\
 &  &  & 25 & 100 & 1.170 & 1.209 & 1.128 & {\bf 1.123} & 1.127 & 1.221 \\
\bottomrule
\end{tabular}

\end{center}
\end{table}

\begin{table}[p]
\scriptsize
\renewcommand{\arraystretch}{0.8}
\begin{center}
\caption{Reverse KL divergence (half the Stein loss) at a matched rank, averaged over $10^5$ simulated data sets with equal per-component signal-to-noise ratio $\rho$ and $N\in\{2K,10K\}$. Bold marks the smallest value in each row. The largest Monte Carlo standard error is $0.0006$.}
\label{tab:supp:highdim:rKL}
\begin{tabular}{ccccccccccc}
\toprule
$K$ & $N$ & $\rho$ & $J$ & Reg.\ (\%) & MLE & MML87 & BC & PLY & DGJ-E & DGJ-S \\
\cmidrule{1-11}
20 & 40 & 10 & 1 & 100 & 0.278 & {\bf 0.209} & 0.263 & 0.257 & 0.246 & {\bf 0.209} \\
 &  &  & 2 & 100 & 0.534 & {\bf 0.402} & 0.500 & 0.489 & 0.469 & 0.403 \\
 &  &  & 5 & 98 & 1.261 & 0.996 & 1.139 & 1.120 & 1.092 & {\bf 0.990} \\
 &  & 20 & 1 & 100 & 0.276 & {\bf 0.216} & 0.263 & 0.263 & 0.257 & {\bf 0.216} \\
 &  &  & 2 & 100 & 0.531 & {\bf 0.418} & 0.501 & 0.502 & 0.491 & {\bf 0.418} \\
 &  &  & 5 & 100 & 1.251 & 1.016 & 1.143 & 1.147 & 1.131 & {\bf 1.012} \\
 & 200 & 5 & 1 & 100 & 0.053 & {\bf 0.049} & 0.053 & 0.052 & 0.051 & {\bf 0.049} \\
 &  &  & 2 & 100 & 0.102 & {\bf 0.094} & 0.100 & 0.099 & 0.097 & {\bf 0.094} \\
 &  &  & 5 & 100 & 0.233 & {\bf 0.216} & 0.227 & 0.225 & 0.223 & 0.219 \\
 &  & 20 & 1 & 100 & 0.053 & {\bf 0.050} & 0.053 & 0.053 & 0.052 & {\bf 0.050} \\
 &  &  & 2 & 100 & 0.101 & {\bf 0.096} & 0.100 & 0.100 & 0.100 & {\bf 0.096} \\
 &  &  & 5 & 100 & 0.232 & {\bf 0.221} & 0.228 & 0.228 & 0.228 & 0.222 \\
\midrule
50 & 100 & 10 & 1 & 100 & 0.262 & {\bf 0.195} & 0.256 & 0.246 & 0.234 & {\bf 0.195} \\
 &  &  & 5 & 100 & 1.259 & {\bf 0.945} & 1.202 & 1.155 & 1.106 & 0.948 \\
 &  &  & 13 & 100 & 3.114 & 2.449 & 2.842 & 2.753 & 2.683 & {\bf 2.444} \\
 &  & 20 & 1 & 100 & 0.261 & {\bf 0.203} & 0.255 & 0.251 & 0.246 & {\bf 0.203} \\
 &  &  & 5 & 100 & 1.252 & {\bf 0.982} & 1.203 & 1.186 & 1.161 & 0.983 \\
 &  &  & 13 & 100 & 3.098 & 2.520 & 2.865 & 2.837 & 2.796 & {\bf 2.510} \\
 & 500 & 5 & 1 & 100 & 0.052 & {\bf 0.047} & 0.051 & 0.050 & 0.049 & {\bf 0.047} \\
 &  &  & 5 & 100 & 0.244 & {\bf 0.224} & 0.241 & 0.236 & 0.232 & 0.225 \\
 &  &  & 13 & 100 & 0.584 & {\bf 0.542} & 0.570 & 0.563 & 0.558 & 0.550 \\
 &  & 20 & 1 & 100 & 0.051 & {\bf 0.048} & 0.051 & 0.051 & 0.051 & {\bf 0.048} \\
 &  &  & 5 & 100 & 0.243 & {\bf 0.230} & 0.241 & 0.240 & 0.239 & {\bf 0.230} \\
 &  &  & 13 & 100 & 0.582 & {\bf 0.556} & 0.573 & 0.573 & 0.571 & 0.558 \\
\midrule
100 & 200 & 10 & 1 & 100 & 0.257 & {\bf 0.191} & 0.254 & 0.242 & 0.230 & {\bf 0.191} \\
 &  &  & 10 & 100 & 2.482 & {\bf 1.862} & 2.385 & 2.278 & 2.180 & 1.867 \\
 &  &  & 25 & 100 & 5.929 & {\bf 4.632} & 5.452 & 5.252 & 5.106 & 4.634 \\
 &  & 20 & 1 & 100 & 0.256 & {\bf 0.199} & 0.253 & 0.248 & 0.242 & {\bf 0.199} \\
 &  &  & 10 & 100 & 2.468 & {\bf 1.935} & 2.385 & 2.340 & 2.289 & 1.937 \\
 &  &  & 25 & 100 & 5.900 & 4.776 & 5.492 & 5.412 & 5.330 & {\bf 4.765} \\
 & 1000 & 5 & 1 & 100 & 0.051 & {\bf 0.046} & 0.051 & 0.050 & 0.049 & {\bf 0.046} \\
 &  &  & 10 & 100 & 0.483 & {\bf 0.443} & 0.478 & 0.469 & 0.460 & 0.445 \\
 &  &  & 25 & 100 & 1.119 & {\bf 1.038} & 1.095 & 1.080 & 1.069 & 1.051 \\
 &  & 20 & 1 & 100 & 0.051 & {\bf 0.048} & 0.051 & 0.050 & 0.050 & {\bf 0.048} \\
 &  &  & 10 & 100 & 0.481 & {\bf 0.456} & 0.478 & 0.476 & 0.474 & {\bf 0.456} \\
 &  &  & 25 & 100 & 1.115 & {\bf 1.065} & 1.100 & 1.098 & 1.095 & 1.068 \\
\bottomrule
\end{tabular}

\end{center}
\end{table}

\clearpage

\bibliographystyle{IEEEtran}
\bibliography{bibliography}

\end{document}